\documentclass[11pt]{article}
\usepackage{axodraw}
\usepackage{epsfig}
\usepackage{amsfonts}
\usepackage{amsmath}
\usepackage{wasysym}
\usepackage{bbm}
\allowdisplaybreaks[1]

\input pix.sty

\renewcommand{\eq}{eq.~}
\renewcommand{\eqs}{eqs.~}
\renewcommand{\se}{sec.~}

\renewcommand{\fig}{fig.~}

\newcommand{\tinymsbar}{{\overline{\mbox{\tiny\rm{MS}}}}}
\newcommand{\Lambdamsbar}{{\Lambda_\tinymsbar}}

\newcommand{\Nc}{N_{\rm c}}

\newcommand{\Tc}{T_{\rm c}}
\newcommand{\gB}{g_\rmii{B}}
\newcommand{\mE}{m_\rmii{E}}

\newcommand{\gammaE}{\gamma_\rmii{E}}

\newcommand{\rmO}{{\mathcal{O}}}
\newcommand{\bmu}{\bar\mu}
\newcommand{\CA}{\Nc}

\def\lsi{\raise0.3ex\hbox{$<$\kern-0.75em\raise-1.1ex\hbox{$\sim$}}}
\def\gsi{\raise0.3ex\hbox{$>$\kern-0.75em\raise-1.1ex\hbox{$\sim$}}}
\newcommand{\lsim}{\apprle} 
\newcommand{\gsim}{\apprge} 

\newcommand{\nF}{n_\rmii{F}}
\newcommand{\nB}{n_\rmii{B}}
 \renewcommand{\nF}[1]{n_\rmii{F{#1}}}
 \renewcommand{\nB}[1]{n_\rmii{B{#1}}}
\newcommand{\rmii}[1]{{\mbox{\tiny\rm{#1}}}}
\newcommand{\re}{\mathop{\mbox{Re}}}
\newcommand{\im}{\mathop{\mbox{Im}}}

\newcommand{\Tint}[1]{{\hbox{$\sum$}\!\!\!\!\!\!\!\int\,}_{\!\!\!\!\raise-0.9ex\hbox{$\scriptstyle{#1}$}}}
\newcommand{\Tinti}[1]{{{\Sigma}\!\!\!\!\raise0.3ex\hbox{$\int$}_\rmii{${#1}$}}}

\newcommand{\unit}{{\mathbbm{1}}} 
\newcommand{\bi}{\begin{itemize}}
\newcommand{\ei}{\end{itemize}}

\newcommand{\hide}[1]{ }

\newcommand{\iv}{iii} 
\newcommand{\ii}{iv} 
\def\TAsc(#1,#2)(#3,#4,#5)%
{\SetWidth{2.0}\CArc(#1,#2)(#3,#4,#5)\SetWidth{1.0}}
\def\Lwidth{3}

\def\TAgl(#1,#2)(#3,#4,#5){\SetWidth{2.0}\PhotonArc(#1,#2)(#3,#4,#5){\Lwidth}%
{6.283 #3 mul 360 div #4 #5 sub #4 #5 sub mul sqrt mul Tdensity mul}%
\SetWidth{1.0}}
\def\TLgl(#1,#2)(#3,#4){\SetWidth{2.0}\Photon(#1,#2)(#3,#4){\Lwidth}
{#1 #3 sub #1 #3 sub mul #2 #4 sub #2 #4 sub mul add sqrt Tdensity mul}%
\SetWidth{1.0}}
\newcommand{\piC}[1]{\;\parbox[c]{40pt}{\begin{picture}(120,60)(0,-20)
\SetWidth{1.0}\SetScale{0.35} #1 \end{picture}}\;}
\def\ConnectedA(#1,#2,#3){\piC{#1(60,-15)(75,34,146) #2(60,75)(75,214,326)%
 #3(60,60)(20,190,350)%
 \GBoxc(0,30)(10,10){1} \GBoxc(120,30)(10,10){1}%
  }}
\def\ConnectedB(#1,#2,#3){\piC{#1(60,-15)(75,34,146) #2(60,75)(75,214,326)%
 #3(60,60)(60,0)%
 \GBoxc(0,30)(10,10){1} \GBoxc(120,30)(10,10){1}%
  }}
\def\ConnectedC(#1,#2){\piC{#1(60,-15)(75,34,146) #2(60,75)(75,214,326)%
 \GBoxc(0,30)(10,10){1} \GBoxc(120,30)(10,10){1}%
  }}
\def\ConnectedD(#1,#2){\piC{#1(60,-15)(75,34,146) #2(60,75)(75,214,326)%
 \GBoxc(0,30)(10,10){1} \GBoxc(120,30)(10,10){1}%
 \SetWidth{2.0} 
 \Line(55,55)(65,65)%
 \Line(55,65)(65,55)
  }}

\def\Lwidth{1.3}
\newcommand{\picu}[1]{\;\parbox[c]{60pt}{\begin{picture}(60,30)(0,0)
\SetWidth{1.0}\SetScale{1.0} #1 \end{picture}}\; }
\def\EleA{\picu{%
 \Agl(30,5)(22.3,27,153)%
 \Agl(30,25)(22.3,207,333)%
 \COval(10,15)(2,2)(0){Black}{Black}%
 \COval(50,15)(2,2)(0){Black}{Black}%
}}
\def\EleB{\picu{%
 \Agl(30,5)(22.3,27,153)%
 \Agl(30,25)(22.3,207,333)%
 \COval(10,15)(2,2)(0){Black}{Black}%
 \COval(50,15)(2,2)(0){Black}{Black}%
 \Agl(58,15)(8,0,360)%
}}
\def\EleC{\picu{%
 \Agl(30,5)(22.3,27,153)%
 \Agl(30,25)(22.3,207,333)%
 \COval(10,15)(2,2)(0){Black}{Black}%
 \COval(50,15)(2,2)(0){Black}{Black}%
 \Lgl(10,15)(50,15)%
}}
\def\EleD{\picu{%
 \Agl(30,5)(22.3,90,153)%
 \Agl(30,25)(22.3,207,333)%
 \COval(10,15)(2,2)(0){Black}{Black}%
 \COval(50,15)(2,2)(0){Black}{Black}%
 \Agl(43,27)(12,180,300)%
 \Agl(38,16)(12,0,120)%
}}
\def\EleE{\picu{%
 \Agl(30,5)(22.3,27,153)%
 \Agl(30,25)(22.3,207,333)%
 \COval(10,15)(2,2)(0){Black}{Black}%
 \COval(50,15)(2,2)(0){Black}{Black}%
 \GCirc(30,27.3){4}{0.5}
}}
\def\EleF{\picu{%
 \Agl(20,10)(11.15,27,153)%
 \Agl(20,20)(11.15,207,333)%
 \Agl(40,10)(11.15,27,153)%
 \Agl(40,20)(11.15,207,333)%
 \COval(10,15)(2,2)(0){Black}{Black}%
 \COval(50,15)(2,2)(0){Black}{Black}%
}}
\def\EleG{\picu{%
 \Agl(30,5)(22.3,27,153)%
 \Agl(30,25)(22.3,207,333)%
 \COval(10,15)(2,2)(0){Black}{Black}%
 \COval(50,15)(2,2)(0){Black}{Black}%
 \Lgl(30,2.7)(30,27.3)%
}}

\def\ScaR{\picu{%
 \Line(-20,30)(20,30)%
 \Line(20,30)(60,60)%
 \Line(20,30)(60,0)%
 \Line(40,45)(60,30)%
 \Text(-10,40)[c]{$(\omega,\vec{0})$}%
 \Text(65,60)[l]{$(q,+\vec{q})$}%
 \Text(65,0)[l]{$(r,-\vec{r})$}%
 \Text(26,47)[c]{$\vec{r}$}%
 \Text(65,30)[l]{$(E_{qr},\vec{r-q})$}%
 \SetWidth{0.5}%
 \LongArrow(20,35)(40,50)%
}}
\def\ScaV{\picu{%
 \Line(-20,30)(20,30)%
 \Line(20,30)(60,60)%
 \Line(20,30)(60,0)%
 \Line(40,45)(40,15)%
 \Text(-10,40)[c]{$(\omega,\vec{0})$}%
 \Text(65,60)[l]{$(q,+\vec{q})$}%
 \Text(65,0)[l]{$(q,-\vec{q})$}%
 \Text(26,47)[c]{$\vec{r}$}%
 \Text(26,13)[c]{$\vec{r}$}%
 \Text(50,30)[l]{$\vec{r-q}$}%
 \SetWidth{0.5}%
 \LongArrow(20,35)(40,50)%
 \LongArrow(45,43)(45,18)%
 \LongArrow(40,10)(20,25)%
}}

\makeatletter \@addtoreset{equation}{section} \makeatother
\renewcommand{\theequation}{\arabic{section}.\arabic{equation}}
\makeatletter
\renewcommand\section{\@startsection {section}{1}{\z@}%
                                   {-5.5ex \@plus -1ex \@minus -.2ex}
                                   {2.3ex \@plus.2ex}%
                                   {\normalfont\large\bfseries}}
\renewcommand\subsection{\@startsection{subsection}{2}{\z@}%
                                     {-3.25ex\@plus -1ex \@minus -.2ex}%
                                     {1.5ex \@plus .2ex}%
                                     {\normalfont\normalsize\bfseries}}
\renewcommand\thesection {\@arabic\c@section}
\renewcommand\thesubsection   {\thesection.\@arabic\c@subsection}
\renewcommand{\@seccntformat}[1]{%
\csname the#1\endcsname.\hspace{1.0em}}
\makeatother

\begin{document}

\begin{titlepage}
\begin{flushright}
BI-TP 2011/18\\
\vspace*{1cm}
\end{flushright}
\begin{centering}
\vfill

{\Large{\bf
 Next-to-leading order thermal spectral functions \\[2mm]
 in the perturbative domain
}} 

\vspace{0.8cm}

M.~Laine, A.~Vuorinen, Y.~Zhu

\vspace{0.8cm}

{\em
Faculty of Physics, University of Bielefeld, 
D-33501 Bielefeld, Germany\\}

\vspace*{0.8cm}

\mbox{\bf Abstract}
 
\end{centering}

\vspace*{0.3cm}
 
\noindent
Motivated by applications in thermal QCD and cosmology, 
we elaborate on a general method for computing next-to-leading order 
spectral functions for composite operators at vanishing spatial
momentum, accounting for real, virtual as well as thermal corrections. 
As an example, we compute these functions (together with 
the corresponding imaginary-time correlators which can be 
compared with lattice simulations) for scalar and 
pseudoscalar densities in pure Yang-Mills theory. Our results
may turn out to be helpful in non-perturbative 
estimates of the corresponding transport coefficients, 
which are the bulk viscosity in the scalar channel and the rate 
of anomalous chirality violation in the pseudoscalar channel.  
We also mention links to cosmology, although the most useful 
results in that context may come from a future generalization 
of our methods to other correlators.

\vfill

 
\vspace*{0.5cm}
  
\noindent
August 2011

\vfill

\end{titlepage}

%
\section{Introduction}
\la{se:intro}

A ``spectral function'', $\rho$, which has a Minkowskian four-momentum,
$\mathcal{P}$, as its argument, can formally be defined as a Fourier transform 
of the thermal expectation value 
($
\langle\ldots\rangle^{ }_T \equiv \frac{1}{\mathcal{Z}} 
\tr [ e^{-\beta\hat H}(\ldots) ]
$)
of the commutator of a given 
gauge-invariant local operator:
\be
 \rho(\mathcal{P}) \equiv 
 \int_{-\infty}^{\infty} \! {\rm d}t
 \int_\vec{x} \, e^{i \mathcal{P}\cdot\mathcal{X}}
 \Bigl\langle 
  \fr12 [\hat O(t,\vec{x}) \, , \, \hat O(0,\vec{0})]
 \Bigr\rangle^{ }_T
 \;, \la{rho_def}
\ee
where $\mathcal{X} \equiv (t,\vec{x})$, the operator
is defined in the Heisenberg picture, and 
$
 \mathcal{P}\cdot\mathcal{X} = \omega\, t - \vec{p}\cdot\vec{x}
$. 
All other orderings (retarded, time-ordered, 
Wightman, etc.)\ can be expressed in terms of the spectral 
function~\cite{leb,kg}. Therefore various spectral functions play
an important role in theoretical analyses of thermal systems.
In particular, in the ``linear response'' regime, 
the production rate of a  weakly-coupled particle
species from the medium is proportional to the spectral function 
of the operator that the particle in question couples 
to~\cite{leb,kg}.  

Because of the fundamental role of spectral functions, 
they make an appearance in many different areas of many-body physics; 
in the following, selected examples having a connection 
to our present study are outlined in some more detail. 

\subsection*{QCD at high temperatures}

Various ``transport coefficients'', which characterize
the relaxation of local excesses in conserved currents (momentum; 
number densities; electromagnetic current) to their equilibrium 
values, can be viewed as the ``low-energy constants'' of 
an interacting hot plasma (cf.\ e.g.\ ref.~\cite{ga_rev}). 
They can be defined 
through the limit $\lim_{\omega\to 0^+} \rho(\omega,\vec{0})/\omega$
of an appropriate spectral function.
In recent years considerable efforts have been devoted to 
determining these quantities at temperatures around 
a few hundred MeV, relevant for the current experimental
heavy ion collision program; an extensive review, 
with a particular perspective on non-perturbative lattice 
measurements, can be found in ref.~\cite{hbm_rev}.

Unfortunately, as becomes clear from ref.~\cite{hbm_rev}, 
a reliable non-perturbative determination of transport coefficients
represents a formidable challenge, even in the limit of high temperatures
where it can be argued that pure Yang-Mills theory should already yield 
a qualitatively correct description. The problem 
is that direct lattice measurements concern Euclidean correlators, 
whereas transport coefficients are related to the longest Minkowskian
time scales characterizing the system. Even though the two 
situations can in principle be related to each other through an analytic 
continuation~\cite{cuniberti}, a necessary requirement appears to be 
the subtraction of short-distance divergences~\cite{analytic}, 
and even then the problem remains numerically ill-posed  
(see e.g.\ ref.~\cite{hbm_rev} and references therein). 
It therefore seems pertinent to obtain as much information 
as possible by analytic means, in order to carry out 
a subtraction leaving over a simple function 
containing (mostly) only infrared physics. 
Thanks to asymptotic freedom, it is possible to address analytically 
not only the temperature-independent short-distance divergences but also, 
at high enough temperatures, thermal modifications originating from 
the large-frequency range $\omega \sim \pi T$, and this is one
of the goals of the present work.

\subsection*{QCD at low temperatures}

At low temperatures, where
chiral symmetry is spontaneously broken, numerical 
simulations are particularly demanding; on the other hand, 
chiral perturbation theory~\cite{gl1}, 
if pursued to a sufficient order, may provide
for a reasonably accurate description of this regime. Since mesons are bosons,
the analytic computations encountered
are technically not unlike those
at very high temperatures in pure Yang-Mills theory, and many of the 
same methods may turn out to be useful. This then serves
as an additional motivation for developing analytic methods. Recent computations 
of various transport coefficients within chiral perturbation theory
(or extensions thereof) can be 
found in refs.~\cite{had0}--\cite{had4}; 
we envisage that our techniques and generalizations thereof
may allow to address
some of the open issues outlined in ref.~\cite{had0}.

\subsection*{Cosmology}

In cosmology, particle production rates are relevant for 
certain Dark Matter scenarios, as well as for leptogenesis 
computations in which a non-thermal distribution of unstable 
particles serves as an intermediate stage. Typically, one 
is then interested in the whole non-equilibrium spectrum of the particles
produced, proportional to a spectral function evaluated at 
$\mathcal{P} = (E_p,\vec{p})$, where $\vec{p}$ is the 
spatial momentum and $E_p\equiv \sqrt{p^2+ M^2}$ 
(cf.\ e.g.\ ref.~\cite{nuMSM}).   
However, if the particles produced are non-relativistic, with 
a mass $M \gg T$, then it can be argued that the spectral
function at $\mathcal{P} = (M,\vec{0})$ already yields
some of the information. An example of a recent analysis
in this spirit, in which a subset of the 2-loop 
topologies studied in the present paper were computed
to leading non-trivial order in an expansion in $T^2/M^2$
(we refer to this approximation as the ``OPE-limit''), 
can be found in ref.~\cite{sls}. Since our techniques work beyond 
the OPE-limit, we expect them to allow for an improvement 
of such computations. A few other  
cosmological applications are suggested in the Conclusions. 

\subsection*{Previous work}

The very first spectral function computed at next-to-leading order (NLO) 
in the domain $\omega \sim \pi T$, $\vec{p}=\vec{0}$ was that related 
to the electromagnetic current associated with massless 
quarks~\cite{spectral1}--\cite{spectral3}. These results 
have subsequently been generalized to the case of a finite (heavy) quark 
mass~\cite{nlo}, and another spectral function related
to this case, correlating Lorentz forces felt by the nearly 
static heavy quarks, was derived at NLO with similar 
methods~\cite{rhoE}. A general analysis of many spectral functions
in the domain $\omega \gg \pi T$, which can be systematized
through Operator Product Expansion (OPE) techniques, 
was carried out in ref.~\cite{sch}. The current study
aims to generalize the approach presented in refs.~\cite{nlo,rhoE}.
Note that there is a huge body of literature concerning the domain
$\omega \lsim gT$, $\vec{p}=\vec{0}$ (here $g = \sqrt{4\pi\alpha_\rmi{s}}$), 
in which perturbation 
theory breaks down; upon this we touch only briefly 
in secs.~\ref{ss:IR}, \ref{ss:XIR}.

\subsection*{Outline of this paper}

The observables that we are specifically concerned with 
are the 2-point correlators of the scalar (``$FF$'') and pseudoscalar
(``$F\tilde F$'') densities in pure SU($\Nc$) Yang-Mills theory. 
These quantities have  recently
been studied particularly by Meyer, who has addressed  
both the Minkowskian and Euclidean 
domains~\cite{hbm_a}--\cite{hbm_d}. 
Noteworthy is also a general discussion concerning the shape of 
the scalar channel spectral function at small 
frequencies~\cite{ms} as well as 
the mentioned OPE representation of 
its ultraviolet asymptotics~\cite{sch}. In addition, sum rules
have been analyzed~\cite{rs}, improving on earlier
findings~\cite{ellis,hbm_f}, and further insight has 
been sought from AdS/CFT inspired models~\cite{hbm_c,sj,kv,hs}.
The corresponding transport coefficients
(this yields the bulk viscosity
for $FF$, cf.\ e.g.\ ref.~\cite{hbm_0}, and the rate 
of anomalous chirality changing transitions for $F\tilde{F}$, 
cf.\ ref.~\cite{ss}), 
together with possible applications, have been the subject 
of much further work; we restrict here to mentioning 
determinations of the transport coefficients 
in the weak-coupling limit~\cite{dogan,mt}.

The current work is a continuation of two recent papers which 
also addressed the same general problems. In ref.~\cite{ope},  
the scalar and pseudoscalar correlators were studied 
within the OPE framework, refining the previous determination
of certain Wilson coefficients and also correcting results concerning
the pseudoscalar channel~\cite{hbm_c}. 
In ref.~\cite{rdep}, which went beyond the OPE-limit, 
the full $|\vec{x}|$-dependences of the time-averaged Euclidean
correlators were determined,  
and compared with related lattice measurements carried out
in ref.~\cite{hbm_c}. One specific purpose of the present paper 
is to ``complement'' ref.~\cite{rdep}, by generalizing the results 
beyond the OPE-limit also in the Minkowskian frequency domain. 

As already mentioned our current analysis borrows techniques from
refs.~\cite{nlo,rhoE} but goes beyond these works in its generality. 
We hope that we can thereby pave the way for a number of 
further generalizations, such as the inclusion 
of the Lorentz non-invariant structures appearing e.g.\ in the  
``shear'' channel of the energy-momentum tensor; 
of a non-zero spatial momentum, relevant for relativistic 
particle production rates in cosmology; and of effects from 
a non-trivial mass spectrum, relevant e.g.\ in chiral 
perturbation theory applications. 

The paper is organized as follows. The observables are defined
in \se\ref{se:setup}, the general method for their determination
is discussed in \se\ref{se:method}, 
and the basic results are presented
in \se\ref{se:results}. The results are analyzed in some detail 
in \se\ref{se:analysis}, which also contains numerical evaluations. 
We conclude and offer an outlook in \se\ref{se:concl}. Three appendices, 
\ref{se:app1}, \ref{se:app2} and \ref{se:app3}, 
contain a detailed investigation 
of one of the sum-integrals, concise results for the others, 
and an account of a resummation relevant for 
$
 g^2 T /\pi \ll \omega \ll \pi T
$, 
respectively. 

%
\section{Specific setup}
\la{se:setup}

Our basic notation follows ref.~\cite{ope}, so we 
discuss the specific setup only briefly. We consider pure SU($\Nc$)
Yang-Mills theory (with $\Nc = 3$ in numerical estimates), 
dimensionally regularized by analytically continuing 
to $D=4-2\epsilon$ space-time dimensions
(the Euclidean action reads $S_\rmi{E} = \int_0^{\beta} \! {\rm d}\tau 
\int \! {\rm d}^{3-2\epsilon}\vec{x}\, \fr14\! F^a_{\mu\nu} F^a_{\mu\nu}$). 
The operators considered are 
\be
 \theta \equiv c^{ }_\theta\, \gB^2 F^a_{\mu\nu}F^a_{\mu\nu}
 \;, \quad
 \chi \equiv c_\chi\, \epsilon_{\mu\nu\rho\sigma} 
 \gB^2 F^a_{\mu\nu}F^a_{\rho\sigma}
 \;, \la{ops}
\ee
where $\gB^2$ is the bare gauge coupling squared; these operators
require no renormalization at the order of our computation.
We normally leave the coefficients $c^{ }_\theta$, $c_\chi$ unspecified, 
but note that often the values
$
  c^{ }_\theta \approx  
 -\frac{b_0}{2}  - \frac{b_1 g^2}{4}
 \;, 
 c_\chi \equiv \frac{1}{64\pi^2} 
$
are chosen, where $b_0 \equiv \frac{11\Nc}{3(4\pi)^2}$ and 
$b_1 \equiv \frac{34\Nc^2}{3(4\pi)^4}$. 
The bare gauge coupling squared can be expanded in terms of 
the renormalized one, $g^2$, as 
\be
 \gB^2 = g^2 \mu^{2\epsilon} \biggl[
 1 - \frac{g^2 b_0}{\epsilon} + \ldots 
 \biggr] 
 \;, \la{gB}
\ee
where $\mu$ denotes a scale parameter. The 
$\msbar$ scheme renormalization scale is denoted by 
$ 
 \bmu^{2} \equiv {4\pi} \mu^2  
 {e^{-\gammaE}}
$.
In order to avoid unnecessary clutter
all appearances of $\mu^{2\epsilon}$, which play
no role in the final renormalized results, will be suppressed. 

The observables considered are related to the 2-point
correlators of $\theta$ and $\chi$. We define 
the quantities 
($X\equiv (\tau,\mathbf{x})$; 
$P\equiv (p_n,\mathbf{p})$; 
$p_n\equiv 2\pi T n$, 
$n\in\mathbbm{Z}$) 
\be
 G_\theta(X) \equiv  
 \langle\, \theta(X)\, \theta(0) \,\rangle^{ }_{T}
 \;, \quad 
 G_\chi(X) \equiv 
 \langle\, \chi(X)\, \chi(0) \,\rangle^{ }_T
\;, 
\ee
as well as the corresponding Fourier transforms
\be
 \tilde G_\theta(P) \equiv \int_X e^{- i P\cdot X} G_\theta(X)
 \;, \quad
 \tilde G_\chi(P) \equiv \int_X e^{- i P\cdot X} G_\chi(X)
 \;, \la{GP}
\ee
where 
$
 \int_X \equiv \int_0^\beta \! {\rm d}\tau \int_\vec{x}
$.
No hats appear above the operators because Euclidean correlators
can be evaluated with standard path integral techniques. It can be 
shown (cf.\ e.g.\ refs.~\cite{leb,kg}) that the corresponding
spectral functions, 
defined as in \eq\nr{rho_def} with $\hat O\to \hat\theta, \hat\chi$, 
are then obtained (at vanishing spatial momentum) from 
\be
 \rho (\omega) 
 = \im \Bigl[ \tilde G(P) 
 \Bigr]_{P \to (-i[\omega + i 0^+],\vec{0})}
 \;. \la{rho_general}
\ee
It is these quantities (denoted by $\rho^{ }_\theta$, $\rho^{ }_\chi$)
that we are mostly interested in. 

%
\section{General method}
\la{se:method}

%
\subsection{Wick contractions}

%
\begin{figure}[t]
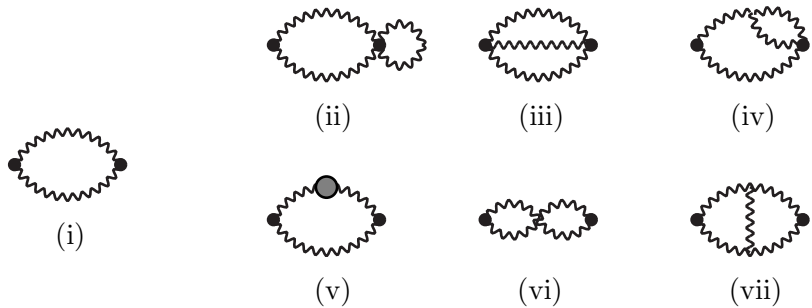


\hspace*{1.5cm}%
\begin{minipage}[c]{3cm}
\begin{eqnarray*}
&& 
 \hspace*{-1cm}
 \EleA 
\\[1mm] 
&& 
 \hspace*{0.0cm}
 \mbox{(i)} 
\end{eqnarray*}
\end{minipage}%
\begin{minipage}[c]{10cm}
\begin{eqnarray*}
&& 
 \hspace*{-1cm}
 \EleB \quad\; 
 \EleC \quad\; 
 \EleD \quad\; 
\\[1mm] 
&& 
 \hspace*{0.0cm}
 \mbox{(ii)} \hspace*{2.2cm}
 \mbox{(iii)} \hspace*{2.2cm}
 \mbox{(iv)} 
\\[5mm] 
&& 
 \hspace*{-1cm}
 \EleE \quad\; 
 \EleF \quad\; 
 \EleG \quad 
\\[1mm] 
&& 
 \hspace*{0.0cm}
 \mbox{(v)} \hspace*{2.2cm}
 \mbox{(vi)} \hspace*{2.2cm}
 \mbox{(vii)} 
 \\[3mm]
\end{eqnarray*}
\end{minipage}

\caption[a]{\small 
The graphs contributing to the 2-point 
correlators up to 2-loop order. 
The wiggly lines denote gluons; the small dots 
the operators $\theta$ or $\chi$ (cf.\ \eq\nr{ops}); 
and the grey blob the 1-loop gauge field self-energy. 
``Disconnected'' contractions only arise at higher orders.
} 
\la{fig:graphs}
\end{figure}
%

Having defined the correlators, the first step in their determination
is to carry out the Wick contractions 
(for our specific example 
the corresponding graphs are shown in \fig\ref{fig:graphs}), 
and to reduce the functions 
$\tilde G_\theta(P)$, $\tilde G_\chi(P)$ to sums over various types 
of integrals. We strongly advocate taking this step 
in Euclidean signature, whereby the Feynman rules are elementary
(no excessive $i$'s or doublings of degrees of freedom appear), 
and by using bare parameters, whereby counterterm vertices are 
avoided. An additional strength of the imaginary-time formalism is 
that in terms of physical processes, all ``real'' and ``virtual''
corrections related to each other
are automatically captured by a single Matsubara sum-integral. 
Before carrying out the contractions, it is helpful to make
use of translational invariance in order to write 
\ba
 \tilde G_\theta(P)
 & = & \int_X e^{- i P\cdot X} 
  \langle\, \theta(X)\, \theta(0) \,\rangle^{ }_{T} 
  =  
 \int_X e^{- i P\cdot (X-Y)} 
  \langle\, \theta(X)\, \theta(Y) \,\rangle^{ }_{T} 
 \nn 
 & = & \frac{1}{\int_Y}
  \Bigl\langle\, 
     \int_X e^{- i P\cdot X} \theta(X)\, 
     \int_Y e^{i P\cdot Y} \theta(Y)
  \,\Bigr\rangle^{ }_{T} 
  \;, 
\ea
so that the full analysis can be carried out in momentum space
(the factor $\int_Y = \beta V = \beta\delta^{ }_{p_n=0}
(2\pi)^{d}\delta_{ }^{(d)}(\vec{p}=\vec{0})$, with $d=D-1$,  
cancels out because of momentum conservation). 

%
\subsection{Scalarization}

In the next step, the goal is to ``scalarize'' the sum-integrals, 
i.e.\ to turn them into ones that also appear in scalar field theories. 
This can be achieved by contracting Lorentz indices 
(using $\delta_{\mu\mu} = D$) and carrying out Dirac traces
(with rules following from 
$
 \{ \gamma_\mu, \gamma_\nu\} = 2 \delta_{\mu\nu} \unit_\rmii{$4\times 4$}
$; in the case of chiral fermions the usual issues with $\gamma_5$
need to be faced and we have nothing to add on this topic).
Subsequently, substitutions of sum-integration variables,  
completions of squares
(e.g.\ $Q\cdot R = \tfr12 [Q^2+R^2-(Q-R)^2]$), and identities
following from partial integrations with respect to spatial
momenta, can be employed in order to remove as many scalar products
as possible from the numerators. The goal is 
to express the result in terms of a minimal number of 
independent ``master'' sum-integrals. (If the set is
minimal indeed, then all dependence on the gauge fixing parameter must have
cancelled after this step.)

For the specific correlators 
$ \tilde G_\theta(P)$ 
and 
$ \tilde G_\chi(P)$, the reduction to master sum-integrals
was carried out already in ref.~\cite{ope}. 
Re-expressing a sum-integral ``$\mathcal{I}^{ }_\rmi{i}$''
of ref.~\cite{ope} in terms
of ``$\mathcal{I}^{ }_\rmi{i'}$'' (cf.\ \eq(A.14) there)
the results from \eqs(3.1), (3.2) of ref.~\cite{ope} become
($d_A \equiv \Nc^2-1$) 
\ba
 && \hspace*{-1cm} \frac{\tilde G_\theta(P)}{4 d_A c_\theta^2 } = 
 \gB^4 (D-2) \biggl[ -\mathcal{J}^{ }_\rmi{a}
 + \fr12 \mathcal{J}^{ }_\rmi{b} \biggr]
 \nn 
 & + & \gB^6 \Nc \biggl\{ 
 2 (D-2) \biggl[ - (D-2)\mathcal{I}^{ }_\rmi{a}
 + (D-4) \mathcal{I}^{ }_\rmi{b} \biggr]
 + (D-2)^2 \biggl[ \mathcal{I}^{ }_\rmi{c} - \mathcal{I}^{ }_\rmi{d} \biggr]
 \nn & & \quad + \, 
 \frac{34-13D}{3} \mathcal{I}^{ }_\rmi{f}
 - \frac{(D-4)^2}{2} \mathcal{I}^{ }_\rmi{g}
 + (D-2) 
 \biggl[ 
   -  \mathcal{I}^{ }_\rmi{e} + 3 \mathcal{I}^{ }_\rmi{h}
 + 2\mathcal{I}^{ }_\rmi{i'}
    - \mathcal{I}^{ }_\rmi{j}
 \biggr] \biggr\} 
 \;, \la{Gtheta_bare} \\
 && \hspace*{-1cm} \frac{\tilde G_\chi(P)}{-16 d_A c_\chi^2 (D-3)} = 
 \gB^4 (D-2) \biggl[ -\mathcal{J}^{ }_\rmi{a}
 + \fr12 \mathcal{J}^{ }_\rmi{b} \biggr]
 \nn 
 & + & \gB^6 \Nc \biggl\{ 
 2 (D-2)  (D-4) \mathcal{I}^{ }_\rmi{b}
 + (D-2)^2 \biggl[ \mathcal{I}^{ }_\rmi{c} - \mathcal{I}^{ }_\rmi{d} \biggr]
 \nn & & \quad - \, 
 \frac{2 D^2-11D+30}{3} \mathcal{I}^{ }_\rmi{f}
 - 2 (D-4) \mathcal{I}^{ }_\rmi{g}
 + (D-2) 
 \biggl[ 
    - \mathcal{I}^{ }_\rmi{e} + 3 \mathcal{I}^{ }_\rmi{h}
 + 2\mathcal{I}^{ }_\rmi{i'}
    - \mathcal{I}^{ }_\rmi{j}
 \biggr] \biggr\} 
 \;. \la{Gchi_bare}
\ea 
Here, for brevity, structures containing $\Tinti{Q} 1$, which 
vanishes exactly in dimensional regularization, have been omitted 
($\Tinti{Q} \equiv T \sum_{q_n} \int_\vec{q}$).
The master sum-integrals are defined as 
\ba 
 \mathcal{J}^{ }_\rmi{a}(P) & \equiv &
 \Tint{Q} \frac{P^2}{Q^2}
 \;,  \la{m_first} \\
 \mathcal{J}^{ }_\rmi{b}(P) & \equiv &
 \Tint{Q} \frac{P^4}{Q^2(Q-P)^2}
 \;,  \la{m_Jb} \\
 \mathcal{I}^{ }_\rmi{a}(P) & \equiv &
 \Tint{Q,R} \frac{1}{Q^2R^2}
 \;,\\
 \mathcal{I}^{ }_\rmi{b}(P) & \equiv &
 \Tint{Q,R} \frac{P^2}{Q^2R^2(R-P)^2}
 \;, \\
 \mathcal{I}^{ }_\rmi{c}(P) & \equiv &
 \Tint{Q,R} \frac{P^2}{Q^2R^4}
 \;, \\
 \mathcal{I}^{ }_\rmi{d}(P) & \equiv &
 \Tint{Q,R} \frac{P^4}{Q^2R^4(R-P)^2}
 \;, \\
 \mathcal{I}^{ }_\rmi{e}(P) & \equiv &
 \Tint{Q,R} \frac{P^2}{Q^2R^2(Q-R)^2}
 \;, \\
 \mathcal{I}^{ }_\rmi{f}(P) & \equiv &
 \Tint{Q,R} \frac{P^2}{Q^2(Q-R)^2(R-P)^2}
 \;, \\
 \mathcal{I}^{ }_\rmi{g}(P) & \equiv &
 \Tint{Q,R} \frac{P^4}{Q^2(Q-P)^2R^2(R-P)^2}
 \;, \\
 \mathcal{I}^{ }_\rmi{h}(P) & \equiv &
 \Tint{Q,R} \frac{P^4}{Q^2R^2(Q-R)^2(R-P)^2}
 \;, \\
%
%
 \mathcal{I}^{ }_\rmi{i'}(P) & \equiv &
 \Tint{Q,R} \frac{4(Q\cdot P)^2}{Q^2R^2(Q-R)^2(R-P)^2}
 \;, \la{m_penultimate} \\
 \mathcal{I}^{ }_\rmi{j}(P) & \equiv &
 \Tint{Q,R} \frac{P^6}{Q^2R^2(Q-R)^2(Q-P)^2(R-P)^2}
 \;. \la{m_last}
\ea

%
\subsection{Matsubara sums and discontinuities}

The master sum-integrals defined in \eqs\nr{m_first}--\nr{m_last}
contain a 1- or 2-fold Matsubara sum. Both sums can be carried out
explicitly. In the literature various recipes can be found for this, 
most notably ``cutting rules''; we have carried out the sums with 
the method described in some detail in appendix A.1 of ref.~\cite{nlo}
and illustrated with simple examples 
also in \eqs\nr{mat_sum}--\nr{Ipn_n} and 
\nr{rho_rho_der1}--\nr{rho_rho_der2} below. 
Partial fractioning the result in the variable $p_n$, there may be
polynomial terms, as well as fractions of the type
\be
 \frac{1}{ip_n + \sum_k \sigma_k E_k} 
 \;, 
\ee
where $\sigma_k = \pm 1$ and 
$E_k \in \{ E_\vec{q},E_\vec{q-p}, E_\vec{r}, E_\vec{r-p},E_\vec{q-r} \}$,
$E_\vec{q} \equiv |\vec{q}|$.
According to \eq\nr{rho_general}, 
the corresponding spectral function follows from 
\be
 \im \biggl[ \frac{1}{\omega + i 0^+  + \sum_k \sigma_k E_k} \biggr] 
 = - \pi \delta(\omega + \sum_k \sigma_k E_k)
 \;.
\ee
The spectral functions corresponding to all the master sum-integrals
after this step (omitting terms $\propto \omega^n \delta(\omega)$ after 
setting $\vec{p}\to\vec{0}$, cf.\ \se\ref{ss:tau}) 
are listed in appendices A and B.

%
\subsection{Spatial integrals}

The ``hard work'' of the problem, where little automatization seems
possible, is to carry out the remaining spatial integrals, constrained
by the Dirac-$\delta$'s. There are two classes of integrals, referred
to in the literature as ``virtual'' and ``real'' corrections; we often 
refer to them as ``factorized'' and ``phase space'' integrals, respectively. 
Within the former class, some integrals are ultraviolet divergent, 
necessitating a careful handling by e.g.\ dimensional regularization.  
All classes of integrals are described in detail in appendix A for
the specific spectral function denoted 
by $\rho^{ }_{\mathcal{I}^{ }_\rmii{j}}$,
corresponding to the sum-integral defined in \eq\nr{m_last},  
and more concise results are collected in appendix B for 
all the other cases.

%
\subsection{Renormalization}
\la{ss:renorm}

After all master spectral functions are known, these are inserted
into (the imaginary parts of) \eqs\nr{Gtheta_bare}, \nr{Gchi_bare}.
Simultaneously, $D=4-2\epsilon$ is inserted into the coefficients, 
and the bare gauge coupling is re-expanded 
in terms of the renormalized one according to \eq\nr{gB}. 
If the computation has been carried out correctly, all 
$1/\epsilon$-divergences must cancel at this stage, 
and we can subsequently set $\epsilon\to 0$.

Let us illustrate the procedure with the example at hand. 
A number of the sum-integrals defined 
have no imaginary part 
(${\mathcal{J}^{ }_\rmi{a}}$, 
${\mathcal{I}^{ }_\rmi{a}}$,
${\mathcal{I}^{ }_\rmi{c}}$,
${\mathcal{I}^{ }_\rmi{e}}$), and therefore lead to a vanishing
spectral function. Among the non-zero spectral functions, 
only three have an $1/\epsilon$-divergence
($\rho^{ }_{\mathcal{I}^{ }_\rmii{g}}$,
$\rho^{ }_{\mathcal{I}^{ }_\rmii{h}}$,
$\rho^{ }_{\mathcal{I}^{ }_\rmii{i'}}$). Taking these
facts into account, 
\eqs\nr{rho_general}, \nr{Gtheta_bare}, \nr{Gchi_bare} can be converted into 
\ba
 \frac{\rho^{ }_\theta(\omega)}{4 d_A c_\theta^2 } & = &  
 (1-\epsilon)\Bigl\{
   \gB^4\, \rho^{ }_{\mathcal{J}^{ }_\rmii{b}} (\omega)
  + 2 \gB^6\, \Nc \bigl[ 
     3 \rho^{ }_{\mathcal{I}^{ }_\rmii{h}} (\omega)
    + 
     2 \rho^{ }_{\mathcal{I}^{ }_\rmii{i'}} (\omega)
    \bigr]
 \Bigr\}
 \nn && \hspace*{1cm} - \, 
 2 \gB^6 \Nc \bigl[ 
     2 \rho^{ }_{\mathcal{I}^{ }_\rmii{d}} (\omega)
 + 
     3 \rho^{ }_{\mathcal{I}^{ }_\rmii{f}} (\omega)
  + 
     \rho^{ }_{\mathcal{I}^{ }_\rmii{j}} (\omega)
 \bigr] + \rmO(\gB^6\epsilon,\gB^8)
 \;,  \la{rhobare_theta}
 \\ 
 \frac{-\rho^{ }_\chi(\omega)}{16 d_A c_\chi^2 } & = &  
 (1-2\epsilon)(1-\epsilon)\Bigl\{
   \gB^4\, \rho^{ }_{\mathcal{J}^{ }_\rmii{b}} (\omega)
  + 2 \gB^6\, \Nc \bigl[ 
     3 \rho^{ }_{\mathcal{I}^{ }_\rmii{h}} (\omega)
    + 
     2 \rho^{ }_{\mathcal{I}^{ }_\rmii{i'}} (\omega)
    \bigr]
 \Bigr\}
 \nn && \hspace*{1cm} - \, 
 2 \gB^6 \Nc \bigl[ 
     2 \rho^{ }_{\mathcal{I}^{ }_\rmii{d}} (\omega)
 + 
     3 \rho^{ }_{\mathcal{I}^{ }_\rmii{f}} (\omega)
  + 
     \rho^{ }_{\mathcal{I}^{ }_\rmii{j}} (\omega)
  -  2\epsilon 
     \rho^{ }_{\mathcal{I}^{ }_\rmii{g}} (\omega)
 \bigr] + \rmO(\gB^6\epsilon,\gB^8)
 \;. \nn \la{rhobare_chi}
\ea
The structures within the curly brackets on the first rows 
turn out finite after the insertion of the bare gauge coupling
from \eq\nr{gB} and the re-expansion of the result
in terms of the renormalized
coupling; therefore the prefactors 
$1-\epsilon$, 
$1-2\epsilon$
can actually be set to unity.

%
\subsection{Final expression and its limits}

After renormalization and setting $\epsilon\to 0$
as outlined in \se\ref{ss:renorm}, 
we obtain our final results for the spectral functions. Furthermore, 
if they are only needed in the OPE-regime $\omega \gg \pi T$, then 
all master spectral functions can be determined in a closed form
(cf.\ ref.~\cite{ope} for general methods as well as 
appendices A and B for specific results for the structures
in \eqs\nr{m_first}--\nr{m_last}). Moreover, the general outcome
can be given an interpretation in terms of thermal contributions to  
``condensates'', cf.\ ref.~\cite{sch}, which 
guarantees the cancellation of terms $\propto T^2$, 
if the lowest-dimensional gauge-invariant condensate 
has dimensionality four, as is the case in our study.
(If scalar fields are present, then there can be a contribution
proportional to the condensate $\sim \langle \phi^\dagger \phi \rangle$, 
implying that terms $\propto T^2$ need not cancel; 
see e.g.\ ref.~\cite{sls}).

The situation is considerably more subtle in the infrared 
regime, $\omega \ll \pi T$. In fact, at small frequencies,
$\omega\sim gT$, the NLO
corrections become as large as the leading-order (LO) terms, 
indicating a breakdown of the perturbative series; this 
will be discussed with our specific examples in \se\ref{ss:IR}. 
In this situation perturbation theory needs to be resummed
through effective field theory techniques. For $\omega \sim gT$
the relevant framework is that of the Hard Thermal Loop effective 
theory~(ref.~\cite{htl} and references therein), 
whereas for smaller frequencies still, needed
e.g.\ for transport coefficients, a further re-organization is 
required (ref.~\cite{dogan} and references therein). 
The different infrared frequency scales playing
a role in the scalar channel have been elaborated upon in ref.~\cite{ms}.

%
\section{Results}
\la{se:results}

Inserting the spectral functions 
$\rho^{ }_{\mathcal{J}^{ }_\rmii{b}}$, 
$\rho^{ }_{\mathcal{I}^{ }_\rmii{h}}$, 
$\rho^{ }_{\mathcal{I}^{ }_\rmii{i'}}$ from 
\eqs\nr{Jb_final}, \nr{Ih_final}, \nr{Iip_final}, 
respectively, as well as the bare gauge coupling from \eq\nr{gB}, 
all divergences are seen to cancel in \eqs\nr{rhobare_theta}, 
\nr{rhobare_chi}, as already advertised. 
Making also use of the finite functions  
$\rho^{ }_{\mathcal{I}^{ }_\rmii{d}}$, 
$\rho^{ }_{\mathcal{I}^{ }_\rmii{f}}$, 
$\rho^{ }_{\mathcal{I}^{ }_\rmii{j}}$
 from 
\eqs\nr{Id_final}, \nr{If_final}, \nr{Ij_final}, 
respectively, as well as the divergent part of 
$\rho^{ }_{\mathcal{I}^{ }_\rmii{g}}$ from \eq\nr{Ig_final}, 
the final results can be expressed as  
\ba
 \frac{\rho^{ }_\theta(\omega)}{4 d_A c_\theta^2 } & = &  
 \frac{\pi\omega^4}{(4\pi)^2}
 \bigl( 1 + 2 n_{\frac{\omega}{2}} \bigr)
 \biggl\{
   g^4 + \frac{g^6\Nc}{(4\pi)^2}
    \biggl[
       \frac{22}{3} \ln\frac{\bmu^2}{\omega^2} + \frac{73}{3} 
     + 8\, \phi^{ }_T(\omega) 
    \biggr] 
 \biggr\}
 + \rmO(g^8)
 \;, \la{rhofinal_theta}
 \\ 
 \frac{-\rho^{ }_\chi(\omega)}{16 d_A c_\chi^2 } & = &  
 \frac{\pi\omega^4}{(4\pi)^2}
 \bigl( 1 + 2 n_{\frac{\omega}{2}} \bigr)
 \biggl\{
   g^4 + \frac{g^6\Nc}{(4\pi)^2}
    \biggl[
       \frac{22}{3} \ln\frac{\bmu^2}{\omega^2} + \frac{97}{3} 
     + 8\, \phi^{ }_T(\omega) 
    \biggr] 
 \biggr\}
 + \rmO(g^8)
 \;, \la{rhofinal_chi}
\ea
where we have introduced 
\be
 n_E \equiv \nB{}(E)
 \;, 
\ee
with $\nB{}(E)\equiv 1/(e^{\beta E}-1)$ denoting the Bose distribution. 
To this order the only difference between the two channels is in 
the constant next to the logarithm. The logarithms and the constants
next to them
agree with those found within the OPE-regime in ref.~\cite{ope}. 
(In principle they were computed already
in ref.~\cite{old}, however in a different regularization scheme.) 
The ``non-trivial'' $T$-dependence  
resides inside the function $\phi^{ }_T$, which is given by
(we have substituted $q = \omega\sigma/2$, $r = \omega \tau/2$
in the expressions of appendix A and B, and defined
$\hat n_x \equiv n_{\frac{\omega x}{2}}$)
\ba
 \phi^{ }_T(\omega)  & = &   
 \int_0^{ \frac{1}{2} } \! { {\rm d}\sigma } \, 
   \hat n_{\sigma} \; 
   \biggl\{
      \biggl[\frac{1}{\sigma} -\frac{1}{\sigma-1}  
      -2 + \sigma -\sigma^2
      \biggr]
       \ln\bigl( 1 - \sigma \bigr)
 \nn & & \hspace*{2cm} + \,
      \biggl[\frac{1}{\sigma} -\frac{1}{\sigma+1} + 
      2 + \sigma + \sigma^2
      \biggr]
       \ln\bigl( 1 + \sigma \bigr)
 \nn & & \hspace*{2cm} + \,
  \frac{11}{12} 
   \biggl[
     \frac{1}{\sigma+1} 
   + \frac{1}{\sigma-1} \biggr] 
   + \fr{5\sigma}6 
   \; \biggr\}  
 \nn 
 & + & 
 \int_{ \frac{1}{2} }^{ 1 } 
 \! { {\rm d}\sigma } \, 
     \hat n_{\sigma}  \; 
   \biggl\{
      \biggl[\frac{1}{\sigma} -\frac{2}{\sigma - 1} 
      - \fr{11}{4} + \sigma - \frac{3\sigma^2}{2}
      \biggr]
       \ln\bigl( 1 - \sigma \bigr)
 \nn & & \hspace*{2cm} + \,
      \biggl[\frac{1}{\sigma} -\frac{1}{\sigma+1}  
     + 2  + \sigma + \sigma^2
      \biggr]
       \ln\Bigl( 1 + \sigma \Bigr)
 \nn & & \hspace*{2cm} + \,
      \biggl[\frac{1}{\sigma-1}  
      + \fr34 + \frac{\sigma^2}{2}
      \biggr]
       \ln\bigl( \sigma \bigr)
 \nn & & \hspace*{2cm} + \,
  \frac{11}{12} 
   \frac{1}{\sigma+1} 
   - \fr13 - 2\sigma - \frac{\sigma^2}{3}
   \;\biggr\}  
 \nn 
 & + & 
 \int_{ 1 }^{ \infty } 
 \! { {\rm d}\sigma } \, 
    \hat n_{\sigma} \; 
   \biggl\{
      2\biggl[\frac{1}{\sigma} -\frac{1}{\sigma-1} 
      - 2 + \sigma - \sigma^2
       \biggr]
       \ln\bigl(  \sigma - 1 \bigr)
 \nn & & \hspace*{2cm} + \,
      \biggl[\frac{1}{\sigma} -\frac{1}{\sigma+1}  
      +2+\sigma + \sigma^2
      \biggr]
       \ln\bigl( 1 + \sigma \bigr)
 \nn & & \hspace*{2cm} + \,
      \biggl[\frac{1}{\sigma-1} - \frac{1}{\sigma}  
      + 2 - \sigma + \sigma^2
      \biggr]
       \ln\bigl( \sigma \bigr)
 \nn & & \hspace*{2cm} + \,
  \frac{11}{12} 
   \biggl[ \frac{1}{\sigma+1 } - \frac{1}{\sigma} \biggr] 
  + \frac{23}{12} - \frac{13\sigma}{4} - \frac{11\sigma^2}{12}
   \;\biggr\}  
 \nn & + & 
 \int_0^{ 1 } 
 \! { {\rm d}\sigma } 
 \int_0^{ \frac{1}{2} - |\sigma-\frac{1}{2}| } 
 \! { {\rm d}\tau }  \,
  \frac{  
   \hat n_{1-\sigma} \,
   \hat n_{\sigma+\tau} 
  (1+\hat n_{1-\tau}) }
  {\hat n_{ \tau }^2}
  \; \times 
 \nn & & \hspace*{2cm} \times \,
  \biggl[
  \frac{1}{\sigma\tau} 
  - \frac{5 - 4 \tau + 2\tau^2}{4\sigma}
  - \frac{5 - 4 \sigma + 2 \sigma^2}{4 \tau}
  + \frac{3}{2}
  \biggr] 
 \nn & + & 
 \int_{ 1 }^{\infty}
 \! {{\rm d}\sigma}
 \int_{0}^{ \sigma - 1 }
 \! { {\rm d}\tau } \,
    \frac{  
    \hat n_{\sigma - 1} 
    (1+\hat n_{\sigma - \tau} )
    (\hat n_{ \sigma } 
    - \hat n_{\tau + 1 } )
    }
    { \hat n_{ \tau } \hat n_{- 1} }
  \; \times 
 \nn & & \hspace*{2cm} \times \,
  \biggl[
  \frac{1}{\sigma\tau} 
  + \frac{5 + 4 \tau + 2\tau^2}{4\sigma}
  - \frac{5 - 4 \sigma + 2 \sigma^2}{4 \tau}
  - \frac{3}{2}
  \biggr] 
 \nn & + & 
 \int_0^\infty \! {{\rm d}\sigma}  
 \int_{0}^{ \sigma }
 \! { {\rm d}\tau } \,
 \frac{ 
  ( 1 + \hat n_{ \sigma + 1 } )
  \,\hat n_{ \sigma + \tau }
  \,\hat n_{ \tau + 1 } }
  {\hat n_{ \tau }^2}
  \; \times 
 \nn & & \hspace*{2cm} \times \,
  \biggl[ 
  \frac{1}{\sigma\tau} 
  + \frac{5 + 4 \tau + 2\tau^2}{4\sigma}
  + \frac{5 + 4 \sigma + 2 \sigma^2}{4 \tau}
  + \frac{3}{2}
  \biggr] 
 \;. \hspace*{5mm} \la{phi_T} 
\ea

It is conceivable that some of the structures in \eq\nr{phi_T} could be 
written in a simpler form; given that such rewritings do not allow to reduce
the dimensionality of the integration, however, we prefer
to display the result in the current form, suitable for numerical 
evaluation. On this point we should mention that \eq\nr{phi_T} is
defined in the sense of principal value integration; 
this implies that the 2nd terms from the 2nd and 3rd structures
could be combined into
\ba
 && \hspace*{-1cm}
  \int_{ \frac{1}{2} }^{ 1 } 
 \! { {\rm d}\sigma } \, 
     \hat n_{\sigma}  \; 
   \biggl\{
  -\frac{2}{\sigma - 1}
        \ln\bigl( 1 - \sigma \bigr)
   \;\biggr\}  
 + 
 \int_{ 1 }^{ \infty } 
 \! { {\rm d}\sigma } \, 
    \hat n_{\sigma} \; 
   \biggl\{
 -\frac{2}{\sigma-1}
       \ln\bigl(  \sigma - 1 \bigr)
   \;\biggr\}  
 \nn && = \, 
 -2 \biggl\{ 
  \int_{ 0 }^{ \frac{1}{2} }
  \! \frac{ {\rm d}\sigma }{\sigma} \, 
    \bigl( \hat n_{1+\sigma} - \hat n_{1-\sigma} \bigr)  \; 
    \ln \bigl(\sigma\bigr)
   + 
  \int_{ \frac{3}{2} }^{ \infty }
  \! \frac{ {\rm d}\sigma }{\sigma - 1} \, 
  \hat n_{\sigma} \, 
  \ln \bigl( \sigma - 1 \bigr)
 \biggr\} 
 \;, \la{I23}
\ea
or, alternatively, that the range $(1,\infty)$ in the 3rd structure 
could be reflected to $(1,0)$ through $\sigma\to 1/\sigma$ and 
then combined with the 1st and 2nd structures. In any case
the integral is rapidly convergent; the result
of \eq\nr{phi_T} is illustrated in \fig\ref{fig:phiT}.

\begin{figure}[t]


\centerline{%
 \epsfysize=7.5cm\epsfbox{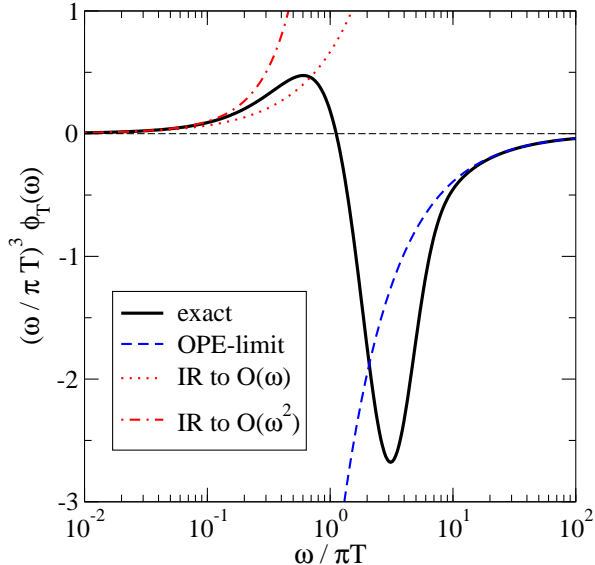}%
}

\caption[a]{\small
The function $\phi^{ }_T$ from \eq\nr{phi_T}, multiplied
with $(\frac{\omega}{\pi T})^3$, as a function of $\frac{\omega}{\pi T}$.
Also shown are the ultraviolet (``OPE'') limit from 
\eq\nr{phi_T_as} as well as the infrared (``IR'') limit 
from \eq\nr{phi_IR}.
}

\la{fig:phiT}
\end{figure}

%
\section{Analysis of the result}
\la{se:analysis}

%
\subsection{Ultraviolet limit ($\omega \gg \pi T$)}
\la{ss:UV}

As has already been mentioned the general results simplify considerably
in the OPE-limit $\omega \gg \pi T$ and can be expressed in a closed
form. Indeed, according to refs.~\cite{sch,ope}, the asymptotic 
$\omega\gg \pi T$ behaviour of the function $\phi^{ }_T$ from 
\eq\nr{phi_T} should read
\be
 \phi^{ }_T(\omega) 
 \;\; \stackrel{\omega\gg \pi T}{=} \;\;
 - \frac{176}{45} \frac{\pi^4 T^4}{\omega^4}
 + \rmO\Bigl( \frac{\pi^6 T^6}{\omega^6} \Bigr)
 \;. \la{phi_T_as}
\ee
Let as stress again that, as required by OPE~\cite{sch}, all contributions 
of $\rmO(\pi^2T^2/\omega^2)$ from individual graphs need to cancel in the 
sum, because no gauge-invariant
condensate of dimensionality GeV$^2$ is available. 
It is easy to verify numerically that 
the form of \eq\nr{phi_T_as} is indeed approached, cf.\ \fig\ref{fig:phiT}. 
The powerlike
asymptotic behaviour comes solely from the first integral in \eq\nr{phi_T},
the other terms falling off exponentially at $\omega \gg \pi T$.

%
\subsection{Infrared limit ($g^2T/\pi  \ll  \omega \ll \pi T$)}
\la{ss:IR}

Another interesting limit is the infrared one, $\omega \ll \pi T$, 
which shows a much richer structure than the ultraviolet limit. 
Denoting the six separate 
parts of \eq\nr{phi_T} by $I^{ }_1, \ldots, I^{ }_6$, 
respectively, 
with the combination appearing in \eq\nr{I23} taken apart and denoted 
by $I^\rmi{div}_{23}$, 
some work leads to the small-$\omega$ expansions 
\ba
 I^{ }_1&=&
 \Bigl\{ -2\,{\rm Li}_2\(\tfr{2}{3}\)-2\ln 3\, {\rm arcoth}\,7
 -\tfr{67\ln3}{12}-\ln^2\! 3+6\ln2+\tfr{\pi^2}{2}
 -\tfr{1}{6}\Bigr\}\fr{T}{\omega} + \rmO(\omega)\;, \hspace*{5mm} \\
 I^{ }_2&=&\Bigl\{{\rm Li}_2\(\tfr{1}{4}\)+2\ln 3\, {\rm arcoth}\,7
 +\tfr{67\ln3}{12}+3\ln^2\! 2-8\ln 2+\tfr{5\pi^2}{8}-3\Bigr\}\fr{T}{\omega}
 + \rmO(\omega)\;, \\
 I^{ }_3&=&-\fr{44\zeta(3)T^3}{3\omega^3}
 +\Bigl\{ -8\ln (\tfr{\omega}{T})-96\ln\! A+8\ln(4\pi)+7 \Bigr\} 
  \fr{\pi^2 T^2}{6\omega^2}\nn
 & & + \,
 \Bigl\{
 -\tfr{17}{6}\ln(\tfr{\omega}{T})
 -\ln^2\! 2+\tfr{14\ln2}{3}
 +\tfr{11\pi^2}{6}+\tfr{37}{12} \Bigr\} \fr{T}{\omega}
 +\rmO(\ln\omega)\;,\\
 I_{23}^\rmi{div}&=& \Bigl\{ 2\,{\rm Li}_2\(\tfr{1}{4}\)
 -4\,{\rm Li}_2\(\tfr{2}{3}\)-2\ln^2\! 3+4\ln^2\! 2
 -\tfr{2\pi^2}{3}\Bigr\}\fr{T}{\omega}+\rmO(\ln\omega)
 \;,\\
 I^{ }_4&=& \Bigl\{ 
   \tfr{\ln 2}{6}-\tfr{\pi^2}{8}+\tfr{19}{12} 
 \Bigr\} \fr{T}{\omega}
 +\rmO(\omega) \;,\\
 I^{ }_5&=&
 \Bigl\{ \ln(\tfr{\omega}{T})
  + 12 \ln\! A - \ln(4\pi) 
  -\tfr76 \Bigr\}
  \fr{2\pi^2 T^2}{3\omega^2}
 \nn
 & & + \, 
 \Bigl\{ \ln^2 (\tfr{\omega}{T}) -(2\ln2+\tfr13)
 \ln(\tfr{\omega}{T}) + c_1 \Bigr\}
 \fr{T}{\omega}
 +\rmO(\ln\omega)\;,\\
 I^{ }_6&=&\fr{44\zeta(3)T^3}{3\omega^3}
 +\Bigl\{ \ln(\tfr{\omega}{T} ) + 12\ln\! A - \ln(4\pi)+ \tfr{5}{12} \Bigr\} 
  \fr{2\pi^2T^2}{3\omega^2}
 \nn
 & & + \,  
 \Bigl\{ -\ln^2(\tfr{\omega}{T})+
 \bigl(2\ln2+\tfr{19}6\bigr)\ln(\tfr{\omega}{T})+c_2 \Bigr\}
 \fr{T}{\omega}
 +{\mathcal O}(\ln\omega)\;,
\ea
where $A$ stands for the Glaisher constant, 
$\ln\! A = -\zeta(-1)-\zeta'(-1)$, 
and we have defined the numerical coefficients
\be
 c_1 \approx  -0.17449 \;, \quad 
 c_2 \approx  -9.32085 \;.
\ee
Summing together, terms of $\rmO(T^3/\omega^3)$ as well as 
all the logarithms cancel, and we are left with the expansion
\be
 \phi_T^{ }(\omega)
 \;\; \stackrel{\omega\ll \pi T}{\approx} \;\;
  \frac{2 \pi^2 T^2}{3 \omega^2} + 
 3.31612\, \frac{\pi T}{\omega} 
 + \rmO\Bigl(\ln \frac{\omega}{\pi T}\Bigr)
 \;. \la{phi_IR}
\ee
As can be seen from \fig\ref{fig:phiT}, this does agree with 
a numerical evaluation of $\phi^{ }_T$ for $\omega \ll \pi T$.

Now, even though we can derive a representation
of the function $\phi^{ }_T$ for $\omega \ll \pi T$, 
this does not mean that we know the spectral function
in this regime. The reason is that, as can be seen from 
\eqs\nr{rhofinal_theta}, \nr{rhofinal_chi} and \nr{phi_IR},  
the thermal NLO corrections overtake 
the LO terms for $\omega \ll g T$. Therefore 
the naive perturbative expansion breaks down in this regime. 
The spectral function can be determined only if we 
find a resummation which re-organizes the perturbative 
series in a way that a breakdown is avoided. 

It is strongly believed that the way to resum the 
perturbative series around $\omega \sim g T$ goes through
the use of the Hard Thermal Loop (HTL) effective theory
(ref.~\cite{htl} and references therein). Although caveats
may be difficult to exclude~\cite{gdm}, we demonstrate in the following 
that this framework does indeed reproduce the leading 
divergence of \eq\nr{phi_IR} and, through 
a matching computation, should allow us to postpone 
the breakdown to $\omega \sim g^2 T/\pi$. 

Schematically, matching can be represented as 
\be
 \rho^\rmii{QCD}_\rmii{resummed} 
 \; = \; 
 \rho^\rmii{QCD}_\rmii{resummed} 
 -  \rho^\rmii{HTL}_\rmii{resummed} 
 + \rho^\rmii{HTL}_\rmii{resummed}  
 \; \approx \;
 \rho^\rmii{QCD}_\rmii{naive} 
 -  \rho^\rmii{HTL}_\rmii{naive} 
 + \rho^\rmii{HTL}_\rmii{resummed}  
 \;. \la{master_resum}
\ee
Here we have subtracted and added the HTL part, and subsequently noted
that if the correct effective theory is used, then the difference between
the full and the effective theory computations is infrared safe, so that
no resummation is needed for the difference. 
Since in \se\ref{se:results} a ``naive''
QCD computation was reported, we then need two different HTL 
computations, one naive and one resummed, in order to obtain the 
correctly resummed version for QCD.

Some details concerning the HTL computations are given 
in appendix~\ref{se:app3}. As far as the naive version 
goes we note that, replacing the Bose distributions 
through their ``classical'' limits,  
i.e.\ $n^{ }_q \to T/q$, $1+n^{}_{\omega - q}\to T/(\omega - q)$, etc., 
which is equivalent to the physics of HTL resummation~\cite{cl1,cl2},
the integrals
in \eq\nr{nlo_htl} of appendix~\ref{ss:htl_naive}
are elementary and we get 
\ba
 & & \hspace*{-1cm}
 \left. \frac{\rho_\theta^\rmii{HTL}(\omega)}{4 d_A c_\theta^2}
 \right|_\rmi{naive}
  =
 \left. \frac{-\rho_\chi^\rmii{HTL}(\omega)}{16 d_A c_\chi^2}
 \right|_\rmi{naive}
  =
 \frac{\pi g^4 \bigl( 1 + 2 n_{\frac{\omega}{2} } \bigr)}{(4\pi)^2} 
 \Bigl\{ \omega^4 + \omega^2 \mE^2 
 \Bigr\} 
 + \rmO(g^8) 
 \;. \hspace*{0.5cm} \la{nlo_htl_compact}
\ea
Here the Debye mass parameter is defined by 
\be
  \mE^2 \equiv \frac{ g^2 \CA  T^2}{3}   
  \;. \la{mE}
\ee
The first term 
of \eq\nr{nlo_htl_compact} matches the leading-order 
QCD result in \eqs\nr{rhofinal_theta}, \nr{rhofinal_chi}, 
whereas the second term exactly
corresponds to the leading term of \eq\nr{phi_IR}. 
So, in the difference
$
 \rho^\rmii{QCD}_\rmii{naive} 
 -  \rho^\rmii{HTL}_\rmii{naive} 
$
appearing in \eq\nr{master_resum} the dominant IR divergence drops out. 


\begin{figure}[t]
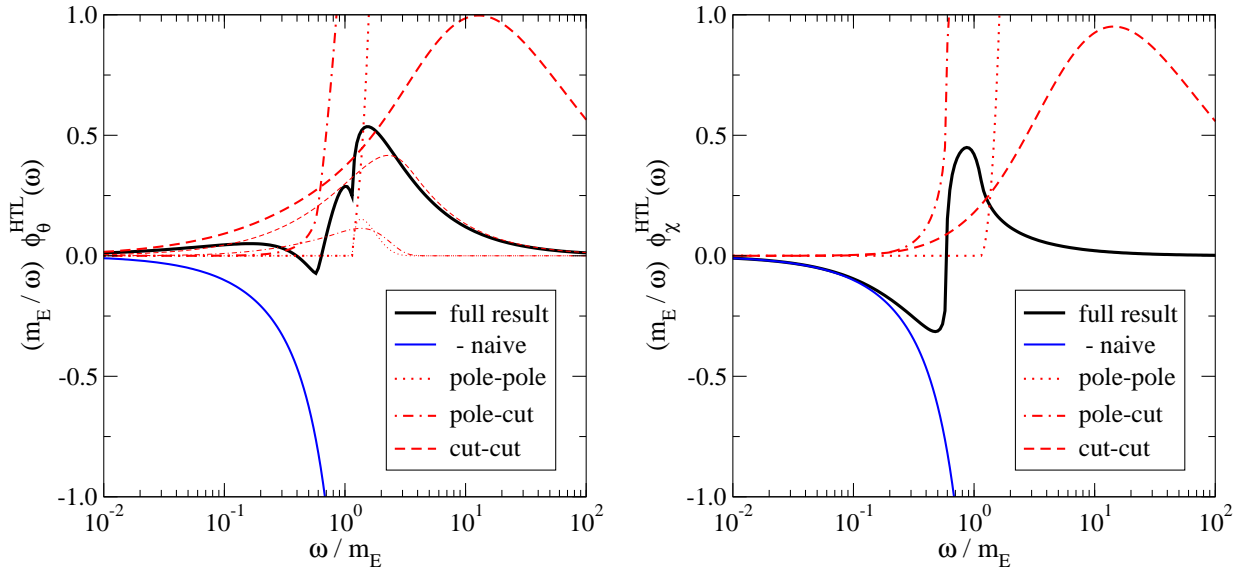



\centerline{%
 \epsfysize=7.5cm\epsfbox{phi_HTL_theta.eps}%
~~~\epsfysize=7.5cm\epsfbox{phi_HTL_chi.eps}
}

\caption[a]{\small
The functions $\phi^\rmii{HTL}_{\theta,\chi}$ from 
\eqs\nr{HTL_resum_theta}, \nr{HTL_resum_chi}, multiplied
with $\frac{\mE}{\omega}$ (so that the same number of 
overall $\omega$'s has been factored out as in \fig\ref{fig:phiT}). 
Also shown are 
partial contributions from various sub-processes
(``\,-\,naive'' refers to the terms $\omega^4 + \omega^2 \mE^2$
taken apart as in \eqs\nr{HTL_resum_theta}, \nr{HTL_resum_chi};
in the left panel, thick lines correspond to ``transverse'' and
thin to ``electric'' modes, respectively).
}

\la{fig:phi_HTL}
\end{figure}

As far as the resummed version of the HTL computation goes, 
it is technically similar to the classic dilepton analysis 
in ref.~\cite{bp_dilepton}. The main results are given in 
\eqs\nr{HTL_theta_final}, \nr{HTL_chi_final}
of appendix~\ref{ss:resum_HTL}. Let us define 
\ba
 \left. \frac{\rho_\theta^\rmii{HTL}(\omega)}{4 d_A c_\theta^2}
 \right|_\rmi{resummed}
 \!\!\! & \equiv & 
  \frac{\pi g^4 \bigl( 1 + 2 n_{\frac{\omega}{2} } \bigr)}{(4\pi)^2} 
  \Bigl\{ \omega^4 + \omega^2 \mE^2 +\mE^4
  \, \phi_\theta^\rmii{HTL}(\omega)
  \Bigr\}
 \;, \la{HTL_resum_theta} \\ 
 \left. \frac{-\rho_\chi^\rmii{HTL}(\omega)}{16 d_A c_\chi^2}
 \right|_\rmi{resummed}
  \!\! & \equiv & 
  \frac{\pi g^4 \bigl( 1 + 2 n_{\frac{\omega}{2} } \bigr)}{(4\pi)^2} 
  \Bigl\{ \omega^4 + \omega^2 \mE^2 +\mE^4
  \, \phi_\chi^\rmii{HTL}(\omega)
  \Bigr\}
  \;, \la{HTL_resum_chi}
\ea
so that the separated terms cancel against the naive version
in the difference appearing in \eq\nr{master_resum}.
The result is shown in \fig\ref{fig:phi_HTL}.
(The first ``cusp'' corresponds physically e.g.\ to a 
2 $\leftrightarrow$ 2 process in which one on-shell ``plasmon''
gets generated; the second to the 2-plasmon threshold.)
For $\omega \lsim \mE$, $\phi^\rmii{HTL}_{\theta,\chi}$ 
modifies the IR behaviour
of the full expression in a qualitative way
(cf.\ \fig\ref{fig:wdep} below). For $\omega \gg \mE$, in contrast,
$\phi^\rmii{HTL}_{\theta,\chi}$
is a constant of $\rmO(g^4 \mE^4) \sim \rmO(g^8 T^4)$ which is vastly 
overshadowed by the unresummed behaviour $\sim g^4\omega^4$; therefore
in this regime, which is the most important one for us, 
HTL resummation plays no role. 

%
\subsection{Extreme infrared limit ($\omega \lsim g^2T/\pi$)}
\la{ss:XIR}

The HTL resummation discussed in \se\ref{ss:IR} is supposed to 
re-organize the perturbative series in a way that it can
formally be defined even in the range $\omega \sim gT$. If, however, 
the frequency is decreased further, down to $\omega \lsim g^2 T/\pi$, 
then a further re-organization is needed in order to 
obtain the correct leading-order result~\cite{ms}. The ``transport 
peak'', whose height yields the transport coefficient, appears within 
this regime; in our computation part of the information concerning
it is hidden in the so-far
omitted structures $\sim \omega^n \delta(\omega)$,
to be discussed in \se\ref{ss:tau}.

%
\subsection{Sum rules}
\la{ss:sum}

Apart from the UV and IR limits, an interesting
crosscheck on the full result is offered by ``sum rules'', which concern
integrals over the spectral function with a certain weight. These 
integrals are related to ``thermodynamic'' Euclidean observables, 
which can be evaluated independently. Let us 
elaborate on the information that can be obtained this way. 

Because of issues of convergence, some terms need
to be subtracted before sum rules can be defined. If we 
formally subtract the zero-temperature parts from both sides, 
denoting the results by ``$\Delta$'' 
(in practice such a subtraction is not without 
problems, as discussed in \se\ref{ss:plots} below) 
then, to the order of our computation, the sum rules boil down to~\cite{ope}
\ba
 - \frac{g^6 \Nc T^4}{18} = 
  \frac{\Delta \tilde G_\theta(0)}{4d_A c_\theta^2} 
   & = & 2 \int_{0}^{\infty} \! \frac{{\rm d}\omega}{\pi\omega} 
 \,  \biggl[ \frac{\Delta\rho^{ }_\theta(\omega)}{4 d_A c_\theta^2 } \biggr]
 + \rmO(g^8) 
 \;, \la{sum_theta} \\
 0 =  \frac{\Delta \tilde G_\chi(0)}{-16d_A c_\chi^2} 
  & = & 2 \int_{0}^{\infty} \! \frac{{\rm d}\omega}{\pi\omega} 
 \, \biggl[ \frac{-\Delta\rho^{ }_\chi(\omega)}{16 d_A c_\chi^2 } \biggr] 
 + \rmO(g^8)
 \;. \la{sum_chi}
\ea
Here, the thermal parts read 
(from \eqs\nr{rhofinal_theta}, \nr{rhofinal_chi})
\ba
 \frac{\Delta\rho^{ }_\theta(\omega)}{4 d_A c_\theta^2 }
  \!\! & = & \!\!  
 \frac{\pi\omega^4}{(4\pi)^2}
 \biggl\{
 2 n_{\frac{\omega}{2}} \biggl[ 
   g^4 + \frac{g^6\Nc}{(4\pi)^2}
    \biggl(
       \frac{22}{3} \ln\frac{\bmu^2}{\omega^2} + \frac{73}{3}
    \biggr) \biggr]
 + 
 \bigl( 1 + 2 n_{\frac{\omega}{2}} \bigr)
 \frac{8 g^6\Nc \phi^{ }_T(\omega) }{(4\pi)^2}  
 \biggr\}
 \;, \hspace*{5mm} \nn \la{Delta_theta}
 \\ 
 \frac{-\Delta\rho^{ }_\chi(\omega)}{16 d_A c_\chi^2 }
  \!\! & = & \!\! 
 \frac{\pi\omega^4}{(4\pi)^2}
 \biggl\{
 2 n_{\frac{\omega}{2}} \biggl[ 
   g^4 + \frac{g^6\Nc}{(4\pi)^2}
    \biggl(
       \frac{22}{3} \ln\frac{\bmu^2}{\omega^2} + \frac{97}{3}
    \biggr) \biggr]
 + 
 \bigl( 1 + 2 n_{\frac{\omega}{2}} \bigr)
 \frac{8 g^6\Nc \phi^{ }_T(\omega) }{(4\pi)^2}  
 \biggr\}
 \;. \hspace*{5mm} \nn \la{Delta_chi}
\ea

Now an immediate ``paradox'', raised in ref.~\cite{ms}, is 
that even though 
\eqs\nr{Delta_theta}, \nr{Delta_chi} have a common positive-definite
thermal part at $\rmO(g^4)$, the integral over this is supposed
to vanish, since the left-hand sides of \eqs\nr{sum_theta}, \nr{sum_chi}
are of $\rmO(g^6)$ or higher. A resolution to this paradox in the 
{\em renormalized case}, in which the coupling constant runs, was given 
in ref.~\cite{sch}. 
Beautiful as the argument is, underlining the strength of 
sum rules in that they allow us to anticipate features of a higher-order
computation without carrying it out, the method is cumbersome in 
our case when we want to use the sum rules as an {\em exact} crosscheck 
of an order-by-order  computation. In the following we therefore 
``freeze'' the gauge coupling, so that different orders do not mix. 
The price to pay is that then the issue of ``contact terms'', 
discussed e.g.\ in ref.~\cite{rs}, needs to be addressed. 
On the other hand, the contact terms necessarily play a 
role in the ``shear'' channel~\cite{sch,hbm_e}, where similar 
structures appear but without $g^4$ and running, so 
the discussion may be useful as an analogue. 

Contact terms arise when there is a part in the Euclidean correlators, 
$\Delta\tilde G_\theta(P)$ and $\Delta\tilde G_\chi(P)$, 
which does not vanish as $p_n^2 \to \infty$. This part is  ``lost''
to the spectral function, and needs to be ``added'' to 
the right-hand side of the sum rule, before a comparison with 
the thermodynamic left-hand side, in which the information
is included, can be carried out. (A more precise 
account of the logic can be found, e.g.,\ in ref.~\cite{rs}.)

{}From \eqs(4.1) and (4.2) of ref.~\cite{ope}, setting
$P=(p_n,\vec{0})$ and inserting 
$
  \int_\vec{q} q \, \nB{}(q) = \pi^2 T^4/30
$, 
$
  \int_\vec{q} \nB{}(q)/q = T^2/12
$, 
the ultraviolet limits read
\ba
 \frac{\Delta \tilde G_\theta(P)}{4d_A c_\theta^2} 
 & \stackrel{p_n \gg \pi T }{=} 
 & - \frac{ 4 \pi^2 g^4 T^4 }{15} 
 \biggl[
       1 + \frac{g^2\Nc}{(4\pi)^2}
  \biggl(
    \frac{22}{3} \ln\frac{\bmu^2}{p_n^2} + \frac{203}{18} 
  \biggr) 
  \biggr] 
 + \frac{g^6 \Nc T^4}{18} + \rmO\Bigl( \frac{g^4}{p_n^2} ,g^8 \Bigr)
 \;, \nn \la{contact_theta} \\ 
 \frac{\Delta \tilde G_\chi(P)}{-16d_A c_\chi^2} 
 & \stackrel{p_n \gg \pi T }{=}
 & - \frac{ 4 \pi^2 g^4 T^4 }{15} 
 \biggl[
       1 + \frac{g^2\Nc}{(4\pi)^2}
  \biggl(
    \frac{22}{3} \ln\frac{\bmu^2}{p_n^2} + \frac{347}{18}
  \biggr) 
  \biggr] 
 + \frac{g^6 \Nc T^4}{9} + \rmO\Bigl( \frac{g^4}{p_n^2} ,g^8 \Bigr)
 \;, \hspace*{6mm} \nn \la{contact_chi} 
\ea
where we have kept separate the terms coupling
to two different operators in the OPE limit. 

At $\rmO(g^4)$, 
we now directly observe that if the contact contributions from 
\eqs\nr{contact_theta}, \nr{contact_chi} are added to the integrals
over the leading-order terms in \eqs\nr{Delta_theta}, \nr{Delta_chi}, 
which evaluate to 
\be 
  2 g^4
 \int_{0}^{\infty} \! \frac{{\rm d}\omega}{\pi\omega}
 \, \frac{\pi\omega^4}{(4\pi)^2}
 \; 2 n_{\frac{\omega}{2}} 
  \stackrel{x = \sfr{\omega}2}{=} \frac{4g^4}{\pi^2} 
 \int_0^\infty \! {\rm d}x \, x^3 n_x 
 = \frac{4 \pi^2 g^4 T^4}{15}
 \;, \la{rhs1}
\ee
then the desired cancellation duly takes place. 

As far as $\rmO(g^6)$ goes, we observe that 
the constants next to the logarithms
in \eqs\nr{contact_theta}, \nr{contact_chi} can be written as
$
 \frac{203}{18} = \frac{73}{3} - \frac{235}{18}
$, 
$
 \frac{347}{18} = \frac{97}{3} - \frac{235}{18} 
$.
By making use of \eq\nr{rhs1}, 
the parts containing $\frac{73}{3}$ and 
$\frac{97}{3}$ are seen to exactly 
cancel against corresponding 
terms from the integrals over the spectral functions
in \eqs\nr{Delta_theta}, \nr{Delta_chi}. 
Therefore, the non-trivial sum rule has a 
common $-\frac{235}{18}$ in both channels. 
We also observe that the difference in the last 
terms of \eqs\nr{contact_theta} and \nr{contact_chi} precisely
explains the difference of the left-hand sides of 
\eqs\nr{sum_theta}, \nr{sum_chi}. So, we are left with  
one non-trivial sum rule to check; we choose \eq\nr{sum_chi}. 

It requires some care to formulate this sum rule. 
The reason is the asymptotics of \eq\nr{phi_T_as} which, for 
a {\em frozen} coupling, implies that the integral in \eq\nr{sum_chi} 
diverges. This is reflected by the appearance 
of the logarithms involving $p_n^2$ on the right-hand 
side of \eq\nr{contact_chi} (the imaginary part of this logarithm 
gives the asymptotics of \eq\nr{phi_T_as}). Let us regulate 
the divergence by restricting $\omega$ to $|\omega| \le \Lambda$.
Noting that 
\be
 \im \biggl( \ln\frac{1}{p_n^2} 
 \biggr)_{p_n \to -i[\omega + i 0^+] } \; = \; \pi 
 \;, \quad
 2 \int_{ }^{\Lambda} \! \frac{{\rm d}\omega}{\pi\omega}\; \pi
 \; = \; \ln \Lambda^2
 \;,
\ee
we deduce that in the contact term  we can  
replace $\ln p_n^2\to \ln \Lambda^2$ in the presence of the cutoff. 
With this logic, the coefficients of $g^6 \Nc$ 
in \eqs\nr{sum_chi}, \nr{Delta_chi} and \nr{contact_chi} imply the
sum rule 
\ba
 \lim_{\Lambda\to\infty}
 \biggl\{ 
 2 \int_0^\Lambda 
 \! \frac{{\rm d}\omega \, \omega^3}{(4\pi)^2}
 \biggl[ 
   2 n_{\frac{\omega}{2}}
    \biggl(
       \frac{22}{3} \ln\frac{\bmu^2}{\omega^2} 
    \biggr)
  + 8  \bigl( 1 + 2 n_{\frac{\omega}{2}} \bigr)
  \phi^{ }_T(\omega)
 \biggr]
 \!\!&-&\!\! \frac{4\pi^2 T^4}{15}
 \biggl( 
       \frac{22}{3} \ln\frac{\bmu^2}{\Lambda^2} 
 \biggr) \biggr\} 
 \nn  
 &=& \frac{4\pi^2 T^4}{15}
 \biggl( 
        - \frac{355}{18}
 \biggr)
 \;, \hspace*{1cm}
\ea
where $- \frac{355}{18} = - \frac{235}{18} - \frac{15\times 4}{9}$.
We have verified numerically (with a relative error smaller
than $\frac{1}{355}$)
that this sum rule is satisfied by our function $\phi^{ }_T$.

Sum rules could also be used in connection with the 
HTL-results in \eqs\nr{HTL_resum_theta}, \nr{HTL_resum_chi}. 
As far as the thermodynamic
left-hand sides are concerned, \eq\nr{Gtheta_HTL} yields 
a contribution $\frac{5}{4\pi}(\frac{\Nc}{3})^{\fr32} g^7 T^4$
to \eq\nr{sum_theta}, which equals the next-to-leading order 
correction to $T^5 \frac{{\rm d}}{{\rm d}T} (\frac{e-3p}{T^4})$, 
whereas \eq\nr{Gchi_HTL} implies that the left-hand side of 
\eq\nr{sum_chi} vanishes even in the presence of HTL-resummation. 
However, there are again contact terms which need to be 
added in order to satisfy the sum rules. Given that HTL-resummation
is of secondary importance for our study, we refrain from a more
detailed discussion here, noting only that a cancellation 
in the $\chi$-channel is also visually suggested 
by \fig\ref{fig:phi_HTL}(right).

%
\subsection{Numerical evaluation}
\la{ss:plots}

After the various crosschecks carried out in secs.~\ref{ss:UV}, \ref{ss:IR} 
and \ref{ss:sum}, 
we now move on to the numerical evaluation
of our results. The goal is to obtain curves
which can in principle be used in the applications
outlined in the introduction. 

In order to evaluate our result 
(\eqs\nr{rhofinal_theta} + \nr{rhofinal_chi} + \nr{master_resum})
numerically, we need to assign a value to the running coupling $g^2$.
It is only in the regime $\omega \gg \pi T$, where $\phi^{ }_T \ll 1$,  
that two subsequent orders with the same functional form are at our 
disposal, so that a scale optimization is possible; 
we then define $\bmu^\rmi{opt}_{\theta,\chi}$ from 
the ``fastest apparent convergence'' criterion based 
on \eqs\nr{rhofinal_theta}, \nr{rhofinal_chi}: 
\be
 \ln(\bmu^\rmi{opt($\omega$)}_{\theta}) \equiv \ln(\omega) 
 -  \frac{73}{44}
 \;, \quad
 \ln(\bmu^\rmi{opt($\omega$)}_{\chi}) \equiv \ln(\omega) 
 -  \frac{97}{44}
 \;. \la{muopt_w}
\ee
In the infrared regime $\omega \ll \pi T$ we choose 
$g^2$ from the ``EQCD'' setup (cf.\ ref.~\cite{gE2} and 
references therein) 
which at 1-loop order amounts to:
\be
 \ln(\bmu^\rmi{opt($T$)}_{\theta,\chi}) \equiv 
 \ln(4\pi T) - \gammaE - \frac{1}{22}
 \;. \la{muopt_T}
\ee 
For a given $\omega$ the larger among these scales is chosen; the 
switch happens at $\omega \approx 11.3\, \pi T$ for $\rho^{ }_\theta$, 
and at $\omega \approx 19.5\, \pi T$ for $\rho^{ }_\chi$. We note 
that the extremely ``late'' transition to the vacuum scaling is 
related to the well-known slow convergence of the perturbative 
expansion for the Wilson coefficients appearing in the OPE regime. 
Indeed the fact that in the full computation
there is another scale present, $\pi T$, which acts as a sort of 
an infrared cutoff, allows us to obtain a numerical evaluation for 
frequencies much smaller than in the OPE regime~\cite{ope}.
There is a certain price to pay, however, which is that it is no longer
possible for us to split the result into a ``zero-temperature'' and a 
``finite-temperature'' part, because they have been mixed by 
the scale choice. (Put another way, the two parts separately 
are quite sensitive to the confinement scale, but their sum, which 
we evaluate, is not.)
 
In order to get an error band for the uncertainty 
related to the choice of $g^2$, we use the 
2-loop running of $g^2$,  with the renormalization scale  
varied in the range  $(0.5 \ldots 2.0)\times \bmu^\rmi{opt}$.
The scale parameter, defined as 
  $
   \Lambdamsbar \equiv \lim_{\bmu\to\infty}
   \bmu \bigl[ b_0 g^2 \bigr] ^{-b_1/2 b_0^2}
   \exp \bigl[ -\frac{1}{2 b_0 g^2}\bigr]
  $, 
is re-expressed through the critical temperature
of the deconfinement transition of pure SU(3) gauge 
theory; we take this to be $\Tc = 1.25 \Lambdamsbar$, 
which could also be viewed as our definition of ``$\Tc$''.

\begin{figure}[t]


\centerline{%
\epsfysize=7.5cm\epsfbox{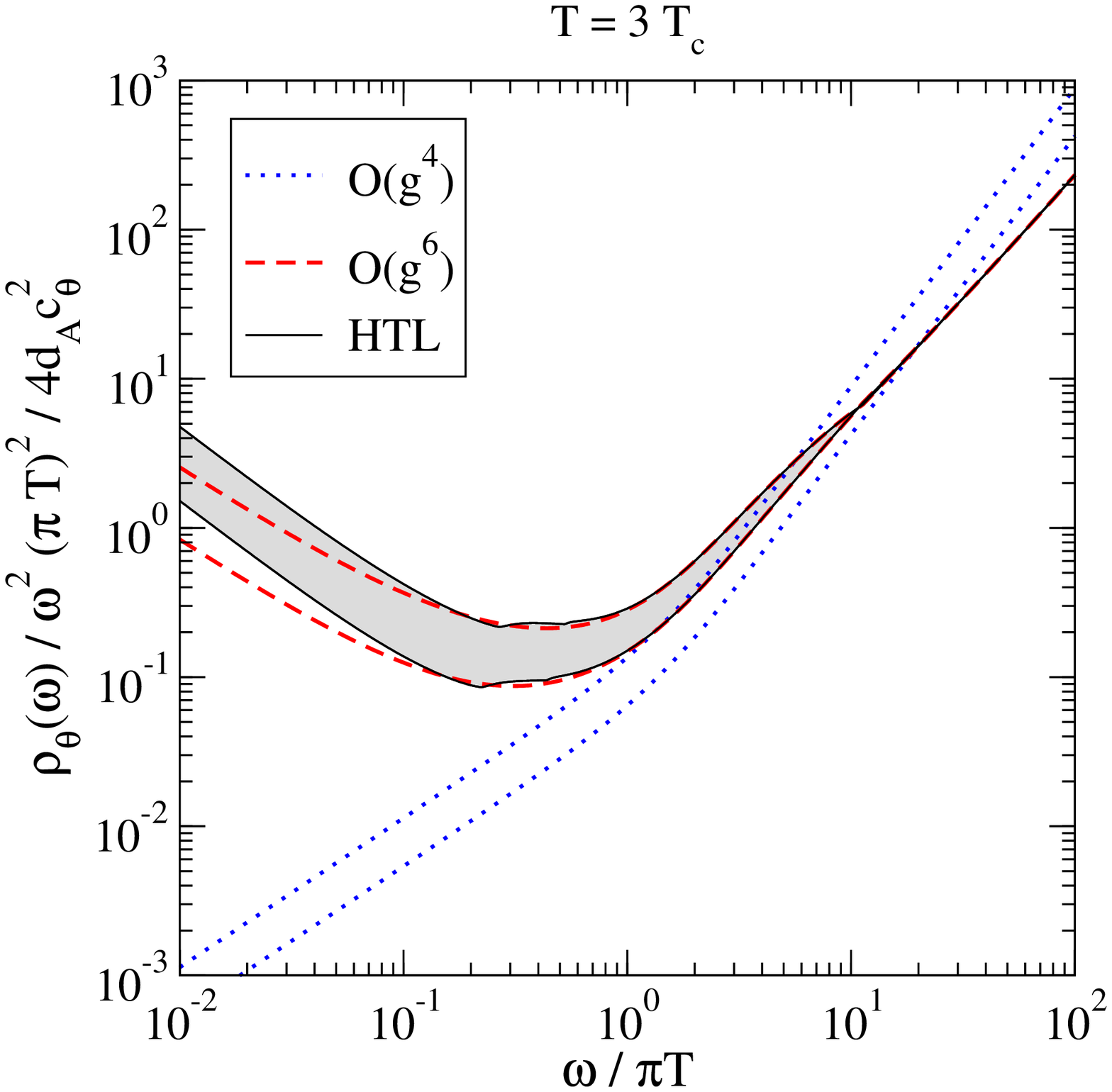}%
~~\epsfysize=7.5cm\epsfbox{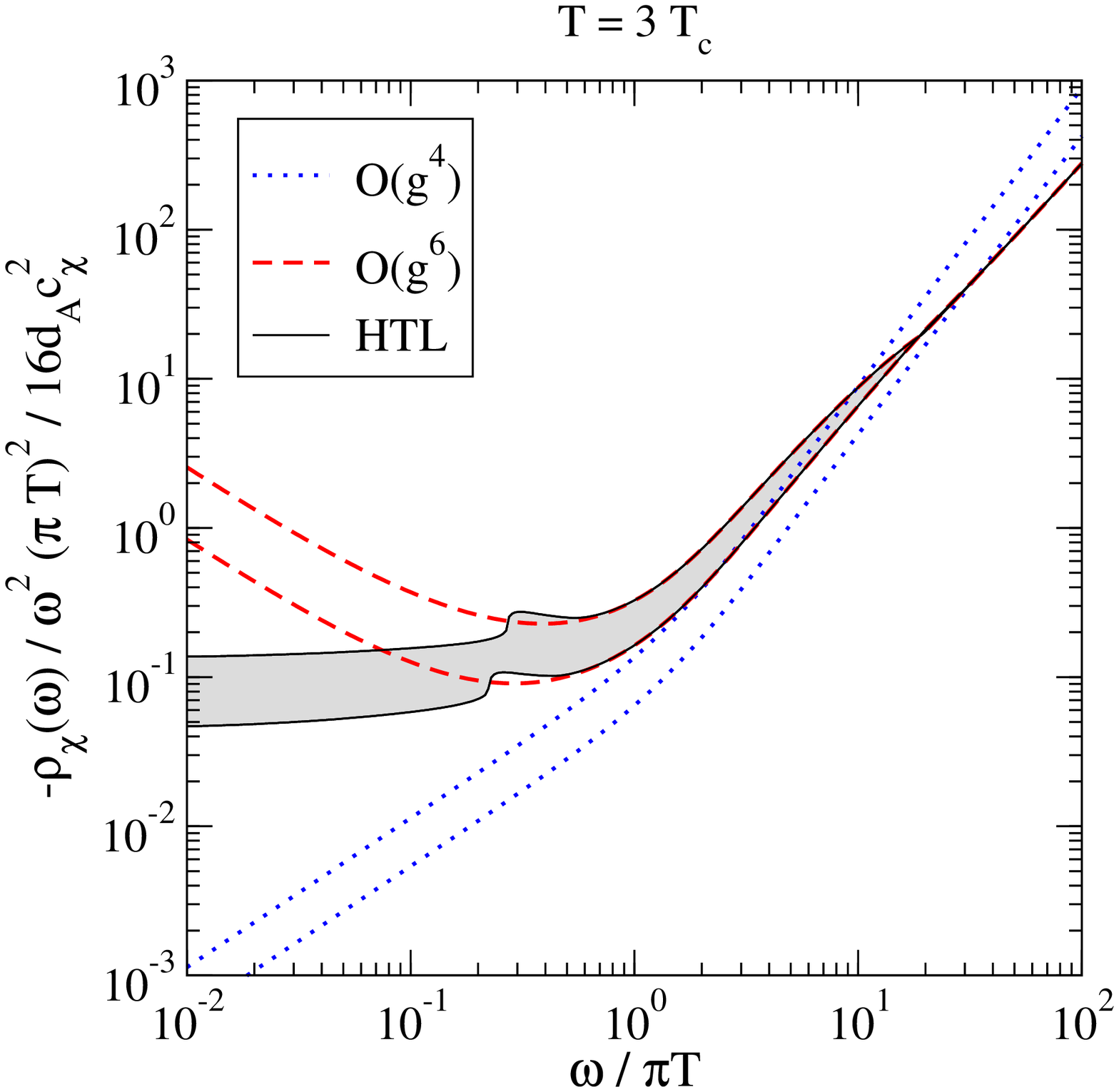}%
}

\caption[a]{\small 
  A numerical evaluation of \eqs\nr{rhofinal_theta},
  \nr{rhofinal_chi}, \nr{master_resum},
  in units of  $\omega^2 (\pi T)^2$, for 
  $T = 3.75 \Lambdamsbar$ 
  corresponding to   
  $T = 3 \Tc$.
  The gauge coupling has been fixed as explained 
  around \eqs\nr{muopt_w}, \nr{muopt_T}, 
  and the gray band reflects the corresponding uncertainty.
  (In the $\rmO(g^4)$ result the ``optimal'' scale 
   is fixed to the thermal value of \eq\nr{muopt_T}, 
   i.e.\ does not change with the frequency.)
 }
\la{fig:wdep}
\end{figure}

The outcome of a numerical evaluation 
of \eqs\nr{rhofinal_theta}, \nr{rhofinal_chi}, \nr{master_resum} 
is plotted in \fig\ref{fig:wdep}, in units of $\omega^2 (\pi T)^2$.
We observe that the dependence on the scale choice practically
disappears from the NLO result 
in the ultraviolet domain, $\omega \gg \pi T$. 
As far as the infrared domain is concerned, we observe 
an enhancement due to HTL-resummation in the $\theta$-channel 
akin to that seen in the vector channel~\cite{bp_dilepton}, 
whereas in the $\chi$-channel the HTL-resummation removes much of the 
spectral weight from small frequencies. 

In ref.~\cite{hbm_d}, a significant ``cancellation'' in  
the {\em thermal part} of the spectral function was advocated for, 
with a positive domain at small frequencies and a negative domain
at large frequencies. For reasons mentioned above, we do not believe
our numerical evaluations to be quantitatively
reliable for the zero-temperature and 
finite-temperature parts separately, 
and therefore abstain from a direct comparison. Nevertheless, 
we note that on the qualitative level the structure of the 
finite-temperature part as discussed in ref.~\cite{hbm_d} 
(cf.\ \fig5 there)
agrees perfectly
with that of the function~$\phi^{ }_T$
as shown in \fig\ref{fig:phiT}.

%
\subsection{Imaginary-time correlators}
\la{ss:tau}

Although the ultimate goal of a lattice study could be 
the determination of a transport coefficient, which is by definition
a Minkowskian object and obtained from the zero-frequency intercept
of $\rho(\omega,\vec{0})/\omega$, 
any practical measurement is carried out in 
Euclidean signature. It is relatively easy to go from a known
Minkowskian spectral function
to a Euclidean correlator, but very difficult to take the inverse step;
therefore, as a first comparison between lattice data and
an analytic computation, it may be helpful to extract
the Euclidean correlator from the latter. The principal tool for this is
the well-known formula 
\be
 G(\tau) =
 \int_0^\infty
 \frac{{\rm d}\omega}{\pi} \rho(\omega,\vec{0})
 \frac{\cosh \left(\frac{\beta}{2} - \tau\right)\omega}
 {\sinh\frac{\beta \omega}{2}} 
 \;. \la{int_rel} 
\ee

As it turns out, there are some problems hidden in \eq\nr{int_rel}. 
One is the issue of ``contact terms'', discussed in 
\se\ref{ss:sum}; as the name says, one may
move them to the Euclidean side of the relation and interpret
as a zero-distance term $\propto \delta(\tau)$, giving a finite 
contribution to $\int_0^\beta \! {\rm d}\tau \, G(\tau)$. 
Here we stay at non-zero $\tau$, and this issue
does not arise. However, there is another issue, namely the 
possible existence of terms of the type $\sim \omega\delta(\omega)$
in the spectral function which, according to \eq\nr{int_rel}
(replacing 
$\int_0^\infty \! {\rm d}\omega \to 
 \fr12 \int_{-\infty}^\infty \! {\rm d}\omega
$)
can give a constant contribution to the Euclidean correlator. 
This can be important 
particularly around the middle of the Euclidean time interval
where the absolute value of the correlator is smallest
(cf.\ e.g.\ ref.~\cite{umeda} and \fig\ref{fig:taudep} below).

Before proceeding with the specific discussion, let us note that
in general the two issues mentioned are somewhat ``complementary''. 
Operators with a higher dimension are more ultraviolet
sensitive; this means that, like in many of the sum-integrals
in \eqs\nr{m_first}--\nr{m_last}, explicit powers of $P^2$ appear 
in the numerator and therefore contact contributions, 
originating from a non-trivial limit of the Euclidean correlators 
at $p_n^2 \gg (\pi T)^2$, play a role; cf.\ \se\ref{ss:sum}.
If the operator has a lower dimension, with  
less or no powers of $P^2$, 
then the ultraviolet regime is safer; on the other hand, 
the issue with zero-mode contributions is more likely to arise. 
(Only in exceptional cases are both problems avoided~\cite{eucl,rhoE}.)

Let us recall a simple illustration of the issue. Consider
the structure defined in \eq\nr{m_Jb}, but {\em without} the $P^4$
in the numerator, 
\ba
 \mathcal{J}^{ }_\rmi{b'}(P) & \equiv &
 \Tint{Q} \frac{1}{Q^2(Q-P)^2}
 \;. \la{Jb_prime}
\ea
Carrying out the Matsubara sum and taking the imaginary part 
leads to (we keep $\vec{p}\neq\vec{0}$ for the moment; 
$E_{qp} \equiv |\vec{q} - \vec{p}|$)
\ba
 \rho^{ }_{\mathcal{J}^{ }_\rmii{b'}}(\omega,\vec{p}) & = & 
 \int_{\vec{q}} \frac{\pi}{4 q E_{qp} }
 \biggl\{ 
 \Bigl[ \delta(\omega - q - E_{qp}) - \delta(\omega + q + E_{qp} ) \Bigr]
 (1+n_q + n_{qp}) 
 \nn & & \hspace*{1.5cm} + \,
 \Bigl[ \delta(\omega - q + E_{qp}) - \delta(\omega + q - E_{qp} ) \Bigr]
 (n_{qp} - n_q ) 
 \biggr\} 
 \;. \la{m_Jb_sum}
\ea
(This is like \eq\nr{Jb_sum} but now with the second term added.)
For $\omega\to 0$ the first row gives no contribution because of
vanishing phase space, whereas on the second row we can write
\ba
 & & \hspace*{-2cm} 
 \Bigl[ \delta(\omega - q + E_{qp}) - \delta(\omega + q - E_{qp} ) \Bigr]
 (n_{qp} - n_q ) 
 \nn
 & = & 
 \delta(\omega - q + E_{qp})
 \bigl( n_{q - \omega} - n_{q} \bigr)
 - 
 \delta(\omega + q - E_{qp} )  
 \bigl( n_{q + \omega} - n_{q} \bigr)
 \nn 
 & \stackrel{\vec{p}\to\vec{0}}{\approx} & 
 - 2\, \omega \delta(\omega) n_{q}'
 \; = \; 
 2 \beta \omega \delta(\omega)\, n_{q} \bigl( 1 + n_{q} \bigr)
 \;. 
\ea
Inserting this into \eq\nr{int_rel} 
(with $\int_0^\infty \! {\rm d}\omega \to 
 \fr12 \int_{-\infty}^\infty \! {\rm d}\omega$) 
yields 
\be
 \delta G(\tau) = 
 \int_{-\infty}^{\infty} \! \frac{{\rm d}\omega}{2\pi}
 \int_\vec{q} \frac{\pi}{4 q^2} \,
 2 \beta\omega\delta(\omega) n_{q} \bigl( 1 + n_{q} \bigr)
 \frac{2}{\beta\omega}
 \; = \; 
 \int_\vec{q} \frac{1}{2 q^2} \,
 n_{q} \bigl( 1 + n_{q} \bigr)
 \;,
\ee
which would yield a constant ($\tau$-independent)
contribution of the type advertised (in this case 
it is even infrared divergent). 

In our situation, however, \eq\nr{Jb_prime} contains an additional $P^4$ or, 
after analytic continuation, $\omega^4$. This kills the contribution 
from $\omega\delta(\omega)$ at $\rmO(g^4)$. At higher orders, the loop 
expansion may introduce additional factors of 
$g^2 n_{\omega} \approx {g^2T} / {\omega}$ or, if there is 
a factorized ``hard thermal loop'' involved, even $g^2T^2 / \omega^2$.
 Indeed it is expected that at $\rmO(g^8)$ a contribution 
 $\sim \omega\delta(\omega)$ arises to $\rho^{ }_\theta$, proportional
 to the heat capacity and related to the fact that $\int_\vec{x}\theta$ 
 has an overlap with a conserved charge (energy)~\cite{ms,rs,hbm_d}. 
At our order, $\rmO(g^6)$, no such contributions are present and this is 
the reason that we have omitted from the outset all 
structures $\sim \omega^n\delta(\omega)$ in appendices A and B.

Even if there are no contributions 
to $G(\tau)$ from $\omega\delta(\omega)$-peaks, then how about
the infrared structures at small but non-zero frequencies, 
$\omega \lsim gT$, 
discussed in secs.~\ref{ss:IR}, \ref{ss:XIR}?
We note that, parametrically, the range 
$\omega \lsim \mE$ gives 
a contribution to \eq\nr{int_rel} of magnitude
\be
 \delta G (\tau,\vec{0}) 
 \sim 
 \int_0^{\mE} \! \frac{{\rm d}\omega}{\pi} 
 \frac{T \rho(\omega,\vec{0})}{\omega}
 \sim 
 g^4 \mE^3 T^2 \sim \rmO(g^7 T^5)
 \;, \la{ir_est}
\ee
where we made use of the estimate 
$
 \rho(\omega,\vec{0}) \sim g^4 \mE^2 T \omega
$
from \se\ref{ss:IR}. This is formally of higher order 
than the ultraviolet contribution, which is $\rmO(g^6T^5)$, 
but nevertheless more important than a full 3-loop analysis, 
of $\rmO(g^8 T^5)$. (The estimate in \eq\nr{ir_est} also represents the 
area delineated by the transport peak and the $\omega$-axis~\cite{ms}.)

\begin{figure}[t]


\centerline{%
\epsfysize=7.5cm\epsfbox{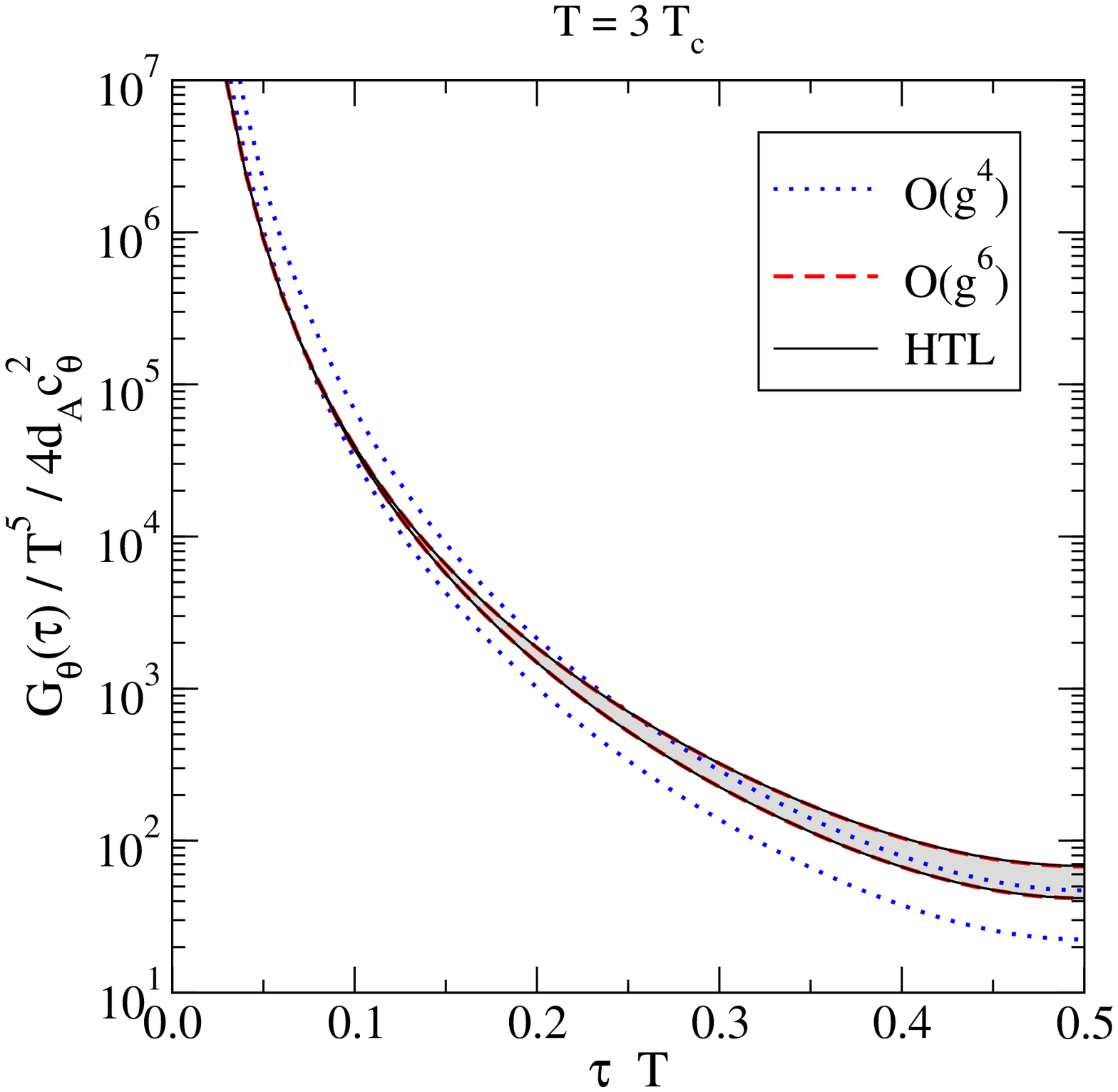}%
~~\epsfysize=7.5cm\epsfbox{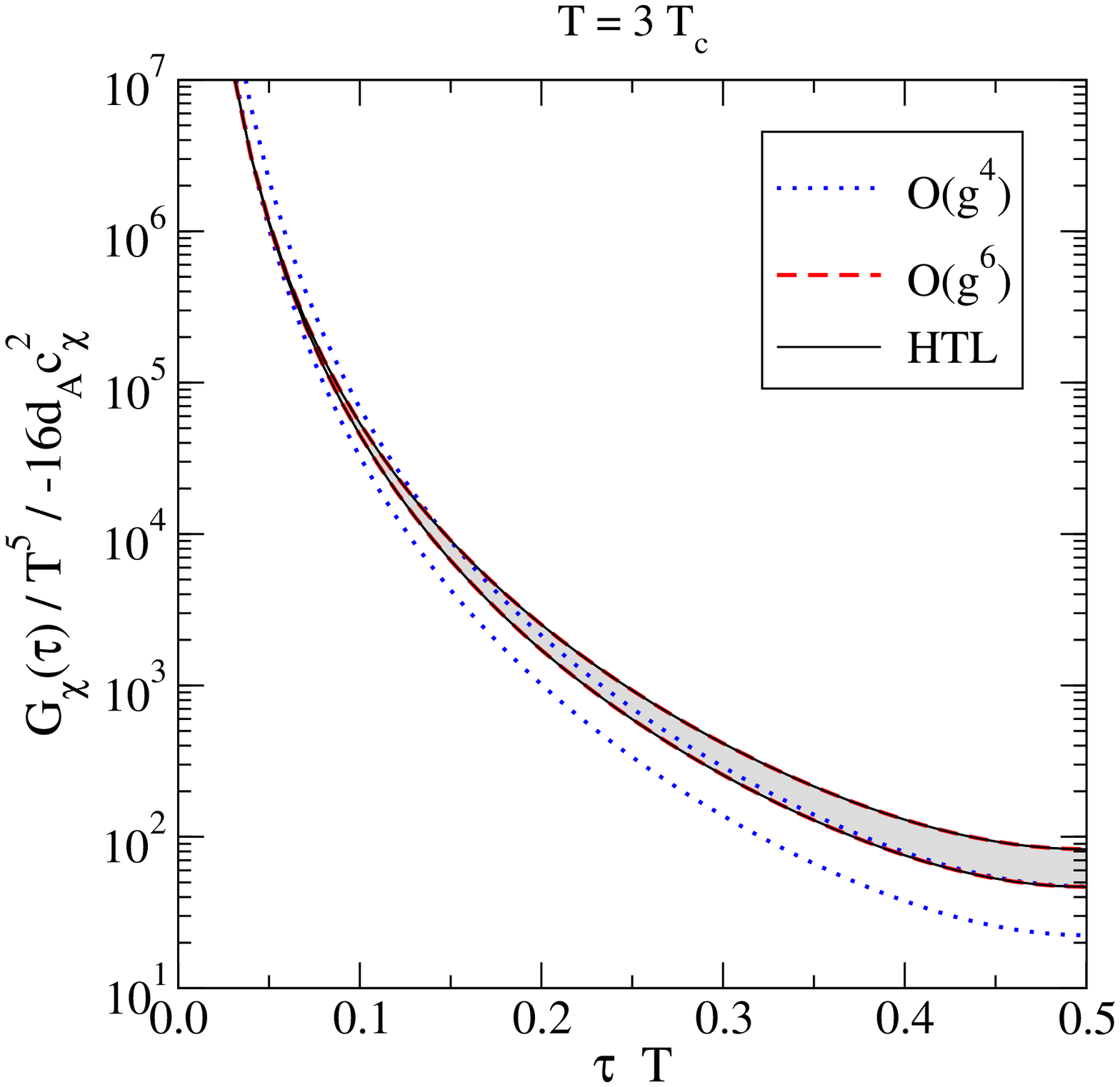}%
}

\caption[a]{\small 
  A numerical evaluation of $G_{\theta}$ (left) and 
  $G_{\chi}$ (right),
  in units of  $T^5$, for 
  $T = 3.75 \Lambdamsbar$
  corresponding to   
  $T = 3 \Tc$.
  The gauge coupling has been fixed as explained 
  in \fig\ref{fig:wdep}, 
  and the gray band reflects the corresponding uncertainty.
 }
\la{fig:taudep}
\end{figure}

After these lengthy qualifications, 
we apply  \eq\nr{int_rel} to the results 
in \eqs\nr{rhofinal_theta}, \nr{rhofinal_chi}, \nr{master_resum}.
In accordance with the discussion around \eq\nr{ir_est}, 
the significant difference of the spectral 
functions at $\rmO(g^4)$ and $\rmO(g^6)$ in the infrared 
regime, visible in \fig\ref{fig:wdep}, converts to a rather 
mild enhancement of the NLO Euclidean correlator around
the center of the Euclidean time interval in \fig\ref{fig:taudep}.
Even though $\rho^{ }_\chi$ is suppressed
relative to $\rho^{ }_\theta$ at small frequencies, 
the enhancement is larger in $G^{ }_\chi$, because of a contribution
from large frequencies.


\begin{figure}[t]


\centerline{%
\epsfysize=7.5cm\epsfbox{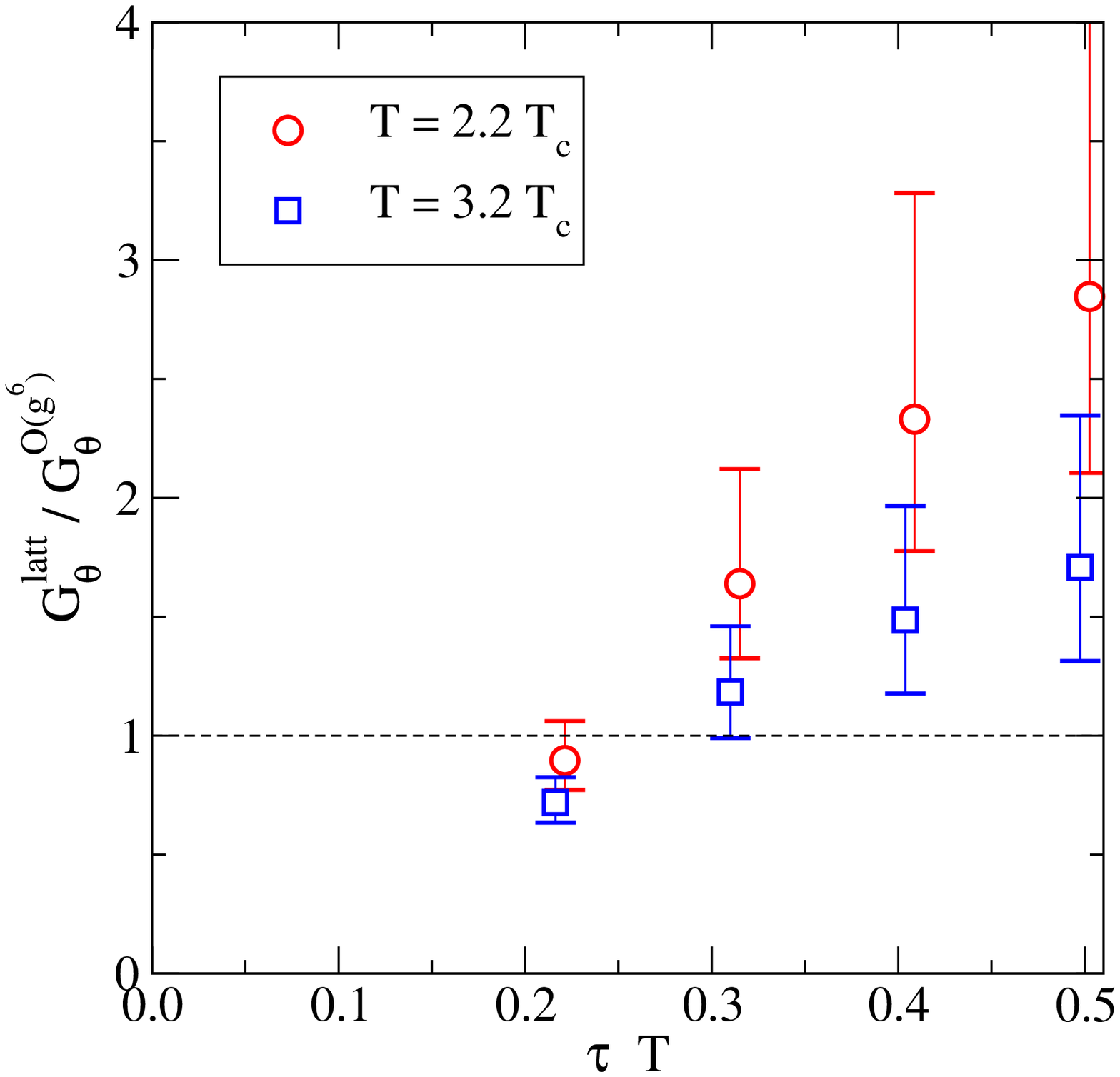}%
~~\epsfysize=7.5cm\epsfbox{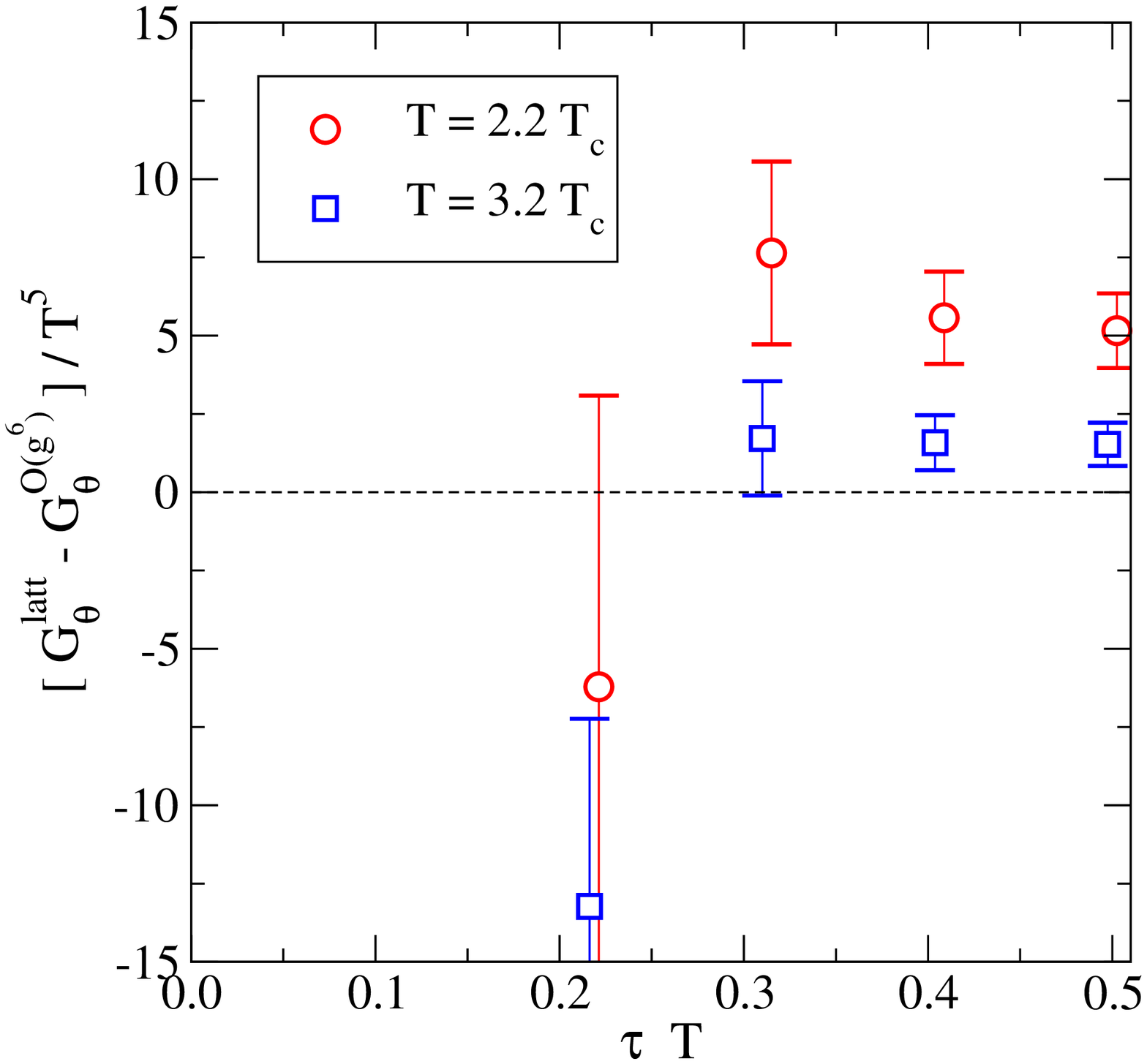}%
}

\caption[a]{\small 
  The ratio (left) and difference (right) of the lattice
  data from \fig3(left) of ref.~\cite{hbm_d} and our results
  from \fig\ref{fig:taudep}(left). The symbols have been slightly
  displaced for better visibility; the error bars have 
  been obtained by varying both results independently 
  within their uncertainties. Lattice errors are 
  statistical only; discretization artifacts could  
  affect the difference at the shortest distance, 
  where a significant cancellation takes place, 
  the absolute value being about $G^{ }_\theta/T^5 \sim 50$ there.  
 }
\la{fig:compare}
\end{figure}

%
%
%

Lattice Monte Carlo results for $G^{ }_\theta / T^5$ in 
pure SU(3) gauge theory have been presented in ref.~\cite{hbm_d}. 
Multiplying our values by 
$
 4 d_A c_\theta^2 \approx 8 (\frac{11}{(4\pi)^2} + \frac{51g^2}{(4\pi)^4})^2 
 \approx 0.043
$, 
we can carry out a direct comparison; the results are shown
in \fig\ref{fig:compare}. (We use the $\rmO(g^6)$ rather 
than the HTL result here, because it gives a simpler 
partial contribution to the transport coefficient, 
cf.\ \eq\nr{phi_IR}; however 
the difference is barely visible, 
cf.\ \fig\ref{fig:taudep}.) 
As far as the ratio is concerned, we are 
quite impressed by the good semi-quantitative 
agreement at short distances. For the difference, 
the good news is that after the subtraction a function is 
obtained which no longer shows a visible short-distance 
divergence; this implies that a model-independent analytic 
continuation could in principle be attempted. 
Obviously, however, the significance loss caused by the 
subtraction implies that it remains a tremendous challenge to obtain 
the per-mille accuracy level that 
might turn out to be necessary for this~\cite{analytic}. 
(There is the further challenge that, as discussed above, 
a constant $\tau$-independent part in $G^{ }_\theta$ corresponds
to a term $\sim \omega\delta(\omega)$ in the spectral function
and does not contribute to the transport coefficient; the resolution
has to be good enough to show a statistically significant 
$\tau$-dependence in the subtracted correlator.)

%
\section{Conclusions and outlook}
\la{se:concl}

Even though our study has centered around 
a specific example, defined in \se\ref{se:setup} and concerning
the ``bulk'' and ``topology'' channels of pure Yang-Mills theory, 
its wider point, as explained in \se\ref{se:method}, 
has been to elaborate on a general method for computing 
next-to-leading order thermal spectral functions for composite operators. 
We have shown how the method described 
yields a relatively simple result, 
\eqs\nr{rhofinal_theta}--\nr{phi_T}, which is well 
suited for a numerical evaluation (\fig\ref{fig:wdep}); as well as how 
the result can be crosschecked, resummed, and applied in 
a number of ways (\se\ref{se:analysis}). 
Hopefully the step-by-step description of the labour-intensive
phase of the procedure (appendix A for the most complicated
``master'' sum-integral appearing at 2-loop level), will also
turn out to lend itself to a number of generalizations; 
the simplest of them may be the inclusion of a non-trivial 
mass spectrum, which in a specific case has 
in fact already been achieved for many intermediate stages. 

In the course of the computation  
we have shown that various divergences, 
sometimes classified as being of ``soft'', ``collinear'', 
or ``thermal'' origin, cancel in every of our ``master'' 
spectral functions separately. That this happens for 
the sum is in accordance with the general OPE analysis of 
ref.~\cite{sch}, showing the absence of infrared divergences 
up to 4-loop level in the regime $\omega \gg \pi T$, and also 
with a NLO analysis of the vector current correlator at $\omega\sim\pi T$, 
where the cancellation of divergences was verified
long ago~\cite{spectral1}--\cite{spectral3}.
It also agrees with the specific discussion of another correlator 
in ref.~\cite{sls} for the limit $\omega \gg \pi T$, 
which found terms proportional 
to $T^2$ to be finite at NLO. However, our results are valid 
down to $\omega\sim \pi T$ and somewhat below it,  
and for any single 2-loop master spectral function separately.

For small frequencies, $\omega \ll \pi T$,  
the naive perturbative series does break down. Nevertheless,  
as long as $g^2 T/\pi \ll \omega \ll \pi T$, it can (probably)
still be ``repaired'' through a Hard Thermal Loop resummation
(\se\ref{ss:IR}). The price for the resummation
is that in this regime only one 
order of the weak-coupling series is available, and NLO corrections 
could well be substantial~\cite{hm}. Once $\omega \lsim g^2 T/ \pi$, 
even the resummed result loses its validity, and other 
techniques are needed (cf.\ \se\ref{ss:XIR}); this regime
is not addressed in the present study.  

Even if the small-frequency regime of a spectral function 
suffers from infrared problems, the corresponding imaginary-time
correlator can still be computed and compared directly with 
lattice simulations, because the contribution emerges dominantly from 
ultraviolet frequencies, $\omega \gsim \pi T$ (cf.\ \se\ref{ss:tau}). 
The ultimate 
use envisaged for such results within thermal QCD is that the perturbative 
imaginary-time correlator can be subtracted from a lattice measurement; 
the remainder could be 
ultraviolet finite and only sensitive to thermal infrared
physics, hence showing relatively 
little ``structure''; and that therefore
the spectral function corresponding to the remainder could be 
reconstructed with less model input than without the 
subtraction. That such a program may be feasible is demonstrated 
by \fig\ref{fig:compare}(right) for the bulk channel, although much further 
work is needed before the quality of the remainder is satisfactory.

As promised in the introduction we finally comment 
on cosmological applications. 
In the present paper the spectral functions were computed at vanishing
spatial momentum, i.e.\ for $\mathcal{P}=(\omega,\vec{0})$ in the notation
of the introduction, or $P = (-i\omega,\vec{0})$ in the Euclidean
notation used elsewhere. If, in contrast,  
the pseudoscalar correlator were computed ``on-shell'', i.e.\ for 
$P = (-i E_p,\vec{p})$, then it would be directly proportional 
to the production rate of axions (see, e.g., ref.~\cite{fds}), which 
are among standard Dark Matter candidates. In the 
case of a scalar density, one could similarly envisage a  
production rate of dilatons, although in 
this case  it might
be natural to assume the dilatons to further decay, perhaps
into particle--antiparticle pairs in a ``hidden sector''. Then the 
dilaton itself could be off-shell, and our result would
yield the production rate of the hidden-sector pair, 
in the same way as the spectral 
function of the electromagnetic current in QCD with 
$ 
 P = (-i\omega,\vec{0})
$ 
yields the production rate of a dilepton pair with 
a total energy equal to $\omega$
(cf.\ ref.~\cite{spectral1}). 
As far as the corresponding transport coefficients
are concerned, that related to the pseudoscalar density yields
for $\Nc = 2$ the baryon number violation rate which is certainly 
important for cosmology~\cite{krs}
(even the case $\Nc = 3$ may have some relevance, 
cf.\ the discussion in ref.~\cite{mt} and references therein), 
whereas bulk viscosity could 
determine e.g.\ a moduli decay rate~\cite{moduli1,moduli2}. 
Beyond this, the sum-integrals that we have considered are similar
to those appearing in other production rates, e.g.\ 
that of right-handed fermions~\cite{db}, and so 
(with the transformation of some lines to fermions, which 
goes simply by changing $\nB{}\to -\nF{}$), we envisage
that our techniques extended to $P = (-i E_p,\vec{p})$
might find further applications. In fact, in ref.~\cite{sls}, 
the results were approximated by setting $P=(-iM,\vec{0})$, 
and then the kinematics is identical to ours.  
We hope to be able to apply our methods 
to some of these problems in future work. 

%
\section*{Acknowledgements}

 We thank L.~Mether for collaboration during initial
 stages of this work, and H.~Meyer for providing us 
 with lattice data from ref.~\cite{hbm_d}. 
 M.L.\ was partly supported by 
 the BMBF under project
 {\em Heavy Quarks as a Bridge between
      Heavy Ion Collisions and QCD},
 A.V.\ by the Sofja Kovalevskaja program 
 of the Alexander von Humboldt foundation, 
 and Y.Z.\ by the DFG International Graduate School 
 {\em Quantum Fields and Strongly Interacting Matter}.

\newpage

%
\appendix
\renewcommand{\thesection}{\Alph{section}} 
\renewcommand{\thesubsection}{\Alph{section}.\arabic{subsection}}
\renewcommand{\theequation}{\Alph{section}.\arabic{equation}}

%
\section{Detailed procedure with the example 
of $\rho_{\mathcal{I}^{ }_\rmii{j}}$}
\la{se:app1}

Let us consider the master sum-integral defined as 
(this is a slight generalization of \eq\nr{m_last})
\ba
 \mathcal{I}^{ }_\rmi{j}(P) & \equiv & 
 \Tint{Q,R} \frac{P^6}{Q^2R^2[(Q-R)^2+\lambda^2](Q-P)^2(R-P)^2}
 \;. \la{Ij}
\ea
The corresponding spectral function, evaluated 
at vanishing spatial momentum, is defined by 
\be
 \rho_{\mathcal{I}^{ }_\rmi{j}}(\omega) 
 \equiv \im \Bigl[ \mathcal{I}^{ }_\rmi{j}(P) 
 \Bigr]_{P \to (-i[\omega + i 0^+],\vec{0})}
 \;. \la{rho_w}
\ee
In the following we give some details 
on how $\rho^{ }_{\mathcal{I}^{ }_\rmi{j}}$
can be given a rapidly convergent 2-dimensional integral
representation.
Since $\rho^{ }_{\mathcal{I}^{ }_\rmi{j}}$ 
is the most complicated structure encountered,
the same techniques will also apply to the other cases.  
(The only major simplification in $\rho_{\mathcal{I}^{ }_\rmi{j}}$ is that  
it is ultraviolet finite so that the analysis can be carried out 
without ultraviolet regularization; nevertheless, the part which
contains ultraviolet divergences in other cases will already
be clearly identified, cf.\ \se\ref{ss:fz_p}.)

%
\subsection{Matsubara sums}

%
\begin{figure}[t]
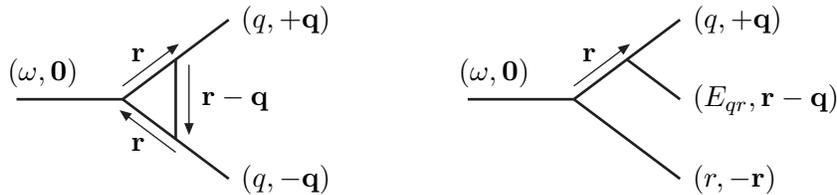


\hspace*{1.5cm}%
\begin{minipage}[c]{6cm}
\begin{eqnarray*}
&& 
 \hspace*{-1cm}
 \ScaV \quad\; 
\end{eqnarray*}
\end{minipage}%
\begin{minipage}[c]{6cm}
\begin{eqnarray*}
&& 
 \hspace*{-1cm}
 \ScaR \quad\; 
\end{eqnarray*}
\end{minipage}

\vspace*{0.5cm}

\caption[a]{\small 
Examples of physical amplitudes contained in \eq\nr{Ij_T}. 
Left: ``virtual corrections'', or ``factorized integrals''. 
Right:  ``real corrections'', or ``phase space integrals''.
The whole spectral function amounts essentially to the absolute 
value squared of an amplitude, and the virtual process contributes
through its interference with a tree-level process. (In the case shown, 
the final answer vanishes in the zero-temperature limit, but remains
finite for $T\neq 0$.) 
} 
\la{fig:processes}
\end{figure}
%

As a first step we carry out the Matsubara sums
in \eq\nr{Ij};\footnote{%
 An alternative strategy might consist
 of performing first the spatial integrals, which can be done for
 general masses~\cite{akr}, and by then re-interpreting (some of) 
 the masses as Matsubara frequencies, over which a summation
 is carried out. 
 }  
take the discontinuity 
leading to the spectral function; and set $\vec{p}\to\vec{0}$. 
The procedure is similar to that described in some detail
in refs.~\cite{nlo,rhoE}.
Like in ref.~\cite{rhoE} it is useful
to ``regulate'' some of the propagators by introducing a small mass parameter
which is set to zero in the end ($\lambda^2$ in \eq\nr{Ij}). 
To keep the situation maximally
symmetric, it is convenient to choose the propagator carrying the
momentum $Q-R$ for this task. Denoting 
\be
 E_q \equiv q \;, \quad E_r \equiv r \;, \quad 
 E_{qr} \equiv \sqrt{(\vec{q}-\vec{r})^2 + \lambda^2}
 \;,
\ee
and not displaying any terms proportional to $\omega^n\delta(\omega)$
(cf.\ \se\ref{ss:tau}), 
the result from \eq(B.33) of ref.~\cite{nlo} 
can be re-interpreted as\footnote{%
 We have also rederived the expression from 
 scratch, using the same techniques as in ref.~\cite{nlo}
 but applied directly to the bosonic case, with all propagators
 taking their tree-level forms.  
 }
\ba
 & & \hspace*{-0.6cm} \rho^{ }_{\mathcal{I}^{ }_\rmii{j}}(\omega) = 
 \int_{\vec{q,r}} 
 \frac{\omega^6 \pi }{4 q r E_{qr}} \biggl\{ 
 \la{Ij_T} \\
 & & \!\!
 \frac{1}{8q^2} 
 \Bigl[\delta(\omega - 2 q) - \delta(\omega+2 q) \Bigr]
 \times 
 \nn & & \times \biggl[
 \biggl( 
 \frac{1}{(q+r-E_{qr})(q+r)} -  
 \frac{1}{(q-r+E_{qr})(q-r)} 
 \biggr)
 (1 + 2 n_q) (n_{qr}-n_r)
 \nn & & \;\; 
 +
 \biggl(
 \frac{1}{(q+r+E_{qr})(q+r)} -  
 \frac{1}{(q-r-E_{qr})(q-r)} 
 \biggr)
 (1 + 2 n_q) (1 + n_{qr}+n_r)
 \biggr]
 \nn & + & \!\!
 \frac{1}{8r^2} 
 \Bigl[\delta(\omega - 2 r) - \delta(\omega+2 r) \Bigr]
 \times 
 \nn & & \times \biggl[
 \biggl( 
 \frac{1}{(q+r-E_{qr})(q+r)} -  
 \frac{1}{(q-r-E_{qr})(q-r)} 
 \biggr)
 (1+2n_r)(n_{qr}-n_q)
 \nn & & \;\; 
 +
 \biggl(
 \frac{1}{(q+r+E_{qr})(q+r)} -  
 \frac{1}{(q-r+E_{qr})(q-r)} 
 \biggr)
 (1+2n_r)(1 + n_{qr}+n_q)
 \biggr]
 \nn & + & \!\!
 \Bigl[\delta(\omega - q - r -E_{qr}) - \delta(\omega+q+r+E_{qr}) \Bigr]
 \frac{(1+n_{qr})(1+n_q+n_r)+n_q n_r}
      {(q+r+E_{qr})^2(q-r+E_{qr})(q-r-E_{qr})}
 \nn & + &  \!\!
 \Bigl[\delta(\omega-q-r+E_{qr}) - \delta(\omega + q + r -E_{qr}) \Bigr]
 \frac{n_{qr}(1+n_q + n_r ) - n_qn_r}
      {(q+r-E_{qr})^2(q-r+E_{qr})(q-r-E_{qr})}
 \nn & + &  \!\!
 \Bigl[\delta(\omega - q + r -E_{qr}) - \delta(\omega+q-r+E_{qr}) \Bigr]
 \frac{n_r(1+n_q+n_{qr})-n_q n_{qr}}
      {(q-r+E_{qr})^2(q+r+E_{qr})(q+r-E_{qr})}
 \nn & + &  \!\!
 \Bigl[\delta(\omega + q - r -E_{qr}) - \delta(\omega-q+r+E_{qr}) \Bigr]
 \frac{n_q(1+n_r+n_{qr})-n_r n_{qr}}
      {(q-r-E_{qr})^2(q+r+E_{qr})(q+r-E_{qr})}
 \biggr\} \;. \nonumber
\ea
The first two structures, with simple $\delta$-function constraints, 
will technically be referred to as ``factorized'' integrals but 
correspond physically to virtual corrections; 
the latter four 
structures, with more complicated $\delta$-constraints, 
are technically labelled
``phase space'' integrals but correspond
physically to real processes
(some examples are illustrated in \fig\ref{fig:processes}). 
Among the factorized integrals a further subdivision
into two classes is possible, namely to ``powerlike'' integrals in
which the (unconstrained) integration proceeds like 
at zero temperature, 
without any Boltzmann suppression at large momenta
(these are virtual vacuum corrections); as well as 
to ``exponential'' integrals, which contain Bose distributions
and are guaranteed to be 
ultraviolet convergent (these are virtual thermal corrections). 
We first discuss these three classes 
separately, and then collect together the results. 

It is important to note that the factorized and phase 
space integrals are both divergent (or ill-defined) for $\lambda\to 0$, 
because of the appearance of poles in some of the denominators. 
For example in ref.~\cite{sls} 
various divergences were classified as being of 
``soft'', ``collinear'', or  ``thermal infrared'' origin 
(the last coming from $n_{E}$ at small $E$).
We do not separately keep track of the nature of the divergences,
but their eventual cancellation, when the real and virtual 
corrections are added together,  
works out like in ref.~\cite{sls}, 
and is the reason that we can take $\lambda\to 0$ in the end; 
however we establish
the cancellation for any value of $\omega / T$, whereas
in ref.~\cite{sls} only terms of relative magnitude 
$T^2 / \omega^2$ in the OPE-limit were analyzed.
(Of course, similar cancellations were seen long ago 
in refs.~\cite{spectral1}--\cite{spectral3}.)
In practice we add as a further regulator 
a principle value prescription for the poles; 
then changes of integration variables, such as $q\leftrightarrow r$, 
are unproblematic and allow to simplify the expressions somewhat.

%
\subsection{Virtual vacuum corrections}
\la{ss:fz_p}

Starting with the factorized part, it is immediately clear that 
the symmetry $q\leftrightarrow r$ allows us to combine the two
first terms of \eq\nr{Ij_T}. 
Let us pick the latter case, with $\delta(\omega - 2r)$ for $\omega > 0$, 
as a representative. The integral over $\vec{r}$ is trivial, 
\be
 \int_{\vec{r}} \pi \delta(\omega - 2 r) = \frac{\omega^2}{16\pi}
 \;,
\ee
and $1/(8r^3)$ from the original ``measure'' goes over into $1/\omega^3$.
In the remaining $\vec{q}$-integral we substitute the angular variable
through $E_{qr}$:
\be
 \int_{\vec{q}} \frac{1}{2q E_{qr}}
 = \frac{1}{4\pi^2\omega} \int_0^\infty \! {\rm d}q 
 \int_{E_{qr}^-}^{E_{qr}^+} \!\! {\rm d}E_{qr}
 \;, \quad
 E_{qr}^\pm \equiv \sqrt{\Bigl(q\pm\frac{\omega}{2}\Bigr)^2 + \lambda^2}
 \;. \la{fz_measure}
\ee
Thereby the factorized powerlike integral can be expressed as 
\ba
 && \hspace*{-1cm} \rho_{\mathcal{I}^{ }_\rmii{j}}^{(\rmi{fz,p})}(\omega)
  \equiv  
  \frac{\omega^4}{(4\pi)^3}  (1 + 2 n_{\frac{\omega}{2}})
  \int_0^\infty \! {\rm d}q 
  \int_{E_{qr}^-}^{E_{qr}^+} \!\! {\rm d}E_{qr} \biggl\{ 
  \nn & & \; \; \;
 \mathbbm{P} \, \biggl[
 \frac{1}{(q+\frac{\omega}{2}+E_{qr})(q+\frac{\omega}{2})} -  
 \frac{1}{(q-\frac{\omega}{2}+E_{qr})(q-\frac{\omega}{2})} 
 \biggr]
  \biggr\}  
  \;. \la{Ij_fzp_1}  
\ea
The integration range is illustrated in \fig\ref{fig:fz_range}.

%
\begin{figure}[t]

\centerline{%
\epsfysize=4.5cm\epsfbox{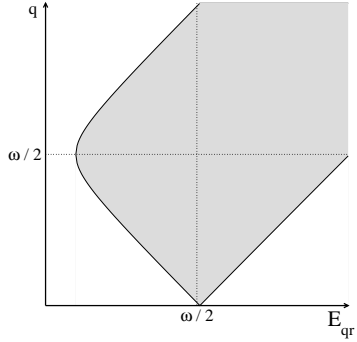}%
}

\caption[a]{\small 
The integration range for the factorized integrals, cf.\ \eq\nr{fz_measure}, 
for $\lambda = \omega/10$. 
} 
\la{fig:fz_range}
\end{figure}
%

Now, we could immediately integrate over the variable $E_{qr}$
in \eq\nr{Ij_fzp_1}. It turns out to be somewhat 
simpler to take the limit $\lambda\to 0$, however, 
if we change the order of integrations (the same change will also
be useful in \se\ref{ss:fei}). This can be achieved through 
\be
 \int_0^\infty \! {\rm d}q 
 \int_{ 
   \sqrt{(q - \frac{\omega}{2})^2 + \lambda^2}
  }^{ 
   \sqrt{(q + \frac{\omega}{2})^2 + \lambda^2}
  } \! {\rm d}E_{qr} 
 = 
 \int_{\lambda}^{\infty} \! {\rm d}E_{qr} 
 \int_{ 
  |\frac{\omega}{2} - \sqrt{E_{qr}^2-\lambda^2}|
 }^{
  \frac{\omega}{2} + \sqrt{E_{qr}^2-\lambda^2}
 } \!\! {\rm d}q
 \;. \la{fz_order}
\ee
Subsequently we rename the inner integration variable to be $r$, 
the outer one to be $q$. Partial fractioning with respect to the new $r$, 
\eq\nr{Ij_fzp_1} becomes 
\ba
 \rho_{\mathcal{I}^{ }_\rmii{j}}^{(\rmi{fz,p})}(\omega) & = & 
  \frac{\omega^4}{(4\pi)^3}   (1 + 2 n_{\frac{\omega}{2}})
 \int_{\lambda}^{\infty} \! \frac{{\rm d}q}{q} 
 \int_{ 
  |\frac{\omega}{2} - \sqrt{ q^2-\lambda^2}|
 }^{
  \frac{\omega}{2} + \sqrt{ q^2-\lambda^2}
 }
  \!\! {\rm d}r \, \biggl\{ 
  \nn & & \; \; \;
  \mathbbm{P} \, \biggl[ 
  \frac{1}{r+\frac{\omega}{2}} - 
  \frac{1}{q+r+\frac{\omega}{2}} -   
  \frac{1}{r-\frac{\omega}{2}} +
  \frac{1}{q+r-\frac{\omega}{2}} 
  \biggr] 
  \biggr\}
  \;. \hspace*{5mm} \la{Ij_fzp_2}
\ea
Carrying out the integrals yields
\ba
 && \hspace*{-1cm} \rho_{\mathcal{I}^{ }_\rmii{j}}^{(\rmi{fz,p})}(\omega) =  
  \frac{\omega^4}{(4\pi)^3} (1+2 n_{\frac{\omega}{2}}) \biggl\{  
  \nn && 
  \int_\lambda^{\sqrt{(\frac{\omega}{2})^2+\lambda^2}} 
  \! \frac{{\rm d}q}{q}  
      \ln\biggl| 
            \frac{q +\sqrt{q^2-\lambda^2}}
                 {q -\sqrt{q^2-\lambda^2}}
         \biggr|
         \biggl| 
            \frac{\omega+\sqrt{q^2-\lambda^2}}
                 {\omega-\sqrt{q^2-\lambda^2}}
         \biggr|
         \biggl| 
            \frac{\omega + q - \sqrt{q^2-\lambda^2}}
                 {\omega + q + \sqrt{q^2-\lambda^2}}
         \biggr|
  \nn & + & 
  \int_{\sqrt{(\frac{\omega}{2})^2+\lambda^2}}^\infty 
  \! \frac{{\rm d}q}{q} 
      \ln\biggl| 
            \frac{(q +\sqrt{q^2-\lambda^2})^2}
                 {{q^2-\lambda^2}}
         \biggr|
         \biggl| 
            \frac{\sqrt{q^2-\lambda^2} + \omega}
                 {q + \sqrt{q^2-\lambda^2} + \omega}
         \biggr|
         \biggl| 
            \frac{\sqrt{q^2-\lambda^2} - \omega}
                 {q + \sqrt{q^2-\lambda^2} - \omega}
         \biggr|
 \biggr\}
  \;, \hspace*{8mm} \la{Ij_fzp_3}
\ea
where $\sqrt{(\frac{\omega}{2})^2+\lambda^2}$ is the value of $q$
for which the lower limit 
$|\frac{\omega}{2} - \sqrt{ q^2-\lambda^2}|$
of the $r$-integration in
\eq\nr{Ij_fzp_2} vanishes.

Whereas \eq\nr{Ij_fzp_3} is still exact, we would 
in the end like to take the limit $\lambda\to 0$. 
In most of the terms of \eq\nr{Ij_fzp_3} this can be achieved through
a Taylor expansion in $\lambda/q$, but if an integration is sensitive to the 
range $q\sim \lambda$ then this is not possible, and the integrand needs
to be kept in its exact form. 
Analyzing the situation carefully, this can be seen to only happen 
with the very first term, in the vicinity of the lower bound. 
Thereby we obtain
\ba
 && \hspace*{-0.5cm}
 \rho_{\mathcal{I}^{ }_\rmii{j}}^{(\rmi{fz,p})}(\omega) \approx 
  \frac{\omega^4}{(4\pi)^3} (1 + 2 n_{\frac{\omega}{2}} ) \biggl\{
 \la{Ij_fzp_5} \\
  & & 
  \int_\lambda^{ \frac{\omega}{2} } 
  \! \frac{{\rm d}q}{q} 
      \ln\biggl| 
            \frac{q +\sqrt{q^2-\lambda^2}}
                 {q -\sqrt{q^2-\lambda^2}}
         \biggr|
  +   \int_0^{ \frac{\omega}{2} } 
   \! \frac{{\rm d}q}{q}
         \ln
         \biggl| 
            \frac{(\omega+q)\,\omega}
                 {(\omega-q)(\omega + 2 q)}
         \biggr|
  +   \int_{ \frac{\omega}{2} }^{\infty} 
  \! \frac{{\rm d}q}{q} 
         \ln
         \biggl| 
            \frac{4(q^2 - \omega^2)}
                 {4q^2 - \omega^2}
         \biggr|
  \biggr\} 
  \;, \hspace{5mm} \nonumber
\ea
where ``$\approx$'' indicates that 
terms of $\rmO(\lambda \ln\lambda)$ and smaller have 
been omitted. Remarkably
the 2nd and 3rd integrals in \eq\nr{Ij_fzp_5} 
cancel against each other, so that in total 
\be
 \rho_{\mathcal{I}^{ }_\rmii{j}}^{(\rmi{fz,p})}(\omega) \approx 
  \frac{\omega^4}{(4\pi)^3} (1 + 2 n_{\frac{\omega}{2}} )
  \int_\lambda^{ \frac{\omega}{2} } 
  \! \frac{{\rm d}q}{q} 
      \ln\biggl| 
            \frac{q +\sqrt{q^2-\lambda^2}}
                 {q -\sqrt{q^2-\lambda^2}}
         \biggr|
 \;. 
 \la{Ij_fzp_6} 
\ee

%
\subsection{Virtual thermal corrections}
\la{ss:fei}

For the factorized exponential integrals we proceed in a similar fashion. 
The change of ordering in \eq\nr{fz_order} and the subsequent
renaming of variables is only carried out in terms containing
the distribution $n_{qr}$.
Thereby the relevant part of \eq\nr{Ij_T} becomes 
\ba
 \rho_{\mathcal{I}^{ }_\rmii{j}}^{(\rmi{fz,e})}(\omega) & \equiv & 
  \frac{\omega^4}{(4\pi)^3} (1 + 2 n_{\frac{\omega}{2}})
  \int_0^\infty \! {\rm d}q \, n_q
  \int_{ 
    \sqrt{(q - \frac{\omega}{2})^2 + \lambda^2}
   }^{ 
    \sqrt{(q + \frac{\omega}{2})^2 + \lambda^2}
   }
   \!\! {\rm d}r \, \biggl\{ 
  \nn & & \; \; \;
  \mathbbm{P} \, \biggl[ 
  \frac{1}{(q+r+\frac{\omega}{2})(q+\frac{\omega}{2})} -  
  \frac{1}{(q+r-\frac{\omega}{2})(q-\frac{\omega}{2})}
  \nn & & \hspace*{1cm} + \, 
  \frac{1}{(q-r-\frac{\omega}{2})(q-\frac{\omega}{2})} -
  \frac{1}{(q-r+\frac{\omega}{2})(q+\frac{\omega}{2})} 
  \biggr] 
  \biggr\}
  \nn & + &
  \frac{\omega^4}{(4\pi)^3} (1 + 2 n_{\frac{\omega}{2}})
 \int_{\lambda}^{\infty} \! {\rm d}q \, n_q 
 \int_{ 
  |\frac{\omega}{2} - \sqrt{ q^2-\lambda^2}|
 }^{
  \frac{\omega}{2} + \sqrt{ q^2-\lambda^2}
 }
  \!\! {\rm d}r \, \biggl\{ 
  \nn & & \; \; \;
  \mathbbm{P} \, \biggl[ 
  \frac{1}{(q+r+\frac{\omega}{2})(r+\frac{\omega}{2})} -  
  \frac{1}{(q+r-\frac{\omega}{2})(r-\frac{\omega}{2})}
  \nn & & \hspace*{1cm} + \, 
  \frac{1}{(q-r+\frac{\omega}{2})(r-\frac{\omega}{2})} -
  \frac{1}{(q-r-\frac{\omega}{2})(r+\frac{\omega}{2})} 
  \biggr] 
 \biggr\}
  \;. \hspace*{5mm} \la{Ij_fz_2}
\ea
The integration over $r$ is again trivial, 
partial fractioning the latter term like in \eq\nr{Ij_fzp_2}. 
In a few steps we obtain
\ba
 && \hspace*{-1cm} \rho_{\mathcal{I}^{ }_\rmii{j}}^{(\rmi{fz,e})}(\omega) =  
  \frac{\omega^4}{(4\pi)^3} (1 + 2 n_{\frac{\omega}{2}}) \biggl\{  
  \nn && 
  \int_0^\infty \! {\rm d}q \, n_q \, \mathbbm{P} \, \biggl[ 
      \frac{1}{q+\frac{\omega}{2}}
      \ln\biggl| \frac{\lambda^2}{2q\omega - \lambda^2} \biggr|
      +  \frac{1}{q-\frac{\omega}{2}} 
      \ln\biggl| \frac{\lambda^2}{2q\omega + \lambda^2} \biggr|
   \; \biggr]
  \nn & + & 
  \int_{\lambda}^\infty \! {\rm d}q \, n_q  \biggl[ 
    \frac{1}{q}
     \ln
         \biggl| 
            \frac{q + \frac{\lambda^2}{\omega}+\sqrt{q^2-\lambda^2}}
                 {q + \frac{\lambda^2}{\omega}-\sqrt{q^2-\lambda^2}}
         \biggr|
    + 
    \frac{1}{q}
     \ln
         \biggl| 
            \frac{q - \frac{\lambda^2}{\omega}+\sqrt{q^2-\lambda^2}}
                 {q - \frac{\lambda^2}{\omega}-\sqrt{q^2-\lambda^2}}
         \biggr|
  \; \biggr] \biggr\}
  \;. \la{Ij_fz_3}
\ea

%
\subsection{Real corrections}

As far as the latter four structures of \eq\nr{Ij_T} 
are concerned, we again
rewrite the integration measure by substituting the angle between 
$\vec{q}$ and $\vec{r}$ through $E_{qr}$:
\be
 \int_{\vec{q},\vec{r}} \frac{\pi}{4qrE_{qr}} 
 = \frac{2}{(4\pi)^3} \int_0^\infty \! {\rm d}q \int_0^\infty \! {\rm d}r
 \int_{E_{qr}^-}^{E_{qr}^+} \! {\rm d}E_{qr}
 \;, \quad
 E_{qr}^\pm \equiv \sqrt{(q\pm r)^2 + \lambda^2}
 \;.
\ee
It is also convenient to factor out a common $n_q n_r n_{qr}$ 
from the Bose distributions, yielding
\ba
 {(1+n_{qr})(1+n_q+n_r)+n_q n_r} & = &  
 n_q n_r n_{qr} \bigl( e^{q+r+E_{qr}} -1 \bigr)
 \;, \\
 {n_{qr}(1+n_q + n_r )-n_q n_r} & = & 
 n_q n_r n_{qr} \bigl( e^{q+r} - e^{E_{qr}} \bigr)
 \;, \\ 
 {n_r(1+n_q+n_{qr})-n_q n_{qr}} & = & 
 n_q n_r n_{qr} \bigl( e^{q+E_{qr}} - e^{r} \bigr)
 \;, \\ 
 {n_q(1+n_r+n_{qr})-n_r n_{qr}} & = & 
 n_q n_r n_{qr} \bigl( e^{r+E_{qr}} - e^{q} \bigr)
 \;, 
\ea
and to make use of the $\delta$-functions
to simplify the denominators. Choosing furthermore dimensionless units
in the following whereby all variables are expressed in terms of $T$, 
and noting that, for $0 < \lambda < \omega$, only four of 
the $\delta$-constraints get realized, we can rewrite 
the phase space part of the integral as 
\ba
 \rho_{\mathcal{I}^{ }_\rmii{j}}^{(\rmi{ps})}(\omega) & \equiv & 
 \frac{2\omega^4}{(4\pi)^3} 
 \int_0^\infty \! {\rm d}q \int_0^\infty \! {\rm d}r
 \int_{E_{qr}^-}^{E_{qr}^+} \! {\rm d}E_{qr}
 \, n_q n_r n_{qr} \, \biggl\{ 
 \nn
 & \mbox{(i)} 
  & \;\;\; 
 \frac{\delta(\omega-q-r-E_{qr})}{(2r-\omega)(2q-\omega)}
 \Bigl(1 - e^{q+r+E_{qr}} \Bigr) 
 \nn 
 & \mbox{(ii)} 
  & + \, 
 \frac{\delta(\omega-q-r+E_{qr})}{(2r-\omega)(2q-\omega)}
 \Bigl(e^{E_{qr}} - e^{q+r}  \Bigr) 
 \nn 
 & \mbox{(\iv)} 
  & + \, 
 \frac{\delta(\omega+q-r-E_{qr})}{(2r-\omega)(2q+\omega)}
 \Bigl(e^{r+E_{qr}} - e^q \Bigr) 
 \nn 
 & \mbox{(\ii)} 
  & + \, 
 \frac{\delta(\omega-q+r-E_{qr})}{(2r+\omega)(2q-\omega)}
 \Bigl(e^{q+E_{qr}} - e^r \Bigr) 
 \biggr\} \;. \la{Ij_ps_1}
\ea

Of the integrals here, the ones labelled (\iv) and (\ii) are related
by the symmetry $q\leftrightarrow r$, so essentially only
three cases remain. 
We choose to regard the integral over $r$ as an ``inner''
integration, to be carried out first. Then, with some work, the following
relations can be established for $\omega > \lambda$:
\ba
 \mbox{(i)} &&
 \int_0^\infty \! {\rm d}q \int_0^\infty \! {\rm d}r
 \int_{E_{qr}^-}^{E_{qr}^+} \! {\rm d}E_{qr} \;
 \delta(\omega-q-r-E_{qr}) \, 
 \phi(q,r,E_{qr}) 
 \nn &  & \hspace*{4cm} = \,  
 \int_0^{\frac{\omega^2-\lambda^2}{2\omega}}
 \! {\rm d}q 
 \int_{ \frac{\omega(\omega-2q)-\lambda^2}{2\omega} }^
      { \frac{\omega(\omega-2q)-\lambda^2}{2(\omega-2q)} } 
 \! {\rm d}r \, 
 \phi(q,r,\omega-q-r)
 \;, \la{range1} \\ 
 \mbox{(ii)} &&
 \int_{0}^\infty \! {\rm d}q \int_0^\infty \! {\rm d}r
 \int_{E_{qr}^-}^{E_{qr}^+} \! {\rm d}E_{qr} \;
 \delta(\omega-q-r+E_{qr}) \, 
 \phi(q,r,E_{qr}) 
 \nn &  & \hspace*{4cm} = \,  
 \int_{\frac{\omega}{2}}^{\infty}
 \! {\rm d}q 
 \int_{ \frac{\omega(2q-\omega)+\lambda^2}{2(2q-\omega)} }^
      { \infty } 
 \! {\rm d}r \, 
 \phi(q,r,-\omega+q+r)
 \;, \\ 
 \mbox{(\iv)} &&
 \int_0^\infty \! {\rm d}q \int_0^\infty \! {\rm d}r
 \int_{E_{qr}^-}^{E_{qr}^+} \! {\rm d}E_{qr} \;
 \delta(\omega+q-r-E_{qr}) \, 
 \phi(q,r,E_{qr}) 
 \nn &  & \hspace*{4cm} = \,  
 \int_0^{\infty}
 \! {\rm d}q 
 \int_{ \frac{\omega(\omega+2q)-\lambda^2}{2(\omega+2q)} }^
      { \frac{\omega(\omega+2q)-\lambda^2}{2\omega} } 
 \! {\rm d}r \, 
 \phi(q,r,\omega+q-r)
 \;, \la{range3}
\ea
where $\phi$ is an arbitrary function. The ranges are 
illustrated in \fig\ref{fig:ranges_old}. 

%
\begin{figure}[t]

\centerline{%
\epsfysize=4.5cm\epsfbox{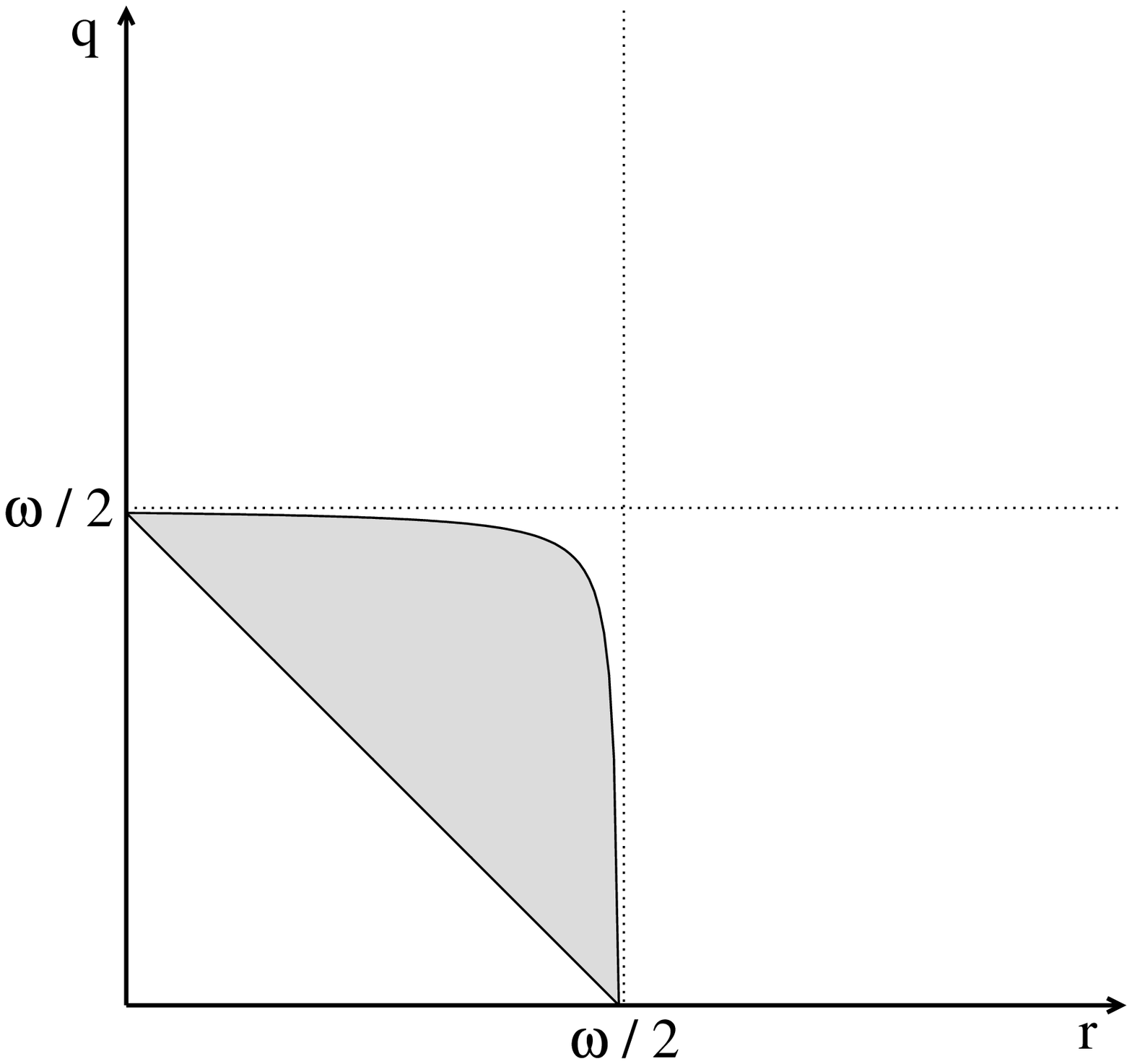}%
~~\epsfysize=4.5cm\epsfbox{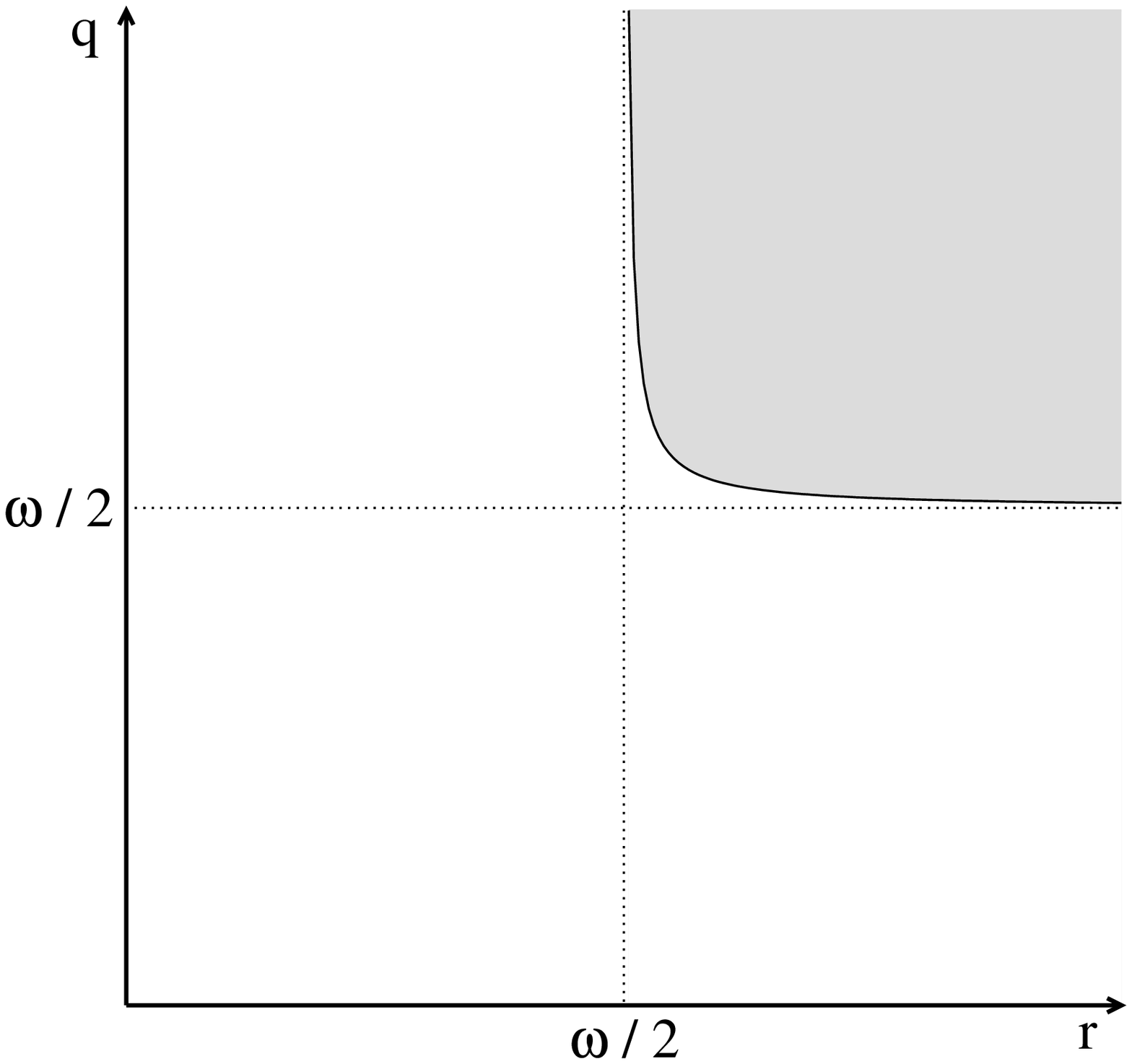}%
~~\epsfysize=4.5cm\epsfbox{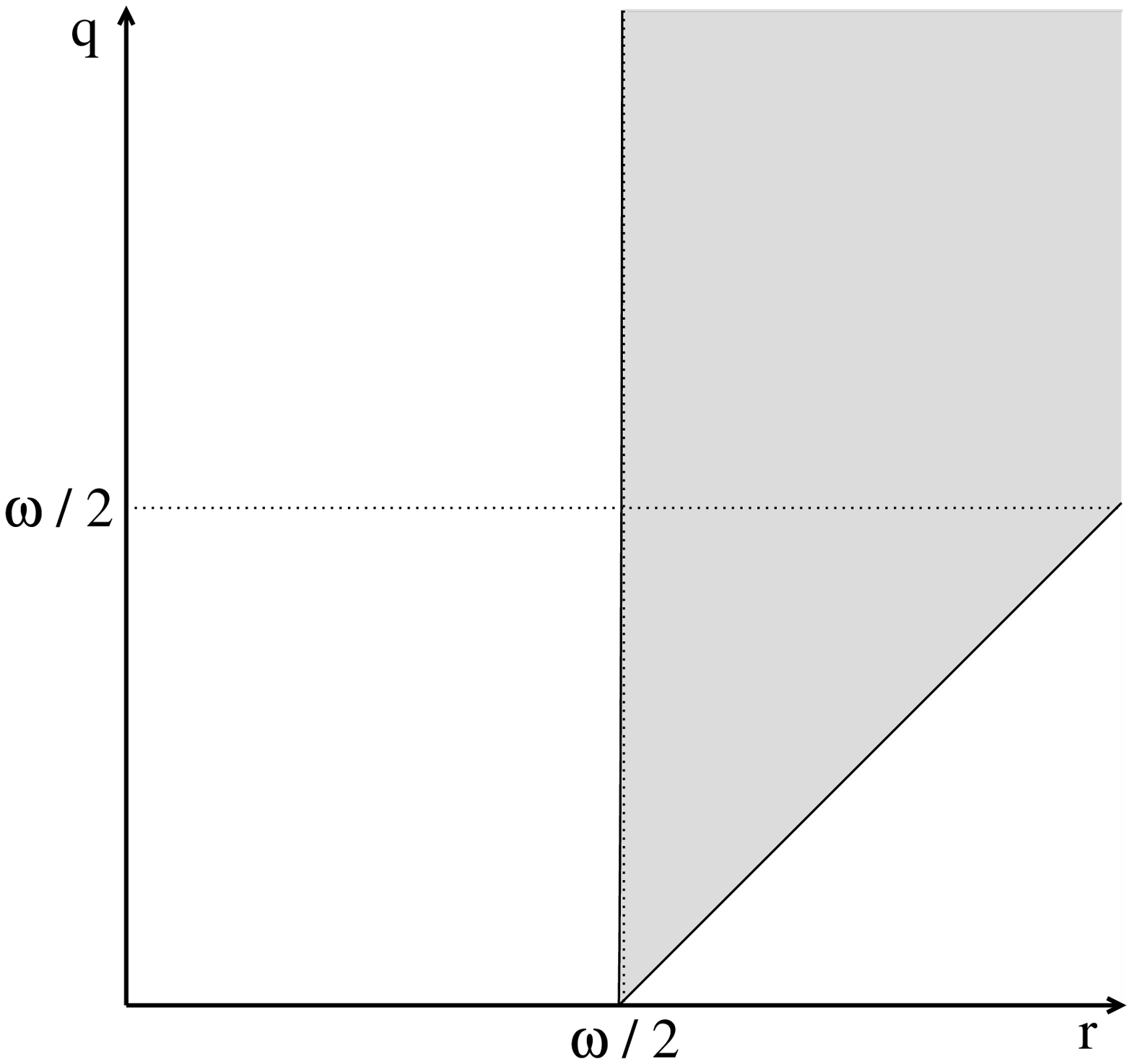}%
}

\hspace*{2.5cm} (i) \hspace*{4.5cm} (ii) \hspace*{4.2cm} (\iv) 

\caption[a]{\small 
Original integration ranges (\eqs\nr{range1}--\nr{range3})
for the phase space integrals, 
for $\lambda = \omega/10$. 
} 
\la{fig:ranges_old}
\end{figure}
%

%
\begin{figure}[t]

\centerline{%
\epsfysize=4.5cm\epsfbox{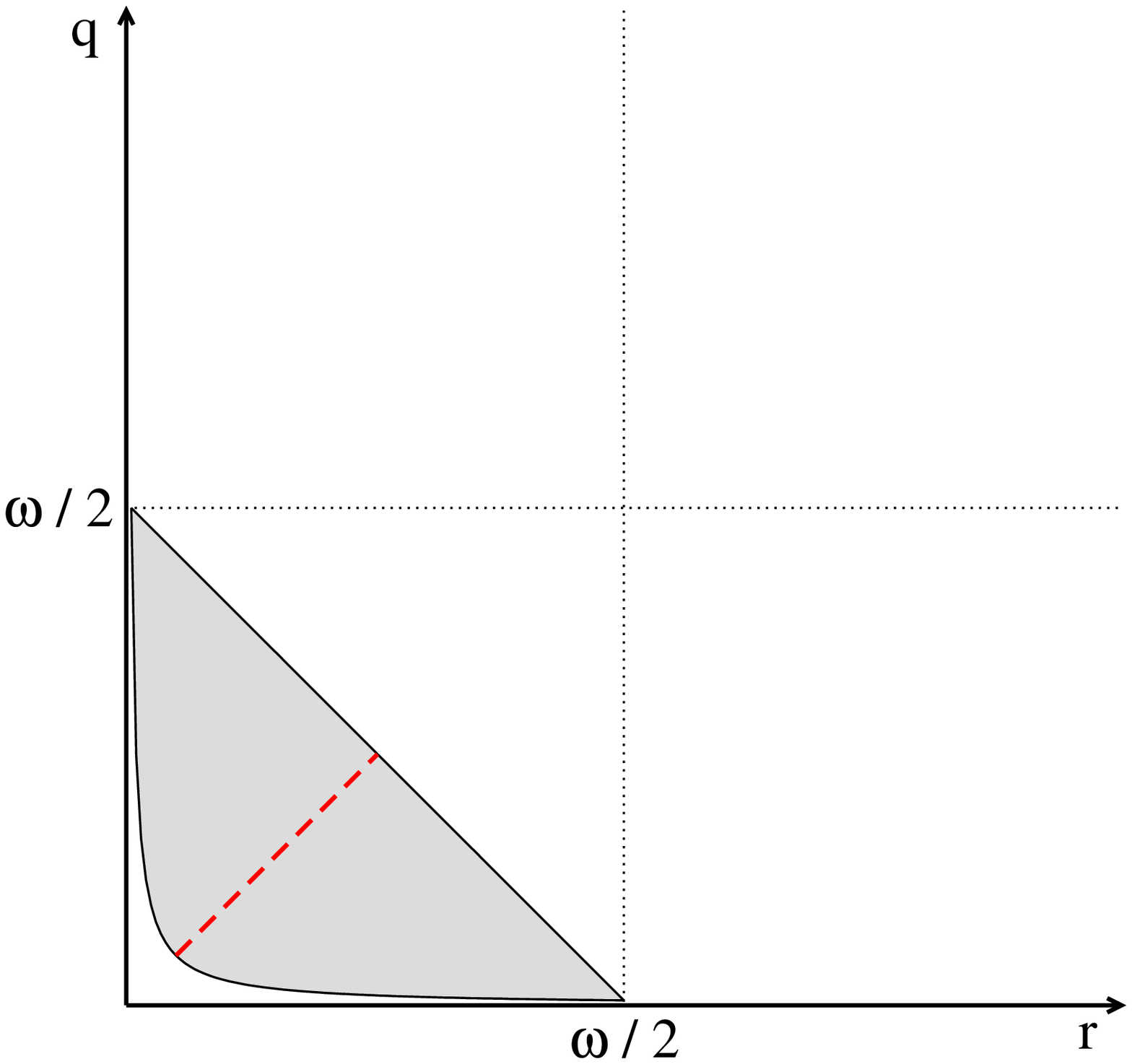}%
~~\epsfysize=4.5cm\epsfbox{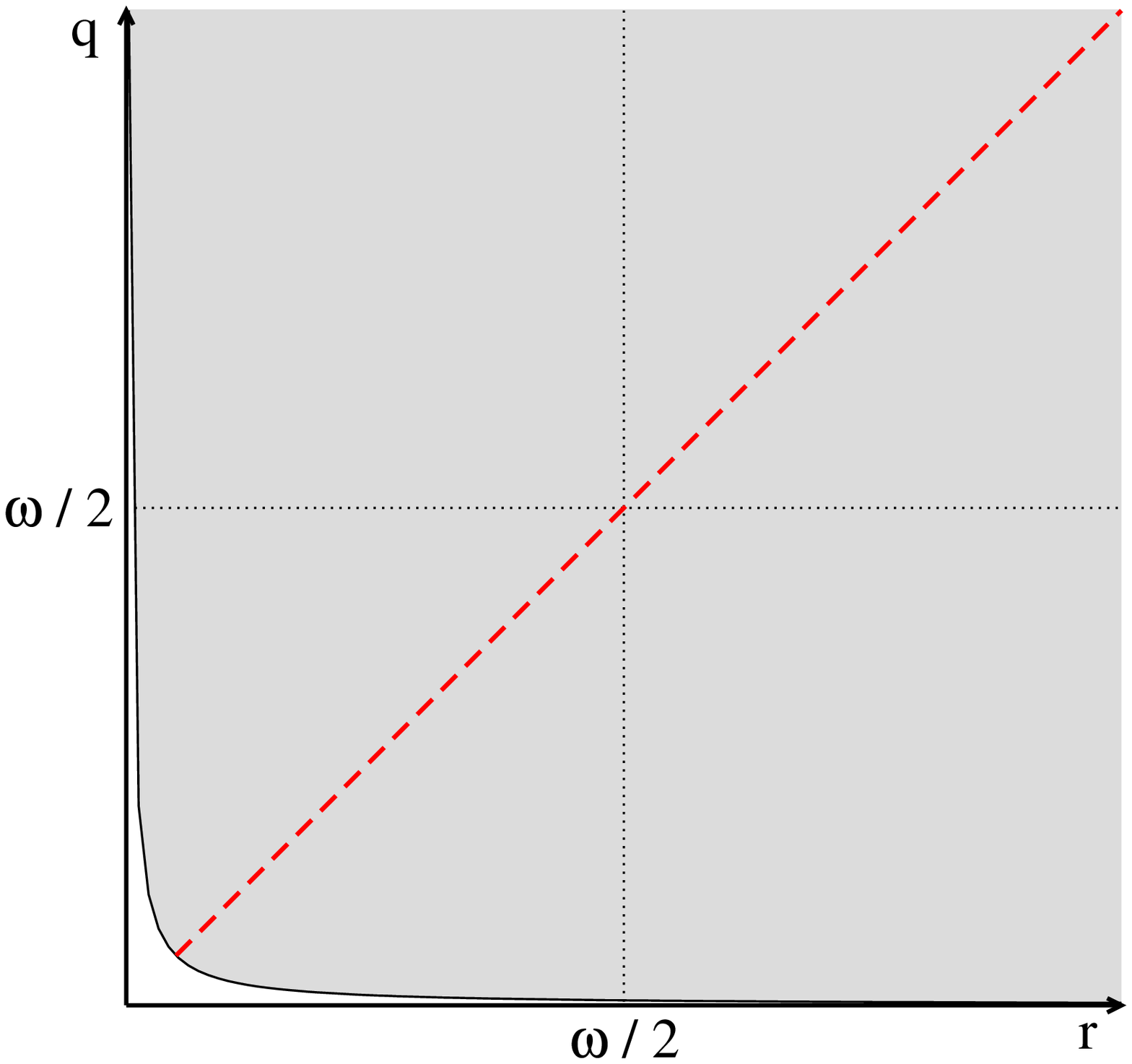}%
~~\epsfysize=4.5cm\epsfbox{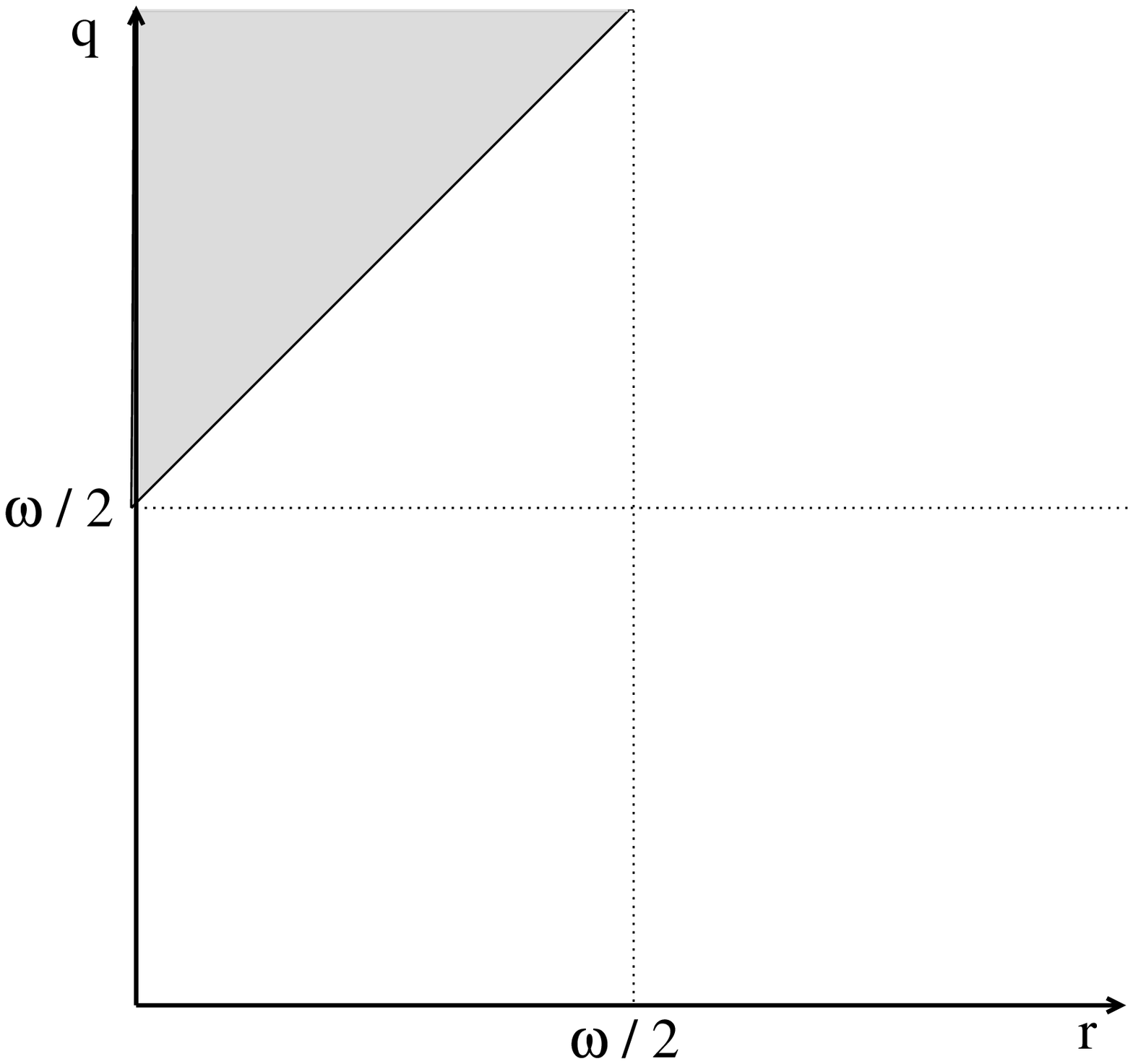}%
}

\hspace*{2.5cm} (i) \hspace*{4.5cm} (ii) \hspace*{4.2cm} (\iv) 

\caption[a]{\small 
Integration ranges after the shifts in
\eqs\nr{nrange1}--\nr{nrange3}, 
for $\lambda = \omega/10$. 
The dashed line indicates the diagonal 
$q=r$, which can be used to reflect all integrations to the octant $q \ge r$. 
} 
\la{fig:ranges_new}
\end{figure}
%

It appears advantageous, however, to shift the ranges somewhat, 
in order to place poles so that their eventual
cancellation is easier to see. To this end we undertake shifts
which turn all denominators of \eq\nr{Ij_ps_1} into a universal 
$1/(4qr)$. The shifts are 
\ba
 \mbox{(i)}: && 
   q \to \frac{\omega}{2} - q \;, \quad r \to \frac{\omega}{2} - r
 \;, \la{shift1} \\ 
 \mbox{(ii)}: && 
   q \to \frac{\omega}{2} + q \;, \quad r \to \frac{\omega}{2} + r
 \;, \\ 
 \mbox{(\iv)}: &&
   q \to -\frac{\omega}{2} +  q \;, \quad r \to \frac{\omega}{2} + r
 \;, \la{shift3}
\ea
and the integration ranges of \eqs\nr{range1}--\nr{range3} convert into
\ba
 \mbox{(i)} &&
 \int_0^{\frac{\omega^2-\lambda^2}{2\omega}}
 \! {\rm d}q 
 \int_{ \frac{\omega(\omega-2q)-\lambda^2}{2\omega} }^
      { \frac{\omega(\omega-2q)-\lambda^2}{2(\omega-2q)} } 
 \! {\rm d}r \, 
 \phi(q,r,\omega-q-r)
 \nn && \hspace*{4cm} = \, 
 \int_{\frac{\lambda^2}{2\omega}}^{\frac{\omega}{2}}
 \! {\rm d}q 
 \int_{ \frac{\lambda^2}{4q} }^
      { \frac{\omega(\omega-2q)+\lambda^2}{2\omega} } 
 \! {\rm d}r \, 
 \phi\Bigl(\frac{\omega}{2}-q,\frac{\omega}{2}-r,q+r\Bigr)
 \;, \la{nrange1} \\ 
 \mbox{(ii)} &&
 \int_{\frac{\omega}{2}}^{\infty}
 \! {\rm d}q 
 \int_{ \frac{\omega(2q-\omega)+\lambda^2}{2(2q-\omega)} }^
      { \infty } 
 \! {\rm d}r \, 
 \phi(q,r,-\omega+q+r)
 \nn &  & \hspace*{4cm} = \,  
 \int_{0}^{\infty}
 \! {\rm d}q 
 \int_{ \frac{\lambda^2}{4q} }^
      { \infty } 
 \! {\rm d}r \, 
 \phi\Bigl(\frac{\omega}{2} + q,\frac{\omega}{2} + r,q+r\Bigr)
 \;, \\ 
 \mbox{(\iv)} &&
 \int_0^{\infty}
 \! {\rm d}q 
 \int_{ \frac{\omega(\omega+2q)-\lambda^2}{2(\omega+2q)} }^
      { \frac{\omega(\omega+2q)-\lambda^2}{2\omega} } 
 \! {\rm d}r \, 
 \phi(q,r,\omega+q-r)
 \nn &  & \hspace*{4cm} = \,  
 \int_{\frac{\omega}{2}}^{\infty}
 \! {\rm d}q 
 \int_{ - \frac{\lambda^2}{4q} }^
      { \frac{\omega(2q-\omega)-\lambda^2}{2\omega} } 
 \! {\rm d}r \, 
 \phi\Bigl(q-\frac{\omega}{2},\frac{\omega}{2}+r,q-r\Bigr)
 \;. \la{nrange3} \hspace*{5mm}
\ea
These are illustrated 
in \fig\ref{fig:ranges_new}. The expression 
of \eq\nr{Ij_ps_1} becomes 
\ba
 \rho_{\mathcal{I}^{ }_\rmii{j}}^{(\rmi{ps})}(\omega) & = & 
 \frac{\omega^4}{2(4\pi)^3} (e^\omega - 1)  \biggl\{ 
 \nn 
 & \mbox{(i)}  &  - \,
 \int_{\frac{\lambda^2}{2\omega}}^{\frac{\omega}{2}}
 \! {\rm d}q
 \int_{\frac{\lambda^2}{4q}}^{\frac{\omega(\omega-2q)+\lambda^2}{2\omega}} 
 \!\!\!\! {\rm d}r \; \mathbbm{P}
 \biggl( \frac{ 
 1 } {qr} \biggr)\;
 n_{\fr{\omega}2-q} \, 
 n_{\fr{\omega}2-r} \, 
 n_{q+r}
 \nn
 & \mbox{(ii)}  & - \, 
 \int_{ 0 }^{\infty}
 \! {\rm d}q
 \int_{\frac{\lambda^2}{4q}}^{ \infty }
 \! {\rm d}r \; \mathbbm{P}
 \biggl( \frac{ 1
 } {qr} \biggr)\;
 n_{\fr{\omega}2+q} \, 
 n_{\fr{\omega}2+r} \, 
 n_{q+r}
 e^{q+r} 
 \nn
 & \mbox{(\iv)}  & + \, 
 \int_{ \frac{\omega}{2} }^{\infty}
 \! {\rm d}q
 \int_{ - \frac{\lambda^2}{4q}}^{
    \frac{\omega(2q-\omega)-\lambda^2}{2\omega}}
 \!\!\!\! {\rm d}r \; \mathbbm{P}
 \biggl( \frac{ 2
  } {qr} \biggr)\;
 n_{q-\fr{\omega}2} \, 
 n_{\fr{\omega}2+r} \, 
 n_{q-r} \,
 e^{q - \frac{\omega}{2}}
 \biggr\} \;. \la{Ij_ps_2}
\ea
Subsequently we can reflect the range $q < r$ to 
$q > r$ in cases (i) and (ii), 
as indicated in \fig\ref{fig:ranges_new}; then the integrations 
only start at $q = \lambda/2$, and explicit divergences are averted.

So far we have made no approximations  concerning $\lambda$ 
in the phase space integrals
(apart from assuming that $\lambda < \omega$), but now we again  
want to send $\lambda \to 0$. As mentioned this is only possible
once the factorized and phase space parts are added together, 
but within each part we can extract the ``asymptotic'' behaviour 
in this limit. To this end we wish to re-arrange the 
integrations in a form in which a simple divergent part 
can be computed analytically and a more complicated but 
finite part is left over for numerical evaluation. 

The divergent parts to be subtracted 
have two separate origins: they are either related to a prefactor $1/r$, 
or to the infrared divergent Bose distribution $n_{q+r}$ 
(or $n_{q-\frac{\omega}{2}}$ for the case (\iv)). To handle 
$n_{q+r}$ we subtract a term of the form $\alpha\, n_{q+r}/(qr)$, 
where $\alpha$ is the ``residue'' of this structure at the origin; the
subtracted term can be integrated analytically, 
by changing variables from $(q,r)$ 
to $(x,y)\equiv (q+r,q-r)$, and carrying out the subsequent
integration over $y$. To handle any remaining
$1/r$-divergences, we subtract a term $\beta(q)/(qr)$, where
$\beta(q)$ is the ``residue'' at $r=0$. After these subtractions
the remainder integral must remain finite even in the limit $\lambda\to 0$.
(It is not clear to us whether these subtractions are the unique or most
elegant ones but they do achieve their goal.)

As far as the mentioned change of integration variables goes it can be 
carried out, in a range illustrated in \fig\ref{fig:ranges_xy}, as
\ba
 \mathcal{I} & \equiv & 
 \int_{\frac{\lambda}{2}}^{q_\rmii{max}} 
 \! \frac{{\rm d}q}{q}
 \int_{\frac{\lambda^2}{4q}}^{r_\rmii{max}(q)}
 \! \frac{{\rm d}r}{r}
 \, \phi(q+r,q-r)
 \nn 
 & = &  
 \int_{\lambda}^{x_\rmii{max}}
 \! \frac{{\rm d}x}{x}
 \int_{0}^{\sqrt{x^2-\lambda^2}} \!\!\!
 {\rm d}y \, 
 \biggl( \frac{1}{x+y} + \frac{1}{x-y} \biggr) 
 \phi(x,y)
 \left. 
 \right|_{x_\rmii{max} = q_\rmii{max}+ r_\rmii{max}(q_\rmii{max})}
 \;. \la{xy}
\ea
The upper limit of the $y$-integration was obtained by eliminating $q$
in favour of $x$ from the parametrization of the left boundary of 
the shaded region in \fig\ref{fig:ranges_xy}: 
$
 (x,y) = (q + \frac{\lambda^2}{4q},q - \frac{\lambda^2}{4q})
$.
If the function $\phi$ does not depend on $y = q-r$, we finally obtain
\be
 \mathcal{I} = \int_{\lambda}^{x_\rmii{max}} 
 \! \frac{{\rm d}x}{x}
 \ln \left| \frac{x+\sqrt{x^2-\lambda^2}}{x-\sqrt{x^2-\lambda^2}} \right|
 \phi(x)
 \;, \la{xy_trick}
\ee
and subsequently rename $x$ to be $q$.

%
\begin{figure}[t]

\centerline{%
~~\epsfysize=4.5cm\epsfbox{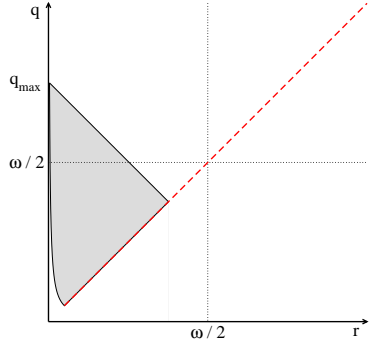}%
}

\caption[a]{\small 
An integration range for which a substitution of the type 
in \eq\nr{xy} can be carried out. 
} 
\la{fig:ranges_xy}
\end{figure}
%

Let us illustrate this procedure first
with the case (ii) of \eq\nr{Ij_ps_2}, which is the simplest.
Symmetrizing in $q\leftrightarrow r$, we cancel the $1/2$ from the 
prefactor (the $q$-integration then starts at $\lambda/2$). 
{}From the integrand, 
$
 n_{\fr{\omega}2+q} \, 
 n_{\fr{\omega}2+r} \, 
 n_{q+r} \,
 e^{q+r}(1   - e^{\omega}) 
$, 
we first subtract $n_{q+r}$ together with its residue, 
i.e.\ 
\be
 n^2_{\frac{\omega}{2}} n_{q+r} (1 - e^{\omega})
 = -(1+2 n_{\frac{\omega}{2}}) n_{q+r}
 \;. \la{sub1}
\ee 
The value of the remainder at $r=0$ can be simplified into 
\be
 n_{\fr{\omega}2 + q} \, 
 n_{\fr{\omega}2} \, 
 n_{q} \, 
 e^q (1  - e^{\omega} ) + 
 (1+2 n_{\frac{\omega}{2}}) n_{q}
 = 
 (1 + 2 n_{\frac{\omega}{2}} )
 n_{q+\frac{\omega}{2}}
 \;. \la{sub2}
\ee
When both \eq\nr{sub1} and \eq\nr{sub2} are subtracted, the final 
remainder simplifies tremendously, obtaining a form 
which vanishes for $r\to 0$ even at small $q$. In total, this 
``partitioning'' amounts to the identity
\ba
 n_{\fr{\omega}2+q} \, 
 n_{\fr{\omega}2+r} \, 
 n_{q+r} \, 
 e^{q+r}
 (1  - e^{\omega}) 
 = 
 (1 + 2 n_{\frac{\omega}{2}} )
 \biggl[
   -n_{q+r} + n_{q+\frac{\omega}{2}} - 
    (1 + n_{q+\frac{\omega}{2}}) 
    \frac{n_{q+r}n_{r+\frac{\omega}{2}}}{n_r^2}
 \biggr] 
 \;. \nn \la{ps_simp_1}
\ea  

The first term of \eq\nr{ps_simp_1} 
can now be integrated with the help 
of \eq\nr{xy_trick}. The second term is independent 
of $r$ so that the principal value integral in \eq\nr{Ij_ps_2} evaluates to 
\be
 \int_{ \frac{\lambda^2}{4q}}^{ q  }
 \! {\rm d}r \; \mathbbm{P} \Bigl( \frac{1}{r} \Bigr)
 = 
 \ln\biggl| 
      \frac{4q^2}{\lambda^2}
    \biggr|
 \;.
\ee
As mentioned the integral over the 
third term remains finite when sending $\lambda\to 0$.
Thereby the whole expression becomes
\ba
 \rho_{\mathcal{I}^{ }_\rmii{j},\rmi{(ii)}}^{(\rmi{ps})}(\omega) & \approx & 
 \frac{\omega^4}{(4\pi)^3}  (1 + 2 n_{\frac{\omega}{2}} )
 \biggl\{ 
 \int_{ {\lambda} }^{\infty}
 \! \frac{{\rm d}q}{q} \, n_q \, 
 \ln\biggl| 
      \frac{q - \sqrt{q^2-\lambda^2}}{q + \sqrt{q^2-\lambda^2}}
    \biggr|
 \nn & & 
  \; + \,  
 \int_{ \frac{\lambda}{2} }^{\infty}
 \! \frac{{\rm d}q}{q} 
  n_{q+\frac{\omega}{2}}
 \ln\biggl| 
      \frac{4q^2}{\lambda^2}
    \biggr|
 - \int_0^\infty \! \frac{{\rm d}q}{q}  
 \bigl( 1 + n_{q+\frac{\omega}{2}} \bigr)
 \int_{0}^{ q }
 \! \frac{ {\rm d}r }{r} 
 \frac{n_{q+r}n_{r+\frac{\omega}{2}}}{n_r^2}
 \biggr\} \;, \hspace*{5mm} \la{Ij_ps_v}
\ea
where ``$\approx$'' is a reminder of having set $\lambda\to 0$
whenever possible. 

As far as case (\iv) goes, the only Bose distribution which 
diverges at the edges of the integration range is $n_{q-\fr{\omega}2}$. 
Cancelling the factor 2
against $1/2$ in the prefactor of \eq\nr{Ij_ps_2}, 
the residue at $q=\frac{\omega}{2},r=0$ can be rewritten as 
\be
 n_{q-\fr{\omega}2} \, 
 n^2_{\fr{\omega}2} \, 
 (e^{{\omega}} - 1 ) 
 = 
  n_{q-\frac{\omega}{2}} 
 \bigl(1 + 2 n_{\frac{\omega}{2}} \bigr)
 \;.
\ee
When this term is subtracted and $r$ is set to zero, we get
\be
 n_{q-\fr{\omega}2} \, 
 n_{\fr{\omega}2} \, 
 n_q \, 
 e^{q-\frac{\omega}{2}} (e^{{\omega}} - 1 )
 -
  n_{q-\frac{\omega}{2}} 
 \bigl(1 + 2 n_{\frac{\omega}{2}} \bigr)
 = 
  - n_{q} 
 \bigl(1 + 2 n_{\frac{\omega}{2}} \bigr)
 \;.
\ee
Subtracting both terms, we are left over with an integrable remainder; 
in total, 
\ba
 && \hspace*{-5mm}
 n_{q-\fr{\omega}2} \, 
 n_{\fr{\omega}2+r} \, 
 n_{q-r} \, 
 e^{q-\frac{\omega}{2}} (e^{{\omega}} - 1 )
 = 
 ( 1 + 2 n_{\frac{\omega}{2}} )
 \biggl[
    n_{q-\frac{\omega}{2}} - n_{q} 
   - n_{q-\frac{\omega}{2}} 
    \frac{(1+n_{q-r})(n_q - n_{r+\frac{\omega}{2}})}
    {n_r n_{-\frac{\omega}{2}}}
 \biggr]
 \;. \nn \la{ps_simp_2}
\ea

Now, the principle value integral over the $1/r$ pole
in \eq\nr{Ij_ps_2} evaluates to 
\be
 \int_{ - \frac{\lambda^2}{4q}}^{
 \frac{\omega(2q-\omega)-\lambda^2}{2\omega}
 }
 \!\!\!\! {\rm d}r \; \mathbbm{P} \Bigl( \frac{1}{r} \Bigr)
 = 
 \ln\biggl| 
      \frac{2q[\omega(\omega-2q)+\lambda^2]}{\omega\lambda^2}
    \biggr|
 \;.
\ee
The whole expression then becomes
\ba
 \rho_{\mathcal{I}^{ }_\rmii{j},\rmi{(\iv)}}^{(\rmi{ps})}(\omega) & \approx & 
 \frac{\omega^4}{(4\pi)^3}  \Bigl(1 + 2 n_{\frac{\omega}{2}} \Bigr)
 \biggl\{ 
 \int_{ \frac{\omega}{2} }^{\infty}
 \! \frac{{\rm d}q}{q} 
 \Bigl(n_{q-\frac{\omega}{2}} - n_q \Bigr)
 \ln\biggl| 
      \frac{2q[\omega(\omega - 2q)+\lambda^2]}{\omega\lambda^2}
    \biggr|
 \nn & & \; - \,   
 \int_{ \frac{\omega}{2} }^{\infty}
 \! \frac{{\rm d}q}{q}
  n_{q-\frac{\omega}{2}} 
 \int_{0}^{ q - \frac{\omega}{2} }
 \! \frac{ {\rm d}r }{r} 
    \frac{(1+n_{q-r})(n_q - n_{r+\frac{\omega}{2}})}
    {n_r n_{-\frac{\omega}{2}}}
 \biggr\} \;. \la{Ij_ps_iv}
\ea

The case (i) is slightly more complicated than (ii) and (\iv)
because of the non-trivial geometry of the integration range
(cf.\ \fig\ref{fig:ranges_new}); nevertheless, the philosophy 
remains the same.  Cancelling the factor $1/2$ in \eq\nr{Ij_ps_2} 
for symmetrization, the residue of $n_{q+r}$ at $q=r=0$ can be 
written as
\be
 n^2_{\fr{\omega}2} \, n_{q+r}(1-e^{\omega})
 = - \bigl(1 + 2 n_{\frac{\omega}{2}} \bigr) n_{q+r}
 \;.
\ee 
It will be convenient to also subtract a ``vacuum'' term
in connection with this one, by replacing $n_{q+r}$ by $1+n_{q+r}$.
When this structure is subtracted and we set $r\to 0$, we get
the second subtraction, which can be simplified into
$
 - (1 + 2 n_{\frac{\omega}{2}} )
 n_{\fr{\omega}2-q}
$. 
The final remainder is again a function
which vanishes as $r\to 0$. The full partitioning
can be expressed as
\ba
 && n_{\fr{\omega}2-q} \, 
 n_{\fr{\omega}2-r} \, 
 n_{q+r} (1-e^{\omega})
 = 
 - \bigl(1 + 2 n_{\frac{\omega}{2}} \bigr)
 \biggl[ 1 + n_{q+r} +  n_{\fr{\omega}2-q}
  +(1+n_{\fr{\omega}2-r})  
  \frac{n_{q+r} n_{\fr{\omega}2-q}}{n_r^2}
 \biggr]
 \;. \nn \la{ps_simp_3}
\ea 

The first two terms of \eq\nr{ps_simp_3}, 
which only depend on $q+r$, are integrated 
as in \eq\nr{xy_trick}. The $r$-integrals over
the third term evaluate to 
\ba
 && 
 \int_{ \frac{\lambda^2}{4q} }^{ q} 
 \! {\rm d}r \; \mathbbm{P} \Bigl( \frac{1}{r} \Bigr)
 = 
 \ln\biggl| 
       \frac{4q^2}{\lambda^2}
    \biggr|
 \;, \\ 
 &&
 \int_{ \frac{\lambda^2}{4q}}^{
 \frac{\omega(\omega-2q)+\lambda^2}{2\omega}
 }
 \! {\rm d}r \; \mathbbm{P} \Bigl( \frac{1}{r} \Bigr)
 = \ln\biggl| 
      \frac{2q[\omega(\omega-2q)+\lambda^2]}{\omega\lambda^2}
    \biggr| 
 \;.
\ea
Setting again  $\lambda\to 0$ whenever possible, 
the whole expression becomes
\ba
 && \hspace*{-1cm}
 \rho_{\mathcal{I}^{ }_\rmii{j},\rmi{(i)}}^{(\rmi{ps})}(\omega)  \approx  
 \frac{\omega^4}{(4\pi)^3}  
  \Bigl(1+ 2 n_{\frac{\omega}{2}} \Bigr)
 \biggl\{ 
 \int_{ {\lambda} }^{ \frac{\omega}{2} }
 \! \frac{{\rm d}q}{q}
 (1+n_q) \ln \biggl| 
      \frac{q - \sqrt{q^2-\lambda^2}}{q + \sqrt{q^2-\lambda^2}}
    \biggr|
 \nn & & \; + \, 
 \int_{ \frac{\lambda}{2} }^{ \frac{\omega}{4} }
 \! \frac{{\rm d}q}{q} n_{\frac{\omega}{2} - q}
   \ln\biggl| 
       \frac{\lambda^2}{4q^2}
    \biggr|
 + 
 \int_{ \frac{\omega}{4} }^{ \frac{\omega}{2} }
 \! \frac{{\rm d}q}{q}
   n_{\frac{\omega}{2} - q} 
  \ln\biggl| 
      \frac{\omega\lambda^2}{2q[\omega(\omega-2q)+\lambda^2]}
    \biggr| 
 \nn & & \; - \,   
 \int_0^{ \frac{\omega}{2} } \! \frac{ {\rm d}q }{q} 
  n_{\frac{\omega}{2} - q} 
 \int_0^{ \frac{\omega}{4} - |q-\frac{\omega}{4}| } \! \frac{ {\rm d}r }{r}  
  \frac{n_{q+r} (1+n_{\fr{\omega}2-r})}{n_r^2}
 \biggr\} \;. \hspace*{5mm} \la{Ij_ps_i}
\ea

%
\subsection{Collecting the pieces}
\la{ss:collect}

Combining now everything together, the first observation is that 
the factorized powerlike term, \eq\nr{Ij_fzp_6}, and the very first
term from \eq\nr{Ij_ps_i}, with unity in the numerator, cancel
against each other. This implies, in particular, that 
$\rho^{ }_{\mathcal{I}^{ }_\rmii{j}}$ vanishes at $T=0$, because these
are the only terms  which are not (necessarily) 
proportional to a Bose distribution.


It is more cumbersome to deal with the 1-dimensional 
integrals containing $n_q$, $n_{|q-\frac{\omega}{2}|}$, 
and $n_{q+\frac{\omega}{2}}$. Summing together the terms from 
\eqs\nr{Ij_fz_3}, \nr{Ij_ps_v}, \nr{Ij_ps_iv} and \nr{Ij_ps_i}, 
we obtain
\ba
 \rho_{\mathcal{I}^{ }_\rmii{j}}^{(\rmi{1d})}(\omega) \!\! & \equiv & \!\! 
 \frac{\omega^4}{(4\pi)^3} (1 + 2 n_{\frac{\omega}{2}}) \biggl\{
 \nn 
  & \mbox{(a)} & \;\quad\;
  \int_{\lambda}^{\frac{\omega}{2}} \! \frac{ {\rm d}q }{q} \, n_q 
     \ln
         \biggl| 
            \frac{(q +\sqrt{q^2-\lambda^2})^2 - \frac{\lambda^4}{\omega^2} }
                 {(q -\sqrt{q^2-\lambda^2})^2 - \frac{\lambda^4}{\omega^2} }
         \biggr|
         \biggl| 
            \frac{q - \sqrt{q^2-\lambda^2}}
                 {q + \sqrt{q^2-\lambda^2}}
         \biggr|^2
  \nn 
  & \mbox{(b)} & \; + \,
  \int_{\frac{\omega}{2}}^{\infty} \! \frac{ {\rm d}q }{q} \, n_q 
     \ln
         \biggl| 
            \frac{(q +\sqrt{q^2-\lambda^2})^2 - \frac{\lambda^4}{\omega^2} }
                 {(q -\sqrt{q^2-\lambda^2})^2 - \frac{\lambda^4}{\omega^2} }
         \biggr|
         \biggl| 
            \frac{q - \sqrt{q^2-\lambda^2}}
                 {q + \sqrt{q^2-\lambda^2}}
         \biggr|
         \biggl| 
            \frac{ \omega \lambda^2 }
                 { 2q[  \omega(\omega - 2q) + \lambda^2 ] }
         \biggr|
  \nn 
  & \mbox{(c)} & 
  \; + \, 
  \int_0^\infty \! {\rm d}q \, n_q \; \mathbbm{P} \, \biggl[ 
      \frac{1}{q+\frac{\omega}{2}}
      \ln\biggl| \frac{\lambda^2}{2q\omega - \lambda^2} \biggr|
      +  \frac{1}{q-\frac{\omega}{2}} 
      \ln\biggl| \frac{\lambda^2}{2q\omega + \lambda^2} \biggr|
   \;\biggr]
  \nn 
  & \mbox{(d)} & 
  \; + \, 
  \int_{\frac{\lambda}{2}}^{\frac{\omega}{4}} \! \frac{{\rm d}q}{q}
    \biggl[ 
      n_{\frac{\omega}{2} + q} 
      \ln\biggl| \frac{4q^2}{\lambda^2} \biggr|
     + n_{\frac{\omega}{2} - q} 
      \ln\biggl| \frac{\lambda^2} {4q^2} \biggr|
    \;\biggr]
  \nn 
  & \mbox{(e)} & 
  \; + \, 
  \int_{\frac{\omega}{4}}^{\frac{\omega}{2}} \! \frac{{\rm d}q}{q}
  \biggl[
      n_{q+\frac{\omega}{2} } 
      \ln\biggl| \frac{4q^2}{\lambda^2} \biggr|
      + 
      n_{\frac{\omega}{2} - q} 
      \ln\biggl| \frac{\omega\lambda^2}
                 {2q[\omega(\omega - 2q) + \lambda^2]} \biggr|
   \;\biggr]
  \nn 
  & \mbox{(f)} & 
  \; + \, 
  \int_{\frac{\omega}{2}}^\infty \! \frac{{\rm d}q}{q}
  \biggl[
      n_{q + \frac{\omega}{2} } 
      \ln\biggl| \frac{4q^2}{\lambda^2} \biggr|
      + 
      n_{q - \frac{\omega}{2}} 
      \ln\biggl| \frac{2q[\omega(\omega - 2q) + \lambda^2]}
            {\omega\lambda^2}
                  \biggr|
   \;\biggr]
  \biggr\} 
  \;. \hspace{5mm} \la{Ij_fz_5} 
\ea
The subsequent step is to take the limit $\lambda\to 0$. 
The most subtle in this respect are the cases (a) and (b).  
The various structures appearing there can be simplified as 
\ba
 && \hspace*{-1cm}
 \biggl| 
            \frac{q +\sqrt{q^2-\lambda^2} + \frac{\lambda^2}{\omega} }
                 {q -\sqrt{q^2-\lambda^2} + \frac{\lambda^2}{\omega} }
 \biggr| 
 \biggl| 
            \frac{q - \sqrt{q^2-\lambda^2}}
                 {q + \sqrt{q^2-\lambda^2}}
 \biggr| 
  = 
 \biggl| 
            \frac{(q +\sqrt{q^2-\lambda^2} + \frac{\lambda^2}{\omega})^2 }
                 {(q + \frac{\lambda^2}{\omega} )^2 -({q^2-\lambda^2})  }
 \biggr| 
 \biggl| 
            \frac{q^2 - ({q^2-\lambda^2})}
                 {(q + \sqrt{q^2-\lambda^2})^2}
 \biggr| 
 \nn 
 && \hspace*{1.5cm} = \,  
            \frac{\lambda^2}
             {\frac{2 q \lambda^2}{\omega} 
            + \frac{\lambda^4}{\omega^2} + \lambda^2 }  
            \biggl( \frac{ q + \sqrt{q^2-\lambda^2}
                              + \frac{\lambda^2}{\omega} }
                         { q + \sqrt{q^2-\lambda^2}} \biggr)^2
 \;  \stackrel{\lambda \ll \omega}{\approx} \; 
 \frac{\omega^2}{\omega(\omega+2q) + \lambda^2}
 \;, \\
 && \hspace*{-1cm}
 \biggl| 
            \frac{q +\sqrt{q^2-\lambda^2} - \frac{\lambda^2}{\omega} }
                 {q -\sqrt{q^2-\lambda^2} - \frac{\lambda^2}{\omega} }
 \biggr| 
 \biggl| 
            \frac{q - \sqrt{q^2-\lambda^2}}
                 {q + \sqrt{q^2-\lambda^2}}
 \biggr| 
  = 
 \biggl| 
            \frac{(q +\sqrt{q^2-\lambda^2} - \frac{\lambda^2}{\omega})^2 }
                 {(q - \frac{\lambda^2}{\omega} )^2 -({q^2-\lambda^2})  }
 \biggr| 
 \biggl| 
            \frac{q^2 - ({q^2-\lambda^2})}
                 {(q + \sqrt{q^2-\lambda^2})^2}
 \biggr| 
 \nn 
 && \hspace*{1.5cm} = \,  
 \biggl| 
            \frac{\lambda^2}
             {- \frac{2 q \lambda^2}{\omega} 
            + \frac{\lambda^4}{\omega^2} + \lambda^2 }  
  \biggr|
            \biggl( \frac{ q + \sqrt{q^2-\lambda^2}
                              - \frac{\lambda^2}{\omega} }
                         { q + \sqrt{q^2-\lambda^2}} \biggr)^2
 \;  \stackrel{\lambda \ll \omega}{\approx} \; 
 \biggl|
 \frac{\omega^2}{\omega(\omega-2q) + \lambda^2}
 \biggr|
 \;, \\
 && \hspace*{-1cm}
 \biggl| 
            \frac{q +\sqrt{q^2-\lambda^2} - \frac{\lambda^2}{\omega} }
                 {q -\sqrt{q^2-\lambda^2} - \frac{\lambda^2}{\omega} }
 \biggr| 
 \biggl| 
            \frac{ \omega \lambda^2 }
                 { 2q[  \omega(\omega - 2q) + \lambda^2 ] }
 \biggr| 
  = 
 \biggl| 
            \frac{(q +\sqrt{q^2-\lambda^2} - \frac{\lambda^2}{\omega})^2 }
                 {(q - \frac{\lambda^2}{\omega} )^2 -({q^2-\lambda^2})  }
 \biggr| 
 \biggl| 
            \frac{ \omega \lambda^2 }
                 { 2q[  \omega(\omega - 2q) + \lambda^2 ] }
 \biggr| 
 \nn 
 && \hspace*{1.5cm} = \,  
 \biggl| 
            \frac{\frac{\omega\lambda^2}{2q}}
             {- \frac{2 q \lambda^2}{\omega} 
            + \frac{\lambda^4}{\omega^2} + \lambda^2 }  
 \biggr| 
 \biggl| 
               \frac{ (q + \sqrt{q^2-\lambda^2}
                              - \frac{\lambda^2}{\omega})^2 }
                         {\omega(\omega - 2q) + \lambda^2 } 
 \biggr| 
 \;  \stackrel{\lambda \ll \omega}{\approx} \; 
 \frac{2 q \omega}{( \omega-2q + \frac{\lambda^2}{\omega})^2}
 \;.
\ea
These then lead to 
\ba
 \mbox{(a)} &\approx &
 \int_0^{\frac{\omega}{2}} 
 \! \frac{{\rm d}q}{q} \, n_q 
 \ln \frac{\omega^2}{(\omega - 2q)(\omega + 2q)}
 \;, \\ 
 \mbox{(b)} &\approx &
 \int_{\frac{\omega}{2}}^{\infty} 
 \! \frac{{\rm d}q}{q} \, n_q 
 \ln \frac{2 \omega^2 q }{(\omega - 2q)^2(\omega + 2q)}
 \;.
\ea
In case (d) we observe that the lower limit of the integration
can safely be set to zero {\em in the sum}. With this knowledge in mind, 
the left-most terms of (d), (e) and (f) can be summed together and 
the integration variable can be shifted as $q\to -\frac{\omega}{2}+q$, 
to obtain 
\ba
 \mbox{(d+e+f)}_\rmii{left} & \approx & 
 \int_{\frac{\omega}{2}}^{\infty} \! {\rm d}q\, n_q 
 \, \biggl[ 
   \frac{1}{q-\frac{\omega}{2}} \ln \frac{(\omega-2q)^2}{\lambda^2}
 \biggr]
 \;. \la{ill}
\ea
In the right-most terms of (d) and (e) we can 
shift $q\to \frac{\omega}{2} - q$, and in the right-most term 
of~(f), we can shift $q\to \frac{\omega}{2} + q$, thereby obtaining
\ba
 \mbox{(d+e)}_\rmii{right} \!\! & \approx & \!\! 
 \int_0^{\frac{\omega}{2}} \! {\rm d}q\, n_q 
 \, \biggl[ 
   \frac{\theta(\frac{\omega}{4}-q)}{q-\frac{\omega}{2}} 
   \ln \biggl| \frac{(\omega-2q)(2\omega q + \lambda^2)}
                    {\omega\lambda^2} \biggr|
 + 
   \frac{\theta(q-\frac{\omega}{4})}{q-\frac{\omega}{2}} 
   \ln \biggl| \frac{(\omega-2q)^2}
                    {\lambda^2} \biggr| \;
 \biggr]
 \;, \nn \\
  \mbox{(f)}_\rmii{right} \!\! & \approx & \!\! 
 \int_0^{\infty} \! {\rm d}q\, n_q 
 \, \biggl[ 
   \frac{1}{q+\frac{\omega}{2}} 
   \ln \biggl| \frac{(\omega+2q)(2\omega q - \lambda^2)}
                    {\omega\lambda^2} \biggr| \;
 \biggr]
 \;.
\ea
Summing all of these together with (c), the $\lambda$'s cancel or become 
harmless, and the regularization observed in (d), momentarily implicit in 
\eq\nr{ill}, is rediscovered as a principle value integral 
over $1/(q-\frac{\omega}{2})$. 
We thereby obtain the first three rows of \eq\nr{Ij_final}
below.

%
\subsection{Final expression}

Putting together the 2-dimensional integrals from 
\eqs\nr{Ij_ps_v}, \nr{Ij_ps_iv} and \nr{Ij_ps_i}, and 
the 1-dimensional integrals from \se\ref{ss:collect},  we obtain
a convergent expression for $\rho^{ }_{\mathcal{I}^{ }_\rmii{j}}$ 
at $\lambda\to 0$:
\ba
 && \hspace*{-0.5cm} 
 \frac{(4\pi)^3\rho^{ }_{\mathcal{I}^{ }_\rmii{j}}(\omega)}
 {\omega^4 (1+2 n_{\frac{\omega}{2}})}  = 
 \nn 
  & & 
 \int_0^{ \frac{\omega}{4} } \! { {\rm d}q } \, n_{q} \; 
   \biggl[  
     \Bigl( \frac{1}{q-\frac{\omega}{2}} - \frac{1}{q}  \Bigr)
     \ln\Bigl( 1 - \frac{2q}{\omega} \Bigr)
   - \frac{\frac{\omega}{2}}{q(q+\frac{\omega}{2})}
     \ln\Bigl( 1 + \frac{2q}{\omega} \Bigr)
   \biggr]  
 \nn 
 & + & 
 \int_{ \frac{\omega}{4} }^{ \frac{\omega}{2} } 
 \! { {\rm d}q } \, n_{q} \; 
   \biggl[  
     \Bigl( \frac{2}{q-\frac{\omega}{2}} - \frac{1}{q}  \Bigr)
     \ln\Bigl( 1 - \frac{2q}{\omega} \Bigr)
   - \frac{\frac{\omega}{2}}{q(q+\frac{\omega}{2})}
     \ln\Bigl( 1 + \frac{2q}{\omega} \Bigr)
  - 
     \frac{1}{q-\frac{\omega}{2}} 
     \ln\Bigl(\frac{2q}{\omega} \Bigr)
   \biggr]  
 \nn 
 & + & 
 \int_{ \frac{\omega}{2} }^{ \infty } 
 \! { {\rm d}q } \, n_{q} \; 
   \biggl[  
     \Bigl( \frac{2}{q-\frac{\omega}{2}} - \frac{2}{q}  \Bigr)
     \ln\Bigl( \frac{2q}{\omega} -1 \Bigr)
   - \frac{\frac{\omega}{2}}{q(q+\frac{\omega}{2})}
     \ln\Bigl( 1 + \frac{2q}{\omega} \Bigr)
  +  \Bigl( \frac{1}{q} 
           - 
     \frac{1}{q-\frac{\omega}{2}} \Bigr)
     \ln\Bigl(\frac{2q}{\omega} \Bigr)
   \biggr]  
%
 \nn & + & 
 \int_0^{ \frac{\omega}{2} } \! { {\rm d}q }  
 \int_0^{ \frac{\omega}{4} - |q-\frac{\omega}{4}| } 
 \! { {\rm d}r }  \;
 \biggl( - \frac{1}{qr }\biggr)\;
  \frac{ n_{\frac{\omega}{2} - q}\, n_{q+r} (1+n_{\fr{\omega}2-r})}{n_r^2}
 \nn & + & 
 \int_{ \frac{\omega}{2} }^{\infty}
 \! {{\rm d}q}
 \int_{0}^{ q - \frac{\omega}{2} }
 \! { {\rm d}r } \;
 \biggl( - \frac{1}{qr }\biggr) \;
    \frac{  n_{q-\frac{\omega}{2}} 
   (1+n_{q-r})(n_q - n_{r+\frac{\omega}{2}})}
    {n_r n_{-\frac{\omega}{2}}}
 \nn & + & 
 \int_0^\infty \! {{\rm d}q}  
 \int_{0}^{ q }
 \! { {\rm d}r } \;
 \biggl( - \frac{1}{qr }\biggr) \;
 \frac{ ( 1 + n_{q+\frac{\omega}{2}} )
 n_{q+r}n_{r+\frac{\omega}{2}}}{n_r^2}
 \; + \; \rmO(\lambda\ln\lambda)
 \;. \hspace*{5mm} \la{Ij_final} 
\ea
For numerical evaluation, it is helpful to combine
the first terms of the 2nd and 3rd rows into
\be
 \int_{ \frac{\omega}{4} }^{ \infty } 
 \! { {\rm d}q } \,   
      \mathbbm{P} \Bigl( \frac{2 n_{q}}{q-\frac{\omega}{2}} \Bigr)
     \ln\Bigl| 1 - \frac{2q}{\omega} \Bigr|
 = 
  \int_0^{ \frac{\omega}{4} }
  \! \frac{{\rm d}q}{q}
  \Bigl( n_{\frac{\omega}{2} + q} - n_{\frac{\omega}{2} - q} \Bigr)
  \ln 
     \frac{4q^2}{\omega^2} 
 +
 \int_{ \frac{3\omega}{4} }^{ \infty } 
 \! { {\rm d}q } \, 
     \frac{2 n_{q} }{q-\frac{\omega}{2}} 
     \ln\Bigl( \frac{2q}{\omega} -1 \Bigr)  
 \;.
\ee

%
\subsection{Ultraviolet asymptotics}

As a check of the result obtained, it is useful to verify that 
it reproduces the known behaviour in the limit $\omega \gg \pi T$.
{}From ref.~\cite{ope}, \eq(A.32), we recall that asymptotically
$\mathcal{I}^{ }_\rmii{j}$ reads
\newcommand{\s}{\mathcal{S}}
\ba
 \mathcal{I}^{ }_\rmi{j} & = &
 \frac{\s_1(1-\s_2)}{\epsilon^3(1-2\epsilon)}
 - \frac{2(3+\epsilon)\s_3}{\epsilon}
 - \frac{2(5+\epsilon)(10+5\epsilon+\epsilon^2)\s_5}
   {3\epsilon}
 \nn & & 
 \; + \; 
 \biggl[ 
 \frac{2(3+\epsilon)}{2-\epsilon} + 
 \frac{20(1-\epsilon)^2}{\epsilon(2-\epsilon)P^2}
 \biggl( \frac{p^2}{3-2\epsilon} - p_n^2 \biggr)
 \biggl]
 \s_6 
 + \rmO\Bigl(\frac{1}{P^2}\Bigr)
 \;, \la{res_Ij}
\ea
with 
\ba
 \s_1 & \equiv & \frac{P^{4-4\epsilon}}{(4\pi)^{4-2\epsilon}}
 \frac{\Gamma^2(1+\epsilon) \Gamma^4(1-\epsilon)}{\Gamma^2(1-2\epsilon)}
 \;, \la{s1_def} \\ 
 \s_2 & \equiv & \frac{\Gamma(1+2\epsilon)\Gamma^2(1-2\epsilon)}
 {\Gamma(1-3\epsilon)\Gamma^2(1+\epsilon)\Gamma(1-\epsilon)}
 \;, \\ 
 \s_3 & \equiv & \frac{P^{2-2\epsilon}}{(4\pi)^{2-\epsilon}} 
 \frac{\Gamma(1+\epsilon) \Gamma^2(1-\epsilon)}{\Gamma(1-2\epsilon)}
 \int_{\vec{q}} \frac{\nB{}(q)}{q}
 \;, \\ 
 \s_4 & \equiv & \frac{(1-2\epsilon)P^2}{2} \int_{\vec{q},\vec{r}}
 \frac{\nB{}(q)}{q} \frac{\nB{}(r)}{r^3} 
 \;, \\ 
 \s_5 & \equiv & 
 \frac{P^{-2\epsilon}}{(4\pi)^{2-\epsilon}} 
 \frac{\Gamma(1+\epsilon) \Gamma^2(1-\epsilon)}{\Gamma(1-2\epsilon)}
 \frac{1}{P^2}\biggl( \frac{p^2}{3-2\epsilon} - p_n^2 \biggr)
 \int_{\vec{q}} q\, \nB{}(q)
 \;, \\ 
 \s_6 & \equiv & 
 \int_{\vec{q},\vec{r}}
 \frac{\nB{}(q)}{q} \frac{\nB{}(r)}{r}
 \;. \la{s6_def}
\ea
Setting $P = (p_n,\vec{0})$, 
analytically continuing to Minkowski signature, 
$p_n \to -i [\omega + i 0^+]$, and 
extracting the spectral function according to \eq\nr{rho_w}, yields
\ba
 \rho^{ }_{\s_1} & = & 
 2\pi\epsilon\, 
 \frac{\omega^4 \mu^{-4\epsilon}}{(4\pi)^4}  
 \Bigl(1 + 2 \epsilon \ln\frac{\bmu^2}{\omega^2} \Bigr) 
 + \rmO(\epsilon^3)
 \;, \\
 \rho^{ }_{\s_2} & = & 
 1  + \rmO(\epsilon^3)
 \;, \\
 \rho^{ }_{\s_3} & = & 
 - \pi\epsilon\, 
 \frac{\omega^2 \mu^{-2\epsilon}}{(4\pi)^2}  
  \frac{T^2}{12} 
 + \rmO(\epsilon^2)
 \;, \\
 \rho^{ }_{\s_4} & = & 
 0
 \;, \\
 \rho^{ }_{\s_5} & = & 
 - \pi\epsilon\, 
 \frac{\mu^{-2\epsilon}}{(4\pi)^2}  
  \frac{\pi^2 T^4}{30} 
 + \rmO(\epsilon^2)
 \;, \\
 \rho^{ }_{\s_6} & = & 
 0
 \;,
\ea
where
$ 
 \mu^{2} =  \bmu^{2}
 {e^{\gammaE}}/{4\pi}
$.
Inserting these into \eq\nr{res_Ij} we notice that the temperature-independent 
part is of $\rmO(\epsilon)$ and therefore vanishes in the continuum 
limit, just as we found at the beginning of \se\ref{ss:collect}. 
The remaining asymptotics reads 
\be
 \rho^{\rmii{OPE}}_{\mathcal{I}^{ }_\rmii{j}}(\omega) = 
 \frac{\omega^2 T^2}{32\pi} + \frac{5\pi T^4}{72} +
 \rmO\Bigl( \frac{T^6}{\omega^2} \Bigr)  \;. \la{rho_Ij_as}
\ee
It is easy to check numerically that \eq\nr{Ij_final} 
indeed approaches this form at $\omega \gg \pi T$.

\newpage

%
\section{Main results for the other cases}
\la{se:app2}

Below we list the spectral functions corresponding 
to the master sum-integrals defined in \eqs\nr{m_first}--\nr{m_penultimate}.
Like in appendix A, all terms containing structures of 
the type $\omega^n \delta(\omega) $ are omitted from the outset although, 
as discussed in \se\ref{ss:tau}, under specific circumstances 
such terms may conspire to give a ``zero-mode'' contribution; 
we do not believe this to be the case for the observables
and perturbative order considered in the present investigation.  

%
\subsection{$\rho^{ }_{\mathcal{J}^{ }_\rmii{a}}$}

The sum-integral $\mathcal{J}^{ }_\rmi{a}$ is defined as 
\ba
 \mathcal{J}^{ }_\rmi{a} & \equiv &
 \Tint{Q} \frac{P^2}{Q^2}
 \;. \la{Ja} 
\ea
Since the dependence on $p_n$ is polynomial, there is no cut, 
and the corresponding spectral function vanishes: 
\be
 \rho^{ }_{\mathcal{J}^{ }_\rmii{a}}(\omega) =  0  \;. \la{Ja_final}
\ee

%
\subsection{$\rho^{ }_{\mathcal{J}^{ }_\rmii{b}}$}

The sum-integral $\mathcal{J}^{ }_\rmi{b}$ is defined as 
\ba
 \mathcal{J}^{ }_\rmi{b} & \equiv &
 \Tint{Q} \frac{P^4}{Q^2(Q-P)^2}
 \;. \la{Jb}
\ea
Carrying out the Matsubara sum, taking the cut, and
setting subsequently $\vec{p}\to \vec{0}$, we get 
\be
 \rho^{ }_{\mathcal{J}^{ }_\rmii{b}}(\omega) =  
 \int_{\vec{q}} \frac{\omega^4\pi}{4 q^2}
 \Bigl[ \delta(\omega - 2 q) - \delta(\omega + 2 q) \Bigr]
 (1+2 n_q)
 \;. \la{Jb_sum}
\ee
Making use of the usual dimensionally regularized integration measure, 
the integral over the Dirac-$\delta$ yields
\be
 \int_{\vec{q}} \pi \delta(\omega - 2q)
 = \frac{\omega^2 \mu^{-2\epsilon}}{16\pi}
 \biggl[ 1 + \epsilon \biggl(
   \ln\frac{\bmu^2}{\omega^2} + 2 \biggr)
 + \rmO(\epsilon^2) \biggr]
 \;, \la{Ja_int}
\ee
and the complete expression reads (for $\omega > 0$)
\be
  \rho^{ }_{\mathcal{J}^{ }_\rmii{b}}(\omega) =  
 \frac{\omega^4 \mu^{-2\epsilon}}{16\pi}
 \bigl( 1 + 2 n_{\frac{\omega}{2}} \bigr)
 \biggl[ 1 + \epsilon \biggl(
   \ln\frac{\bmu^2}{\omega^2} + 2 \biggr)
 + \rmO(\epsilon^2) \biggr]
 \;. \la{Jb_final}
\ee
Thermal effects are seen to be exponentially suppressed here, 
so we nicely match the asymptotic OPE-result as
determined in ref.~\cite{ope}, 
\be
 \rho^{\rmii{OPE}}_{\mathcal{J}^{ }_\rmii{b}}(\omega) =  
 \frac{\omega^4\mu^{-2\epsilon}}{16\pi} 
 \biggl[1 + \epsilon \biggl(
  \ln\frac{\bmu^2}{\omega^2} + 2 \biggr) \biggr]
 +  \rmO\biggl( \frac{T^6}{\omega^2} \biggr)
 \;. \la{rho_Jb_as}
\ee

%
\subsection{$\rho^{ }_{\mathcal{I}^{ }_\rmii{a}}$}

The sum-integral $\mathcal{I}^{ }_\rmi{a}$ is defined as 
\ba
  \mathcal{I}^{ }_\rmi{a} & \equiv &
 \Tint{Q,R} \frac{1}{Q^2R^2}
 \;.
\ea
Since there is no dependence on $P$, the corresponding
spectral function vanishes: 
\be
 \rho^{ }_{\mathcal{I}^{ }_\rmii{a}}(\omega) =  0  \;. \la{Ia_final}
\ee

%
\subsection{$\rho^{ }_{\mathcal{I}^{ }_\rmii{b}}$}

The sum-integral $\mathcal{I}^{ }_\rmi{b}$ is defined as 
\ba
 \mathcal{I}^{ }_\rmi{b} & \equiv &
 \Tint{Q,R} \frac{P^2}{Q^2R^2(R-P)^2}
 \;.
\ea
This factorizes into the same form as $\mathcal{J}^{ }_\rmi{b}$  
in \eq\nr{Jb}, multiplied by the well-known
\be
 \Tint{Q} \frac{1}{Q^2} = \frac{T^2}{12} + \rmO(\epsilon)
 \;. \la{tadpole}
\ee
Within the 2-loop contributions we only need to go to $\rmO(\epsilon^0)$, 
so the result reads 
\be
 \rho^{ }_{\mathcal{I}^{ }_\rmii{b}}(\omega) =  
 - \frac{\omega^2 T^2}{192\pi}
 \bigl( 1 + 2 n_{\frac{\omega}{2}} \bigr)
 \;. \la{Ib_final}
\ee
For $\omega\gg \pi T$ this agrees with the asymptotic OPE-form from 
ref.~\cite{ope}, 
\be
 \rho^{\rmii{OPE}}_{\mathcal{I}^{ }_\rmii{b}}(\omega) = 
 - \frac{\omega^2 T^2}{192\pi} +
 \rmO\biggl( \frac{T^6}{\omega^2} \biggr)  \;. \la{rho_Ib_as}
\ee

%
\subsection{$\rho^{ }_{\mathcal{I}^{ }_\rmii{c}}$}

The sum-integral $\mathcal{I}^{ }_\rmi{c}$ is defined as 
\ba
 \mathcal{I}^{ }_\rmi{c} & \equiv &
 \Tint{Q,R} \frac{P^2}{Q^2R^4}
 \;.
\ea
Since the dependence on $p_n$ is polynomial, there is no cut, 
and the corresponding spectral function vanishes: 
\be
 \rho^{ }_{\mathcal{I}^{ }_\rmii{c}}(\omega) =  0  \;. \la{Ic_final}
\ee

%
\subsection{$\rho^{ }_{\mathcal{I}^{ }_\rmii{d}}$}

The sum-integral $\mathcal{I}^{ }_\rmi{d}$ is defined as 
\ba
 \mathcal{I}^{ }_\rmi{d} & \equiv &
 \Tint{Q,R} \frac{P^4}{Q^2R^4(R-P)^2}
 \;.
\ea
This factorizes into the tadpole in \eq\nr{tadpole}, times a 
1-loop sum-integral. For the latter, a possible trick is to write
\ba
 \Tint{R} \frac{1}{R^4(R-P)^2}
 & = & 
 \lim_{\lambda\to 0} 
 \fr12 
 \Tint{R} \biggl\{ 
  \frac{1}{[R^2+\lambda^2]^2[(R-P)^2+\lambda^2]} + 
  \frac{1}{[R^2+\lambda^2][(R-P)^2+\lambda^2]^2}  
 \biggr\}
 \nn 
 & = & 
 \lim_{\lambda\to 0} 
 \biggl\{
 - \fr12 \frac{{\rm d}}{{\rm d}\lambda^2}
 \Tint{R}  
  \frac{1}{[R^2+\lambda^2][(R-P)^2+\lambda^2]} 
 \biggr\}
 \;. \la{dl_Id}
\ea
The result for the Matsubara sum here (after taking the cut and 
setting $\vec{p}\to\vec{0}$) can be read directly from \eq\nr{Jb_sum},
\ba
 \Tint{R}  
 \frac{1}{[R^2+\lambda^2][(R-P)^2+\lambda^2]}
  & \rightarrow & 
 \int_{\vec{r}} \frac{\pi}{4 E_r^2}
 \Bigl[ \delta(\omega - 2 E_r) - \delta(\omega + 2 E_r) \Bigr]
 \bigl( 1+2 n^{ }_{E_r} \, \bigr)
 \nn & = & 
 \frac{(\omega^2-4\lambda^2)^{\fr12}}{16\pi\omega}
 \bigl( 1 + 2 n_{\frac{\omega}{2}}\bigr)
 \,\theta(\omega - 2\lambda) + \rmO(\epsilon)
 \;, \la{Id_2} 
\ea 
where $E_r^2 \equiv r^2 + \lambda^2$ and the spatial 
integral was performed for $\omega > 0$.
Taking the derivative in \eq\nr{dl_Id}
and setting $\lambda\to 0$ finally yields
\be
  \rho^{ }_{\mathcal{I}^{ }_\rmii{d}}(\omega) =
 \frac{\omega^2 T^2}{192\pi}
 \bigl( 1 + 2 n_{\frac{\omega}{2}} \bigr)
 \; = \; 
 \frac{2\omega^2}{(4\pi)^3}
 \bigl( 1 + 2 n_{\frac{\omega}{2}} \bigr)
 \int_0^\infty \! {\rm d}q\, n_q \, q
 \;, \la{Id_final}
\ee
where we have also given an integral representation, 
in order to facilitate comparison with the other spectral
functions below. 
The result matches the OPE-behaviour from ref.~\cite{ope}, 
\be
 \rho^{\rmii{OPE}}_{\mathcal{I}^{ }_\rmii{d}}(\omega) = 
 \frac{\omega^2 T^2}{192\pi} +
 \rmO\biggl( \frac{T^6}{\omega^2} \biggr)  \;. \la{rho_Id_as}
\ee

%
\subsection{$\rho^{ }_{\mathcal{I}^{ }_\rmii{e}}$}

The sum-integral $\mathcal{I}^{ }_\rmi{e}$ is defined as 
\ba
 \mathcal{I}^{ }_\rmi{e} & \equiv &
 \Tint{Q,R} \frac{P^2}{Q^2R^2(Q-R)^2}
 \;.
\ea
Again the result is a polynomial in $p_n$ and has no cut; 
in fact even before the cut this massless sum-integral vanishes 
exactly in dimensional regularization in any dimension~\cite{gE2}. 
Therefore, 
\be
 \rho^{ }_{\mathcal{I}^{ }_\rmii{e}}(\omega) =  0  \;. \la{Ie_final}
\ee

%
\subsection{$\rho^{ }_{\mathcal{I}^{ }_\rmii{f}}$}

The sum-integral $\mathcal{I}^{ }_\rmi{f}$ is defined as 
\ba
  \mathcal{I}^{ }_\rmi{f} & \equiv &
 \Tint{Q,R} \frac{P^2}{Q^2[(Q-R)^2+\lambda^2](R-P)^2}
 \;.
\ea
After the Matsubara sums the expression for the corresponding
spectral function reads
\ba
 & & \hspace*{-0.6cm} \rho^{ }_{\mathcal{I}^{ }_\rmii{f}}(\omega) = 
 \int_{\vec{q,r}} 
 \frac{\omega^2 \pi }{8 q r E_{qr}} \biggl\{ 
 \la{If} \\
 \nn & - & \!\!
 \Bigl[\delta(\omega - q - r -E_{qr}) - \delta(\omega+q+r+E_{qr}) \Bigr]
 \Bigl[ {(1+n_{qr})(1+n_q+n_r)+n_q n_r} \Bigr]
 \nn & - &  \!\!
 \Bigl[\delta(\omega-q-r+E_{qr}) - \delta(\omega + q + r -E_{qr}) \Bigr]
 \Bigl[ {n_{qr}(1+n_q + n_r )-n_qn_r} \Bigr]
 \nn & - &  \!\!
 \Bigl[\delta(\omega - q + r -E_{qr}) - \delta(\omega+q-r+E_{qr}) \Bigr]
 \Bigl[ {n_r(1+n_q+n_{qr})-n_q n_{qr}} \Bigr]
 \nn & - &  \!\!
 \Bigl[\delta(\omega + q - r -E_{qr}) - \delta(\omega-q+r+E_{qr}) \Bigr]
 \Bigl[ {n_q(1+n_r+n_{qr})-n_r n_{qr}} \Bigr]
 \biggr\} \;. \nonumber
\ea
Evidently this sum-integral only contains a phase-space part, 
and the expression corresponding to \eq\nr{Ij_ps_2} becomes
\ba
 \rho_{\mathcal{I}^{ }_\rmii{f}}^{(\rmi{ps})}(\omega) & = & 
 \frac{\omega^2}{(4\pi)^3}  \bigl(
  1 - e^{\omega} 
 \bigr)
 \biggl\{ 
 \nn 
 & \mbox{(i)}  &  \hspace*{5mm}
 \int_{\frac{\lambda^2}{2\omega}}^{\frac{\omega}{2}}
 \! {\rm d}q
 \int_{\frac{\lambda^2}{4q}}^{\frac{\omega(\omega-2q)+\lambda^2}{2\omega}} 
 \!\!\!\! {\rm d}r \; 
 n_{\fr{\omega}2-q} \, 
 n_{\fr{\omega}2-r} \, 
 n_{q+r}
 \nn
 & \mbox{(ii)}  & + \, 
 \int_{ 0 }^{\infty}
 \! {\rm d}q
 \int_{\frac{\lambda^2}{4q}}^{ \infty }
 \! {\rm d}r \; 
 n_{\fr{\omega}2+q} \, 
 n_{\fr{\omega}2+r} \, 
 n_{q+r}\, 
 e^{q+r}
 \nn
 & \mbox{(\iv)}  & + \, 2
 \int_{ \frac{\omega}{2} }^{\infty}
 \! {\rm d}q
 \int_{ - \frac{\lambda^2}{4q}}^{
    \frac{\omega(2q-\omega)-\lambda^2}{2\omega}}
 \!\!\!\! {\rm d}r \; 
 n_{q-\fr{\omega}2} \, 
 n_{\fr{\omega}2+r} \, 
 n_{q-r} \,
 e^{q - \frac{\omega}{2}}
 \biggr\} \;.  \la{If_ps_2}
\ea
Subsequently we can reflect the range $q < r$ to 
$q > r$ in cases (i) and (ii).
The subtractions of the potentially divergent parts 
are carried out like in \eqs\nr{ps_simp_1}, 
\nr{ps_simp_2}, \nr{ps_simp_3}; the integrals over $y$ 
and $r$ are trivial in these terms. 

Once the pieces are collected together, we can again identify 
a ``vacuum part'', a ``1-dimensional integral'', and 
a ``2-dimensional integral''. The contribution to the vacuum 
part from the phase space integrals reads 
\ba
 & & \hspace*{-2cm}
 \frac{\omega^2}{(4\pi)^3}
  \bigl( 1 + 2 n_{\frac{\omega}{2}} \bigr)
 \biggl\{ - \int_\lambda^{\frac{\omega}{2}}
 \! {\rm d}q \, q \; \biggr\}
  \approx
  - \frac{\omega^4}{8(4\pi)^3}
  \bigl( 1 + 2 n_{\frac{\omega}{2}} \bigr)
 \;.
\ea
This yields the first row of \eq\nr{If_final}.
The 1-dimensional integrals have no delicate divergences here.
Setting $\lambda$ to zero, we obtain
\ba
 && \hspace*{-0.5cm} 
 \frac{(4\pi)^3\rho^{ }_{\mathcal{I}^{ }_\rmii{f}}(\omega)}
 {2 \omega^2 (1+2 n_{\frac{\omega}{2}})}  = 
 -\frac{\omega^2\mu^{-4\epsilon}}{16} 
 \nn 
  & -  & 
 \int_0^{ \frac{\omega}{4} } \! { {\rm d}q } \, n_{q} \;
   \Bigl( 3 q \Bigr) 
  -  
 \int_{ \frac{\omega}{4} }^{ \frac{\omega}{2} } 
 \! { {\rm d}q } \, n_{q} \; 
   \Bigl( q + \frac{\omega}{2} \Bigr) 
 + 
 \int_{ \frac{\omega}{2} }^{ \infty } 
 \! { {\rm d}q } \, n_{q} \; 
   \Bigl( \frac{q}{2} - {\omega} \Bigr) 
 \nn & - & 
 \int_0^{ \frac{\omega}{2} } \! { {\rm d}q } 
 \int_0^{ \frac{\omega}{4} - |q-\frac{\omega}{4}| } 
 \! { {\rm d}r }  \,
  \frac{  n_{\frac{\omega}{2} - q} \,
  n_{q+r} (1+n_{\fr{\omega}2-r})}{n_r^2}
 \nn & + & 
 \int_{ \frac{\omega}{2} }^{\infty}
 \! {{\rm d}q}
 \int_{0}^{ q - \frac{\omega}{2} }
 \! { {\rm d}r } \,
    \frac{  n_{q-\frac{\omega}{2}} 
    (1+n_{q-r})(n_q - n_{r+\frac{\omega}{2}})}
    {n_r n_{-\frac{\omega}{2}}}
 \nn & - & 
 \int_0^\infty \! {{\rm d}q}  
 \int_{0}^{ q }
 \! { {\rm d}r } \,
 \frac{ ( 1 + n_{q+\frac{\omega}{2}} )
  n_{q+r}n_{r+\frac{\omega}{2}}}{n_r^2}
 \; + \; \rmO(\lambda\ln\lambda)
 \;. \hspace*{5mm} \la{If_final} 
\ea
For $\omega \gg \pi T$ it can be checked numerically that this 
goes over to the known~\cite{ope} ultraviolet asymptotics
\be
 \rho^{\rmii{OPE}}_{\mathcal{I}^{ }_\rmii{f}}(\omega) =  
 -\frac{\omega^4\mu^{-4\epsilon}}{8(4\pi)^3} 
 -\frac{\omega^2 T^2}{64\pi} +
 \rmO\biggl( \frac{T^6}{\omega^2} \biggr)  \;. \la{rho_If_as}
\ee

%
\subsection{$\rho^{ }_{\mathcal{I}^{ }_\rmii{g}}$}

The sum-integral $\mathcal{I}^{ }_\rmi{g}$ is defined as 
\ba
 \mathcal{I}^{ }_\rmi{g} & \equiv &
 \Tint{Q,R} \frac{P^4}{Q^2(Q-P)^2R^2(R-P)^2}
 \;.
\ea
After the Matsubara sums the expression for the corresponding
spectral function reads
\ba
 & & \hspace*{-0.6cm} \rho^{ }_{\mathcal{I}^{ }_\rmii{g}}(\omega) = 
 \int_{\vec{q,r}} 
 \frac{\omega^4 \pi }{32 q^2 r^2} (1+2n_q) (1+2n_r)\biggl\{ 
 \nn 
 & & \;\;
 \Bigl[\delta(\omega - 2 r) - \delta(\omega+2 r) \Bigr]
 \biggl( 
 \frac{1}{q-r} +  
 \frac{1}{q+r} 
 \biggr)
 \nn & & 
 + \,
 \Bigl[\delta(\omega - 2 q) - \delta(\omega+2 q) \Bigr]
 \biggl( 
 \frac{1}{r-q} +  
 \frac{1}{r+q} 
 \biggr)
 \biggr\}
 \;. \la{Ig} 
\ea
The two terms can be combined through $q\leftrightarrow r$ if we 
first introduce a principle value prescription. 
Choosing the first one as a representative, the $\vec{r}$-integral 
can be carried out like in \eq\nr{Ja_int}, and we obtain
\ba
 & & \hspace*{-0.6cm} \rho^{ }_{\mathcal{I}^{ }_\rmii{g}}(\omega) = 
 \frac{\omega^4 \mu^{-2\epsilon}}{64\pi}
 \bigl( 1  + 2 n_{\frac{\omega}{2}}\bigr)
  \biggl[ 1 + \epsilon \biggl(
   \ln\frac{\bmu^2}{\omega^2} + 2 \biggr) \biggr] \times
 \nn && \; \times \, 
 \biggl\{
    \int_{\vec{q}} \frac{1}{q^2} 
    \,\mathbbm{P}\, \biggl( \frac{1}{q-\frac{\omega}{2}}
    +  \frac{1}{q+\frac{\omega}{2}} \biggr) 
    + \frac{1}{\pi^2}
   \int_0^\infty \! {\rm d}q \, n_q \,
     \mathbbm{P} \biggl( \frac{1}{q-\frac{\omega}{2}}
    +  \frac{1}{q+\frac{\omega}{2}} \biggr) 
 \biggr\}
 \;.
\ea
The thermal part (with $n_q$) 
is finite as it stands, whereas the vacuum part
may be transformed into a well-known four-dimensional integral:
\ba
    \int_{\vec{q}} \frac{1}{q^2} 
    \,\mathbbm{P}\, \biggl( \frac{1}{q-\frac{\omega}{2}}
    +  \frac{1}{q+\frac{\omega}{2}} \biggr) 
 & = & 
 8      \int_{\vec{q}} \frac{1}{4 q^2} 
 \,\mathbbm{P}\,
 \biggl(
 \frac{1}{\omega+ 2q} +  
 \frac{1}{-\omega+2q} 
 \biggr)
 \nn 
 & = & 8 \re \biggl\{ 
 \left. \int_Q \frac{1}{Q^2(Q-R)^2} 
 \right|_{R=( i {\omega} ,\vec{0} )}
 \biggr\}
 \nn 
 & = & 8 \re
 \biggl\{
   \frac{\mu^{-2\epsilon}}{(4\pi)^2} 
     \biggl[ 
        \frac{1}{\epsilon} + \ln \biggl( \frac{\bmu^2}{-\omega^2} \biggr) + 2
        + \rmO(\epsilon)
     \biggr]   
 \biggr\}
 \;. \la{vacint_g}
\ea
This can be verified by carrying out $\int \frac{{\rm d}q_0}{2\pi}$
on the right-hand side. 
In total, omitting $\rmO(\epsilon)$, 
\ba
 && \hspace*{-0.5cm} 
 \frac{(4\pi)^3\rho^{ }_{\mathcal{I}^{ }_\rmii{g}}(\omega)}
 {\omega^4 (1+2 n_{\frac{\omega}{2}})}  =  
 \frac{\mu^{-4\epsilon}}{2} \biggl(
   \frac{1}{\epsilon} + 2 \ln\frac{\bmu^2}{\omega^2} + 4
 \biggr)
 + 
 \int_0^{ \infty } \! { {\rm d}q } \, n_{q} \; 
   \mathbbm{P} \biggl( 
       \frac{1}{q-\frac{\omega}{2}}
    +  \frac{1}{q+\frac{\omega}{2}}
   \biggr)
   \;. \la{Ig_final}
\ea
For $\omega\gg \pi T$ this approaches the OPE-result~\cite{ope}
\be
 \rho^{\rmii{OPE}}_{\mathcal{I}^{ }_\rmii{g}}(\omega) =  
 \frac{\omega^4\mu^{-4\epsilon}}{2(4\pi)^3} 
 \biggl( 
   \frac{1}{\epsilon} + 2 \ln\frac{\bmu^2}{\omega^2} + 4
 \biggr) 
 -\frac{\omega^2 T^2}{48\pi} - \frac{\pi T^4}{30}
 +\rmO\biggl( \frac{T^6}{\omega^2} \biggr)  \;. \la{rho_Ig_as}
\ee

%
\subsection{$\rho^{ }_{\mathcal{I}^{ }_\rmii{h}}$}

The sum-integral $\mathcal{I}^{ }_\rmii{h}$ is defined as 
\ba
 \mathcal{I}^{ }_\rmi{h} & \equiv &
 \Tint{Q,R} \frac{P^4}{Q^2R^2[(Q-R)^2+\lambda^2](R-P)^2}
 \;.
\ea
After the Matsubara sums the expression for the corresponding
spectral function reads
\ba
 & & \hspace*{-0.6cm} \rho^{ }_{\mathcal{I}^{ }_\rmii{h}}(\omega) = 
 \int_{\vec{q,r}} 
 \frac{\omega^4 \pi }{8 q r E_{qr}} \biggl\{ 
 \nn 
 & & \!\!
 \frac{1}{2r} 
 \Bigl[\delta(\omega - 2 r) - \delta(\omega+2 r) \Bigr]
 \times 
 \nn & & \times \biggl[
 \biggl( 
 \frac{1}{q+r-E_{qr}} +  
 \frac{1}{q-r-E_{qr}} 
 \biggr)
 (1+2n_r)(n_{qr}-n_q)
 \nn & & \;\; 
 +
 \biggl(
 \frac{1}{q+r+E_{qr}} +  
 \frac{1}{q-r+E_{qr}} 
 \biggr)
 (1+2n_r)(1 + n_{qr}+n_q)
 \biggr]
 \nn & - & \!\!
 \Bigl[\delta(\omega - q - r -E_{qr}) - \delta(\omega+q+r+E_{qr}) \Bigr]
 \frac{(1+n_{qr})(1+n_q+n_r)+n_q n_r}
      {(q+r+E_{qr})(q-r+E_{qr})}
 \nn & - &  \!\!
 \Bigl[ \delta(\omega-q-r+E_{qr}) - \delta(\omega + q + r -E_{qr}) \Bigr]
 \frac{n_{qr}(1+n_q + n_r )-n_qn_r}
      {(q+r-E_{qr})(q-r-E_{qr})}
 \nn & - &  \!\!
 \Bigl[\delta(\omega - q + r -E_{qr}) - \delta(\omega+q-r+E_{qr}) \Bigr]
 \frac{n_r(1+n_q+n_{qr})-n_q n_{qr}}
      {(q-r+E_{qr})(q+r+E_{qr})}
 \nn & - &  \!\!
 \Bigl[\delta(\omega + q - r -E_{qr}) - \delta(\omega-q+r+E_{qr}) \Bigr]
 \frac{n_q(1+n_r+n_{qr})-n_r n_{qr}}
      {(q-r-E_{qr})(q+r-E_{qr})}
 \biggr\} \;. \hspace*{1cm} \la{Ih} 
\ea

The main difference
to the procedure described in appendix~\ref{se:app1} is the handling of 
the ``factorized powerlike'' integrals, which are ultraviolet
divergent, like in \eq\nr{vacint_g}. The expression reads 
\be
 \rho^{(\rmi{fz,p})}_{\mathcal{I}^{ }_\rmii{h}}(\omega) \equiv
 \int_{\vec{q,r}} 
 \frac{\omega^4 \pi}{16 q r^2 E_{qr}} 
 \delta(\omega - 2 r)
 \biggl(
 \frac{1}{q+r+E_{qr}} +  
 \frac{1}{-r+q+E_{qr}} 
 \biggr)
 (1+2n_r)
 \;.  \la{Ih_T} 
\ee
The $\vec{r}$-integral can be evaluated like in \eq\nr{Ja_int},  
whereas the $\vec{q}$-integral can be re-expressed as 
a four-dimensional vacuum integral: 
\be
 \int_{\vec{q}} 
 \frac{1}{4 q E_{qr}} 
 \biggl(
 \frac{1}{\frac{\omega}{2}+q+E_{qr}} +  
 \frac{1}{-\frac{\omega}{2}+q+E_{qr}} 
 \biggr)
 = 
 \left. \int_Q \frac{1}{Q^2[(Q-R)^2 + \lambda^2]} 
 \right|_{R=(\frac{\omega}{2} i ,\frac{\omega}{2} \vec{e}_r )}
 \;. \la{vacint_h}
\ee
This can be verified by carrying out $\int \frac{{\rm d}q_0}{2\pi}$
on the right-hand side. The value of the integral is familiar; in fact, 
since O($D$) rotational 
invariance implies that the result depends on $\lambda^2$ 
and $R^2$ only, and $R^2=0$ according to \eq\nr{vacint_h}, the simplest
way to derive it is just by setting $R=0$. In any case, 
\be
 \left. \int_Q \frac{1}{Q^2[(Q-R)^2 + \lambda^2]} 
 \right|_{R=(\frac{\omega}{2} i ,\frac{\omega}{2} \vec{e}_r )}
 = 
 \frac{\mu^{-2\epsilon}}{(4\pi)^2}
 \biggl(
   \frac{1}{\epsilon} + \ln \frac{\bmu^2}{\lambda^2} + 1 + \rmO(\epsilon)
 \biggr) 
 \;,
\ee
and in total (omitting $\rmO(\epsilon)$)
\be
  \rho^{(\rmi{fz,p})}_{\mathcal{I}^{ }_\rmii{h}}(\omega)
 = 
 \frac{\omega^4 \mu^{-4\epsilon}}{4(4\pi)^3}
 \bigl( 1 + 2 n_{\frac{\omega}{2}}\bigr)
 \biggl(
   \frac{1}{\epsilon} + \ln \frac{\bmu^2}{\omega^2}
   + \ln \frac{\bmu^2}{\lambda^2} + 3 
 \biggr) 
 \;. \la{Ih_fz_p}
\ee

The ``factorized exponential'' integrals can be worked out like
before; the expression corresponding to \eq\nr{Ij_fz_3} becomes
\ba
 && \hspace*{-1cm} \rho_{\mathcal{I}^{ }_\rmii{h}}^{(\rmi{fz,e})}(\omega) =  
  \frac{\omega^3}{2(4\pi)^3} (1 + 2 n_{\frac{\omega}{2}}) \biggl\{  
  \int_0^\infty \! {\rm d}q \, n_q \, 
      \ln\biggl| \frac{2q\omega + \lambda^2}{2q\omega - \lambda^2} \biggr|
  \nn & & \qquad + \, 
  \int_{\lambda}^\infty \! {\rm d}q \, n_q 
     \ln
         \biggl| 
            \frac{q + \frac{\lambda^2}{\omega}-\sqrt{q^2-\lambda^2}}
                 {q + \frac{\lambda^2}{\omega}+\sqrt{q^2-\lambda^2}}
         \biggr|
         \biggl| 
            \frac{q - \frac{\lambda^2}{\omega}+\sqrt{q^2-\lambda^2}}
                 {q - \frac{\lambda^2}{\omega}-\sqrt{q^2-\lambda^2}}
         \biggr|
  \;
  \biggr\}
  \;.  \la{Ih_fz_3}
\ea
For the phase space integrals the expression corresponding
to \eq\nr{Ij_ps_2} reads
\ba
 \rho_{\mathcal{I}^{ }_\rmii{h}}^{(\rmi{ps})}(\omega) & = & 
 \frac{\omega^3}{2(4\pi)^3}  \bigl(
  e^{\omega} - 1
 \bigr)
 \biggl\{ 
 \nn 
 & \mbox{(i)}  &  - \,
 \int_{\frac{\lambda^2}{2\omega}}^{\frac{\omega}{2}}
 \! {\rm d}q
 \int_{\frac{\lambda^2}{4q}}^{\frac{\omega(\omega-2q)+\lambda^2}{2\omega}} 
 \!\!\!\! {\rm d}r \; \mathbbm{P}
 \biggl( \frac{ 1 
 } {r} \biggr) 
 n_{\fr{\omega}2-q} \, 
 n_{\fr{\omega}2-r} \, 
 n_{q+r}
 \nn
 & \mbox{(ii)}  & + \, 
 \int_{ 0 }^{\infty}
 \! {\rm d}q
 \int_{\frac{\lambda^2}{4q}}^{ \infty }
 \! {\rm d}r \; \mathbbm{P}
 \biggl( \frac{ 1 
 } {r} \biggr) 
 n_{\fr{\omega}2+q} \, 
 n_{\fr{\omega}2+r} \, 
 n_{q+r}\, 
 e^{q+r}
 \nn
 & \mbox{(\iv)}  & + \, 
 \int_{ \frac{\omega}{2} }^{\infty}
 \! {\rm d}q
 \int_{ - \frac{\lambda^2}{4q}}^{
    \frac{\omega(2q-\omega)-\lambda^2}{2\omega}}
 \!\!\!\! {\rm d}r \; \mathbbm{P}
 \biggl( 
 \frac{ 1 
   } {r} - 
 \frac{ 1 
   } {q} \biggr)
 n_{q-\fr{\omega}2} \, 
 n_{\fr{\omega}2+r} \, 
 n_{q-r} \,
 e^{q - \frac{\omega}{2}}
 \biggr\} \;.  \la{Ih_ps_2}
\ea
Subsequently we can reflect the range $q < r$ to 
$q > r$ in cases (i) and (ii).
The subtractions of the potentially divergent parts 
are carried out like in \eqs\nr{ps_simp_1}, 
\nr{ps_simp_2}, \nr{ps_simp_3}; the integrals over $y$ 
and $r$ are a bit more complicated than with $\mathcal{I}^{ }_\rmii{j}$
but can be carried out. 

Once the pieces are collected together, we can again identify 
a ``vacuum part'', a ``1-dimensional integral'', and 
a ``2-dimensional integral''. The contribution to the vacuum 
part from the phase space integrals reads 
\ba
 & & \hspace*{-2cm}
 \frac{\omega^3}{2(4\pi)^3}
  \bigl( 1 + 2 n_{\frac{\omega}{2}} \bigr)
 \int_\lambda^{\frac{\omega}{2}}
 \! {\rm d}q \,
   \ln 
   \biggl| 
     \frac{q - \sqrt{q^2 - \lambda^2}}{q + \sqrt{q^2 - \lambda^2}}
   \biggr| 
  \approx
 \frac{\omega^4}{4(4\pi)^3}
  \bigl( 1 + 2 n_{\frac{\omega}{2}} \bigr)
  \biggl( \ln\frac{\lambda^2}{\omega^2} + 2 \biggr)
 \;.
\ea
Summing together with \eq\nr{Ih_fz_p}, 
we obtain the first row of \eq\nr{Ih_final}.
The most tedious work concerns the handling of the 
1-dimensional integrals; however in principle everything
proceeds like in \se\ref{ss:collect}, and eventually the 
$\lambda$'s cancel or can safely be put to zero. 
Altogether we obtain
\ba
 && \hspace*{-0.5cm} 
 \frac{2 (4\pi)^3\rho^{ }_{\mathcal{I}^{ }_\rmii{h}}(\omega)}
 {\omega^3 (1+2 n_{\frac{\omega}{2}})}  \; = \;  
 \frac{\omega\mu^{-4\epsilon}}{2} \biggl(
   \frac{1}{\epsilon} + 2 \ln\frac{\bmu^2}{\omega^2} + 5
 \biggr)
 \nn 
  & + & 
 \int_0^{ \frac{\omega}{4} } \! { {\rm d}q } \, n_{q} \; 
   \biggl\{
     - 2
       \ln\Bigl( 1 - \frac{2q}{\omega} \Bigr)
     + 2 
       \ln\Bigl( 1 + \frac{2q}{\omega} \Bigr)
 + \frac{\omega}{2}
   \biggl( \frac{1}{q+\frac{\omega}{2}} 
   + \frac{1}{q-\frac{\omega}{2}} \biggr)
   \biggr\}  
 \nn 
 & + & 
 \int_{ \frac{\omega}{4} }^{ \frac{\omega}{2} } 
 \! { {\rm d}q } \, n_{q} \; 
   \biggl\{
     - 3 
       \ln\Bigl( 1 - \frac{2q}{\omega} \Bigr)
     + 2 
       \ln\Bigl( 1 + \frac{2q}{\omega} \Bigr)
  + \ln \biggl( \frac{2q}{\omega} \biggr)
  + 
    \frac{\omega}{2} \frac{1}{q+\frac{\omega}{2}} 
   - 2 
   \biggr\}  
 \nn 
 & + & 
 \int_{ \frac{\omega}{2} }^{ \infty } 
 \! { {\rm d}q } \, n_{q} \; 
   \biggl\{
     - 4
       \ln\Bigl( \frac{2q}{\omega} - 1 \Bigr)
     + 2 
       \ln\Bigl( 1 + \frac{2q}{\omega} \Bigr)
 + 2 
 \ln \biggl( \frac{2q}{\omega} \biggr)
  + \frac{\omega}{2}
   \biggl( \frac{1}{q+\frac{\omega}{2}} - \frac{1}{q} \biggr)
   - 1 
   \biggr\}  
%
 \nn & + & 
 \int_0^{ \frac{\omega}{2} } \! { {\rm d}q } 
 \int_0^{ \frac{\omega}{4} - |q-\frac{\omega}{4}| } 
 \! { {\rm d}r }  \,
 \biggl( - \frac{1}{q}  
 - \frac{1}{r} \biggr)\;
  \frac{  n_{\frac{\omega}{2} - q} \,
  n_{q+r} (1+n_{\fr{\omega}2-r})}{n_r^2}
 \nn & + & 
 \int_{ \frac{\omega}{2} }^{\infty}
 \! {{\rm d}q}
 \int_{0}^{ q - \frac{\omega}{2} }
 \! { {\rm d}r } \,
 \biggl(\frac{1}{q} - 
 \frac{1}{r} \biggr) \;
    \frac{  n_{q-\frac{\omega}{2}} 
    (1+n_{q-r})(n_q - n_{r+\frac{\omega}{2}})}
    {n_r n_{-\frac{\omega}{2}}}
 \nn & + & 
 \int_0^\infty \! {{\rm d}q}  
 \int_{0}^{ q }
 \! { {\rm d}r } \,
 \biggl(\frac{1}{q} + 
 \frac{1}{r} \biggr) \;
 \frac{ ( 1 + n_{q+\frac{\omega}{2}} )
  n_{q+r}n_{r+\frac{\omega}{2}}}{n_r^2}
 \; + \; 
 \rmO(\lambda\ln\lambda)
 \;. \hspace*{5mm} \la{Ih_final} 
\ea
For $\omega \gg \pi T$ it can be checked numerically that this 
goes over to the known ultraviolet asymptotics~\cite{ope}
\be
 \rho^{\rmii{OPE}}_{\mathcal{I}^{ }_\rmii{h}}(\omega) =  
 \frac{\omega^4\mu^{-4\epsilon}}{4(4\pi)^3} 
 \biggl( 
   \frac{1}{\epsilon} + 2 \ln\frac{\bmu^2}{\omega^2} + 5
 \biggr) 
 +\frac{\omega^2 T^2}{192\pi} - \frac{\pi T^4}{360}
 +\rmO\biggl( \frac{T^6}{\omega^2} \biggr)  \;. \la{rho_Ih_as}
\ee

%
\subsection{$\rho^{ }_{\mathcal{I}^{ }_\rmii{i'}}$}

The sum-integral $\mathcal{I}^{ }_\rmii{i'}$ is defined as 
\ba
 \mathcal{I}^{ }_\rmi{i'} & \equiv &
 \Tint{Q,R} \frac{4(Q\cdot P)^2}{Q^2R^2[(Q-R)^2+\lambda^2](R-P)^2}
 \;.
\ea
At zero spatial momentum, 
$
 4(Q\cdot P)^2 = 4 q_n^2 p_n^2 = 4 (Q^2 - q^2) p_n^2
$, 
so that 
\be
 \mathcal{I}^{ }_\rmi{i'} =  
 \Tint{Q,R} \frac{4 p_n^2}{R^2[(Q-R)^2+\lambda^2](R-P)^2}
 -   
 \Tint{Q,R} \frac{4 p_n^2 q^2}{Q^2R^2[(Q-R)^2+\lambda^2](R-P)^2}
 \;.
\ee
The Matsubara sums here are equivalent to those 
in $\mathcal{I}^{ }_\rmi{b}$ and $\mathcal{I}^{ }_\rmi{h}$, and we obtain
\ba
 & & \hspace*{-0.6cm} \rho^{ }_{\mathcal{I}^{ }_\rmii{i'}}(\omega) = 
 \int_{\vec{q,r}} 
 \frac{\omega^2 q \pi }{2 r E_{qr}} \biggl\{ 
 \nn 
 & & \!\!
 \frac{1}{2r} 
 \Bigl[\delta(\omega - 2 r) - \delta(\omega+2 r) \Bigr]
 \times 
 \nn & & \times \biggl[
 \biggl( -\frac{2}{q} \biggr)
 (1+2n_r)(1 + 2 n_{qr})  
 \nn & & \;\; 
 +
 \biggl( 
 \frac{1}{q+r-E_{qr}} +  
 \frac{1}{q-r-E_{qr}} 
 \biggr)
 (1+2n_r)(n_{qr}-n_q)
 \nn & & \;\; 
 +
 \biggl(
 \frac{1}{q+r+E_{qr}} +  
 \frac{1}{q-r+E_{qr}} 
 \biggr)
 (1+2n_r)(1 + n_{qr}+n_q)
 \biggr]
 \nn & - & \!\!
 \Bigl[\delta(\omega - q - r -E_{qr}) - \delta(\omega+q+r+E_{qr}) \Bigr]
 \frac{(1+n_{qr})(1+n_q+n_r)+n_q n_r}
      {(q+r+E_{qr})(q-r+E_{qr})}
 \nn & - &  \!\!
 \Bigl[\delta(\omega-q-r+E_{qr}) - \delta(\omega + q + r -E_{qr}) \Bigr]
 \frac{n_{qr}(1+n_q + n_r )-n_qn_r}
      {(q+r-E_{qr})(q-r-E_{qr})}
 \nn & - &  \!\!
 \Bigl[\delta(\omega - q + r -E_{qr}) - \delta(\omega+q-r+E_{qr}) \Bigr]
 \frac{n_r(1+n_q+n_{qr})-n_q n_{qr}}
      {(q-r+E_{qr})(q+r+E_{qr})}
 \nn & - &  \!\!
 \Bigl[\delta(\omega + q - r -E_{qr}) - \delta(\omega-q+r+E_{qr}) \Bigr]
 \frac{n_q(1+n_r+n_{qr})-n_r n_{qr}}
      {(q-r-E_{qr})(q+r-E_{qr})}
 \biggr\} \;. \la{Iip} \hspace*{1cm} 
\ea

Like with $\rho^{ }_{\mathcal{I}^{ }_\rmii{h}}$ 
the ``factorized powerlike'' integrals are ultraviolet
divergent; the expression reads 
\be
 \rho^{(\rmi{fz,p})}_{\mathcal{I}^{ }_\rmii{i'}}(\omega) \equiv
 \int_{\vec{q,r}} 
 \frac{\omega^2 q \pi}{4 r^2 E_{qr}} 
 \delta(\omega - 2 r)
 \biggl(
 - \frac{2}{q} + 
 \frac{1}{q+r+E_{qr}} +  
 \frac{1}{-r+q+E_{qr}} 
 \biggr)
 (1+2n_r)
 \;.  \la{Iip_T} 
\ee
The $\vec{r}$-integral is the same as in \eq\nr{Ja_int}, 
whereas the non-trivial $\vec{q}$-integral can be re-expressed as 
a four-dimensional vacuum integral: 
\be
 \int_{\vec{q}} 
 \frac{q}{4 E_{qr}} 
 \biggl(
 \frac{1}{\frac{\omega}{2}+q+E_{qr}} +  
 \frac{1}{-\frac{\omega}{2}+q+E_{qr}} 
 \biggr)
 = 
 \left. \int_Q \frac{q^2}{Q^2[(Q-R)^2 + \lambda^2]} 
 \right|_{R=(\frac{\omega}{2} i ,\frac{\omega}{2} \vec{e}_r )}
 \;. \la{vacint}
\ee
This can be verified by carrying out $\int \frac{{\rm d}q_0}{2\pi}$
on the right-hand side of \eq\nr{vacint}. Once expressed this way, 
we can make use 
of O($D$) rotational covariance, in order to reduce the ``tensor'' structure
in the numerator to scalar integrals. The result is then
a function of $\lambda^2$ and $R^2$ only; given that $R^2=0$, we 
need the leading term in a Taylor-expansion around this point.  
A few steps lead to 
\be
 \left. \int_Q \frac{q^2}{Q^2[(Q-R)^2 + \lambda^2]} 
 \right|_{R=(\frac{\omega}{2} i ,\frac{\omega}{2} \vec{e}_r )}
 = 
 \frac{\omega^2 \mu^{-2\epsilon}}{12(4\pi)^2}
 \biggl(
   \frac{1}{\epsilon} + \ln \frac{\bmu^2}{\lambda^2} + \fr13 + \rmO(\epsilon)
 \biggr) 
 \;,
\ee
and in total 
\be
  \rho^{(\rmi{fz,p})}_{\mathcal{I}^{ }_\rmii{i'}}(\omega)
 = 
 \frac{\omega^4 \mu^{-4\epsilon}}{12(4\pi)^3}
 \bigl( 1 + 2 n_{\frac{\omega}{2}}\bigr)
 \biggl(
   \frac{1}{\epsilon} + \ln \frac{\bmu^2}{\omega^2}
   + \ln \frac{\bmu^2}{\lambda^2} + \fr73 + \rmO(\epsilon)
 \biggr) 
 \;. \la{Iip_fz_p}
\ee

The ``factorized exponential'' integrals can be worked out like
before; the expression corresponding to \eq\nr{Ij_fz_3} becomes
\ba
 && \hspace*{-1cm} \rho_{\mathcal{I}^{ }_\rmii{i'}}^{(\rmi{fz,e})}(\omega) =  
  \frac{2\omega}{(4\pi)^3} (1 + 2 n_{\frac{\omega}{2}}) \biggl\{  
  \int_0^\infty \! {\rm d}q \, n_q \, q^2 
      \ln\biggl| \frac{2q\omega + \lambda^2}{2q\omega - \lambda^2} \biggr|
  + 
  \int_{\lambda}^\infty \! {\rm d}q \, n_q  \times
  \nn & & \qquad \times \,   \biggl[  
     \Bigl(q + \frac{\omega}{2}\Bigr)^2
     \ln
         \biggl| 
            \frac{q + \frac{\lambda^2}{\omega}-\sqrt{q^2-\lambda^2}}
                 {q + \frac{\lambda^2}{\omega}+\sqrt{q^2-\lambda^2}}
         \biggr|
    + 
    \Bigl(q - \frac{\omega}{2}\Bigr)^2
       \ln
         \biggl| 
            \frac{q - \frac{\lambda^2}{\omega}+\sqrt{q^2-\lambda^2}}
                 {q - \frac{\lambda^2}{\omega}-\sqrt{q^2-\lambda^2}}
         \biggr|
  \; \biggr] \biggr\}
  \;. \nn \la{Iip_fz_3}
\ea
For the phase space integrals the expression corresponding
to \eq\nr{Ij_ps_2} reads
\ba
 \rho_{\mathcal{I}^{ }_\rmii{i'}}^{(\rmi{ps})}(\omega) & = & 
 \frac{2\omega}{(4\pi)^3}  \bigl(
  e^{\omega} - 1
 \bigr)
 \biggl\{ 
 \nn 
 & \mbox{(i)}  &  - \,
 \int_{\frac{\lambda^2}{2\omega}}^{\frac{\omega}{2}}
 \! {\rm d}q
 \int_{\frac{\lambda^2}{4q}}^{\frac{\omega(\omega-2q)+\lambda^2}{2\omega}} 
 \!\!\!\! {\rm d}r \; \mathbbm{P}
 \biggl[ \frac{ (q - \frac{\omega}{2})^2
 } {r} \biggr] 
 n_{\fr{\omega}2-q} \, 
 n_{\fr{\omega}2-r} \, 
 n_{q+r}
 \nn
 & \mbox{(ii)}  & + \, 
 \int_{ 0 }^{\infty}
 \! {\rm d}q
 \int_{\frac{\lambda^2}{4q}}^{ \infty }
 \! {\rm d}r \; \mathbbm{P}
 \biggl[ \frac{ (q + \frac{\omega}{2} )^2
 } {r} \biggr] 
 n_{\fr{\omega}2+q} \, 
 n_{\fr{\omega}2+r} \, 
 n_{q+r}\,
 e^{q+r}
 \nn
 & \mbox{(\iv)}  & + \, 
 \int_{ \frac{\omega}{2} }^{\infty}
 \! {\rm d}q
 \int_{ - \frac{\lambda^2}{4q}}^{
    \frac{\omega(2q-\omega)-\lambda^2}{2\omega}}
 \!\!\!\! {\rm d}r \; \mathbbm{P}
 \biggl[ 
 \frac{ (q-\frac{\omega}{2})^2  } {r} - 
 \frac{ (r+\frac{\omega}{2})^2  } {q} \biggr]
 n_{q-\fr{\omega}2} \, 
 n_{\fr{\omega}2+r} \, 
 n_{q-r}\,
 e^{q - \frac{\omega}{2}}
 \biggr\} \;. \nn \la{Iip_ps_2}
\ea
Subsequently we can reflect the range $q < r$ to 
$q > r$ in cases (i) and (ii).
The subtractions of the potentially divergent parts 
are carried out like in \eqs\nr{ps_simp_1}, 
\nr{ps_simp_2}, \nr{ps_simp_3}; the integrals over $y$ 
and $r$ are a bit more complicated than with $\mathcal{I}^{ }_\rmii{j}$
but can be carried out. 

Once the pieces are collected together, we can again identify 
a ``vacuum part'', a ``1-dimensional integral'', and 
a ``2-dimensional integral''. The contribution to the vacuum 
part from the phase space integrals reads 
\ba
 & & \hspace*{-2cm}
 \frac{2\omega}{(4\pi)^3}
  \bigl( 1 + 2 n_{\frac{\omega}{2}} \bigr)
 \int_\lambda^{\frac{\omega}{2}}
 \! {\rm d}q \,
 \biggl[
   \Bigl( q -\frac{\omega}{2} \Bigr)^2 
   \ln 
   \biggl| 
     \frac{q - \sqrt{q^2 - \lambda^2}}{q + \sqrt{q^2 - \lambda^2}}
   \biggr| 
   - \omega q + \fr32 q^2
 \biggr]  \nn 
  & \approx &  
 \frac{\omega^4}{12(4\pi)^3}
  \bigl( 1 + 2 n_{\frac{\omega}{2}} \bigr)
  \biggl( \ln\frac{\lambda^2}{\omega^2} + \frac{13}{6} \biggr)
 \;.
\ea
Summing together with \eq\nr{Iip_fz_p}, 
we obtain the first row of \eq\nr{Iip_final}.
The most tedious work concerns the handling of the 
1-dimensional integrals; however in principle everything
proceeds like in \se\ref{ss:collect}, and eventually the 
$\lambda$'s cancel or can safely be put to zero. 
Altogether we obtain
\ba
 && \hspace*{-0.5cm} 
 \frac{(4\pi)^3\rho^{ }_{\mathcal{I}^{ }_\rmii{i'}}(\omega)}
 {2 \omega (1+2 n_{\frac{\omega}{2}})}  \; = \; 
 \frac{\omega^3\mu^{-4\epsilon}}{24} \biggl(
   \frac{1}{\epsilon} + 2 \ln\frac{\bmu^2}{\omega^2} + \fr92
 \biggr)
 \nn 
  & + & 
 \int_0^{ \frac{\omega}{4} } \! { {\rm d}q } \, n_{q} \; 
   \biggl\{
     - \Bigl[ q^2 + ( q-\frac{\omega}{2})^2  \Bigr]
       \ln\Bigl( 1 - \frac{2q}{\omega} \Bigr)
     + \Bigl[ q^2 + ( q+\frac{\omega}{2})^2  \Bigr]
       \ln\Bigl( 1 + \frac{2q}{\omega} \Bigr)
 \nn & & \hspace*{2cm} + \, \frac{\omega^3}{24}
   \biggl( \frac{1}{q+\frac{\omega}{2}} 
   + \frac{1}{q-\frac{\omega}{2}} \biggr) - \frac{8\omega q}{3}
   \biggr\}  
 \nn 
 & + & 
 \int_{ \frac{\omega}{4} }^{ \frac{\omega}{2} } 
 \! { {\rm d}q } \, n_{q} \; 
   \biggl\{
     - \Bigl[ 2 q^2 + ( q-\frac{\omega}{2})^2  \Bigr]
       \ln\Bigl( 1 - \frac{2q}{\omega} \Bigr)
     + \Bigl[ q^2 + ( q+\frac{\omega}{2})^2  \Bigr]
       \ln\Bigl( 1 + \frac{2q}{\omega} \Bigr)
 \nn & & \hspace*{2cm} + \, 
  q^2 \ln \biggl( \frac{2q}{\omega} \biggr)
  + \frac{\omega^3}{24}
   \biggl( \frac{1}{q+\frac{\omega}{2}} \biggr)
   - \frac{\omega^2}{6} - \frac{5\omega q}{2} - \frac{2 q^2}{3}
   \biggr\}  
 \nn 
 & + & 
 \int_{ \frac{\omega}{2} }^{ \infty } 
 \! { {\rm d}q } \, n_{q} \; 
   \biggl\{
     - 2 \Bigl[ q^2 + ( q-\frac{\omega}{2})^2  \Bigr]
       \ln\Bigl( \frac{2q}{\omega} - 1 \Bigr)
     + \Bigl[ q^2 + ( q+\frac{\omega}{2})^2  \Bigr]
       \ln\Bigl( 1 + \frac{2q}{\omega} \Bigr)
 \nn & & \hspace*{2cm} + \, 
  \Bigl[ q^2 + ( q-\frac{\omega}{2})^2 \Bigr] 
 \ln \biggl( \frac{2q}{\omega} \biggr)
  + \frac{\omega^3}{24}
   \biggl( \frac{1}{q+\frac{\omega}{2}} - \frac{1}{q} \biggr)
   - \frac{\omega^2}{6} - \frac{3\omega q}{2} - \frac{11 q^2}{6}
   \biggr\}  
%
 \nn & + & 
 \int_0^{ \frac{\omega}{2} } 
 \! { {\rm d}q } 
 \int_0^{ \frac{\omega}{4} - |q-\frac{\omega}{4}| } 
 \! { {\rm d}r }  \,
 \biggl[ - \frac{(r-\frac{\omega}{2})^2}{q}  
 - \frac{(q-\frac{\omega}{2})^2}{r} \biggr] \;
  \frac{  n_{\frac{\omega}{2} - q} \,
  n_{q+r} (1+n_{\fr{\omega}2-r})}{n_r^2}
 \nn & + & 
 \int_{ \frac{\omega}{2} }^{\infty}
 \! {{\rm d}q}
 \int_{0}^{ q - \frac{\omega}{2} }
 \! { {\rm d}r } \,
 \biggl[\frac{(r+\frac{\omega}{2})^2}{q} - 
 \frac{(q-\frac{\omega}{2})^2}{r} \biggr] \;
    \frac{  n_{q-\frac{\omega}{2}} 
    (1+n_{q-r})(n_q - n_{r+\frac{\omega}{2}})}
    {n_r n_{-\frac{\omega}{2}}}
 \nn & + & 
 \int_0^\infty \! {{\rm d}q}  
 \int_{0}^{ q }
 \! { {\rm d}r } \,
 \biggl[\frac{(r+\frac{\omega}{2})^2}{q} + 
 \frac{(q+\frac{\omega}{2})^2}{r} \biggr] \;
 \frac{ ( 1 + n_{q+\frac{\omega}{2}} )
  n_{q+r}n_{r+\frac{\omega}{2}}}{n_r^2}
 \; + \; 
 \rmO(\lambda\ln\lambda)
 \;. \hspace*{5mm} \la{Iip_final} 
\ea
Numerically it can checked that for $\omega \gg \pi T$ this 
approaches the known asymptotics~\cite{ope} 
\be
 \rho^{\rmii{OPE}}_{\mathcal{I}^{ }_\rmii{i'}}(\omega) =  
 \frac{\omega^4\mu^{-4\epsilon}}{12(4\pi)^3} 
 \biggl( 
   \frac{1}{\epsilon} + 2 \ln\frac{\bmu^2}{\omega^2} + \fr92
 \biggr) 
 -\frac{\omega^2 T^2}{96\pi} + \frac{\pi T^4}{120}
 +\rmO\biggl( \frac{T^6}{\omega^2} \biggr)  \;. \la{rho_Iip_as}
\ee

\newpage

%
\section{Hard Thermal Loop resummation}
\la{se:app3}

We give here some details for the Hard Thermal Loop (HTL) resummation 
discussed in \se\ref{ss:IR}, which is important when $\omega\sim gT$
and could allow us to postpone the breakdown of the perturbative series 
down to lower frequencies ($\omega \lsim g^2 T / \pi$). 

In general, HTL resummation is needed both for propagators and 
vertices; for instance, the coupling of a photon to quarks and gluons
gets modified by a correction $\propto g^2T^2$ and needs to be taken into 
account at leading order if the photons
are soft~\cite{bp_dilepton}. On the other hand, the coupling
of a heavy quark to gluons does {\em not} get modified even at NLO 
within HTL perturbation theory~\cite{hm}. For us relevant is the 
coupling of a dilaton or an axion to gluons. For ``hard'' axions
it can trivially be argued 
that no vertex modification is needed~\cite{by} but
here we are concerned with ``soft'' dilatons or axions ($\omega \sim gT$, 
$\vec{p} = \vec{0}$). It is nevertheless our belief that 
no vertex modification is needed, neither for dilatons nor
for axions. Concrete indications in this direction 
are that even without a vertex correction we find results 
which exactly cancel the leading infrared divergences from the QCD 
results in both channels (cf.\ \se\ref{ss:IR}), 
and that the results satisfy the expected
sum rules (cf.\ last paragraph of \se\ref{ss:sum}). 
In addition, we have checked the absence of vertex
corrections through explicit 1-loop computations.\footnote{%
 According to ref.~\cite{brpi}, 
 considering a three-leg process in Feynman gauge 
 Hard Thermal Loops could arise from structures of the types 
 $
  \Tinti{K} \frac{K_\mu K_\nu K_\rho}{K^2 (K-P)^2 (K-Q)^2}
 $,  
 $
  \Tinti{K} \frac{K_\mu K_\nu}{K^2 (K-P)^2}
 $,  as well as
 $
  \Tinti{K} \frac{1}{K^2}. 
 $
 In the dilaton case individual graphs do give contributions 
 of the latter two types but these cancel in the sum; in the axion
 case we find no such contributions.  
 }
We assume that simple theoretical arguments could 
also be given for the absence of vertex corrections
but at the time of writing these evade us. 
No further details are provided here because, 
strictly speaking, it would not even be necessary for us to treat 
the infrared regime correctly. 

%
\subsection{Contractions}

With the logic outlined above, the only ingredient needed from 
the HTL theory is the gauge field propagator, which takes the form 
\be
 \bigl\langle
  A^a_\mu(X) \, A^b_\nu(Y)
 \bigr\rangle 
 = 
 \delta^{ab}_{ }\, \Tint{Q}
 e^{i Q\cdot (X - Y)}
 \biggl[
   \frac{\mathbbm{P}^T_{\mu\nu}(Q)}{Q^2 + \Pi^{ }_T (Q)} 
  + \frac{\mathbbm{P}^E_{\mu\nu}(Q)}{Q^2 + \Pi^{ }_E (Q)}
  + \frac{\xi\, Q_\mu Q_\nu}{Q^4} 
 \biggr] 
 \;, \la{prop_HTL}
\ee
where $\xi$ is a gauge parameter. The two independent projection
operators are defined by 
\be
 \mathbbm{P}^T_{\mu\nu}(Q) \equiv \delta_{\mu i} \delta_{\nu j}
 \biggl( \delta_{ij} - \frac{q_i q_j}{q^2} \biggr)
 \;, \quad
 \mathbbm{P}^E_{\mu\nu}(Q) \equiv 
 \delta_{\mu\nu} - \frac{Q_\mu Q_\nu }{Q^2} - 
 \mathbbm{P}^T_{\mu\nu}(Q)
 \;, 
\ee
whereas the self-energies read
\be
  \Pi_T^{ }(Q) = 
  \frac{\mE^2}{D - 2}
  \int_{z} 
  \biggl( 1 - \frac{Q^2}{q^2} 
  \frac{q z}{iq_n + q z}
  \biggr) 
  \;, \quad
  \Pi_E^{ }(Q) =  
  \frac{\mE^2 Q^2}{q^2}
  \int_{z} 
  \frac{q z}{iq_n + q z}
  \;, \la{PiT} 
\ee
with the Debye mass parameter given by
\be
 \mE^2 \equiv g^2 \Nc (D-2)^2 \int_{r} \frac{\nB{}(r)}{r} 
 \;. \la{mmE}
\ee
For $D=4$ this reproduces the expression given in \eq\nr{mE}.\footnote{%
 We have employed a notation valid in an arbitrary 
 dimension, with the integration measures defined as 
 ($d\equiv D-1 = 3-2\epsilon$)
 \be
  \int_r \equiv 
  \frac{2}{(4\pi)^{\frac{d}{2}} \Gamma(\frac{d}{2})}
  \int_0^{\infty} \! {\rm d}r \, r^{d-1} 
  \;, \quad
  \int_z \equiv
  \frac{\Gamma(\frac{d}{2})}{\Gamma(\frac{1}{2})\Gamma(\frac{d-1}{2})} 
  \int_{-1}^{+1} \! {\rm d}z \, (1-z^2)^{\frac{d-3}{2}}
  \;,
 \ee
 such that $\int_\vec{r} = \int_r \int_z$ and $\int_z 1 = 1$.
}

In the main body of the text, a {\em next-to-leading order} 
computation was described.
For $\omega\sim gT$, however, the series representation 
breaks down. Through the HTL
method we then resum infinite orders to a single contribution, 
which is of the {\em leading order} within the HTL power counting. 
At this order, all the integrals met are ultraviolet finite, 
so that we could set $D=4$;
for generality, however, we mostly show $D$ in its original form. 

When the propagator takes the form in \eq\nr{prop_HTL}, 
the contractions leading to \eqs\nr{Gtheta_bare}, \nr{Gchi_bare} 
need to be carried out anew. Because of the non-trivial structure
of \eq\nr{prop_HTL}, this is somewhat tedious even at leading order. 
The problem gets simplified if we set $\vec{p} \to \vec{0}$ from
the outset; then 
\be
 P^{ }_\mu \mathbbm{P}^T_{\mu\nu}(Q) = 
 Q^{ }_\mu \mathbbm{P}^T_{\mu\nu}(Q) = 
 0 \;.
\ee
Also, a combination appearing frequently, particularly 
in the $\chi$-channel, reduces to 
$
 Q^2 P^2 - (Q\cdot P)^2 = q^2 p_n^2
$.
Thereby we obtain (making also use of 
$
 [Q\cdot(Q-P)]^2 = Q^2(Q-P)^2 - 
 [Q^2 P^2 - (Q\cdot P)^2]
$ in order to identify a common part in both channels)
\ba
 \frac{\tilde G_\theta^\rmii{HTL}(P)}{4 d_A c_\theta^2 } & = & 
 - 2 (D-2) g^4 p_n^2 
 \Tint{Q} \frac{q^2}{\Delta^{ }_T(Q) 
 \Delta^{ }_T(Q-P)}
 \nn & & \; + \, 
 2 g^4  \Tint{Q}
 \biggl\{ 
  \frac{Q^2(Q-P)^2}{\Delta^{ }_E(Q) 
 \Delta^{ }_E(Q-P)}
 + 
   \frac{(D-2)Q^2(Q-P)^2}{\Delta^{ }_T(Q) 
 \Delta^{ }_T(Q-P)}
 \biggr\} 
 \;, \la{Gtheta_HTL} \\
 \frac{\tilde G_\chi^\rmii{HTL}(P)}{-16 d_A c_\chi^2 (D-3)} & = & 
 - 2 (D-2) g^4 p_n^2 
 \Tint{Q} \frac{q^2}{\Delta^{ }_T(Q) 
 \Delta^{ }_T(Q-P)}
 \;, \la{Gchi_HTL}
\ea
where we have defined 
$
 \Delta^{ }_T(Q) \equiv Q^2 + \Pi^{ }_T(Q)
$
and
$
 \Delta^{ }_E(Q) \equiv Q^2 + \Pi^{ }_E(Q)
$.

%
\subsection{Naive HTL computation}
\la{ss:htl_naive}

As discussed around \eq\nr{master_resum}, we need both a ``naive'' 
and a ``resummed'' version of the HTL result.  
We start with the naive computation; 
this means that $\mE^2$ is treated as a small quantity, of $\rmO(g^2)$, 
and the HTL-result is ``re-expanded'' in $g^2$ up to NLO.

Now, since the HTL theory (correctly) modifies physics only at soft 
scales, one should be careful not to apply HTL resummation
to parts of the phase space where it is not needed.  
In real-time computations, it is notoriously difficult 
to identify the relevant parts in advance.
Therefore, in the following we carry out the HTL resummation
on all scales, whereby the HTL results show dependence on {\em two}
scales, both the ``soft'' $\mE$ as well as the ``hard'' $\pi T$.
The unwanted dependence on $\pi T$ is easily dropped 
{\em a posteriori}, cf.\ text below \eq\nr{nlo_htl}. 

With this philosophy, we start with the pseudoscalar ($\chi$) channel, 
which is simpler (cf.\ \eq\nr{Gchi_HTL}). It will be checked
afterwards that the additional terms 
in \eq\nr{Gtheta_HTL} give no contribution
in the ``naive'' limit (apart from ``contact 
terms'' of the type discussed in \se\ref{ss:sum}).

First of all, at leading order, $\Delta^{ }_T(Q) = Q^2$. Then the structure
in \eq\nr{Gchi_HTL} is almost the same as the master sum-integral
$ \mathcal{J}^{ }_\rmi{b} $ of \eq\nr{Jb}, and we get
\ba
 \im \biggl\{
 \Tint{Q} \frac{-p_n^2 q^2}{Q^2 (Q-P)^2}
 \biggr\}_{P \to (-i [\omega + i 0^+],\vec{0})}
 & = &  
 \int_{\vec{q}} \frac{\omega^2\pi}{4 }
 \Bigl[ \delta(\omega - 2 q) - \delta(\omega + 2 q) \Bigr]
 (1+2 n_q)
 \nn 
 & = &  
 \frac{\omega^4}{64\pi}
 \bigl( 1 + 2 n_{\frac{\omega}{2}} \bigr)
 + \rmO(\epsilon)
 \;. 
\ea
Adding the factor $2(D-2) g^4$ and setting $D\to 4$ we get 
\be
 \frac{-\rho_\chi^\rmii{HTL}(\omega)}{16 d_A c_\chi^2}
 = \frac{\pi g^4\omega^4}{(4\pi)^2} \bigl( 1 + 2 n_{\frac{\omega}{2} } \bigr)
 + \rmO(g^6) \;, \la{lo_htl}
\ee
which indeed agrees with the 
leading-order term in \eq\nr{rhofinal_chi}.

At NLO, we can expand $\Delta^{ }_T$ to first order in $\Pi^{ }_T$.
Substituting subsequently $Q\to P-Q$ yields
\be
 \left. \frac{\tilde G_\chi^\rmii{HTL}(P)}{-16 d_A c_\chi^2 (D-3)}
 \right|_{\rmO(g^6)}  =  
 4 (D-2) g^4 p_n^2 
 \Tint{Q} \frac{q^2 \Pi^{ }_T(Q)}{Q^4(Q-P)^2}
 \;,
\ee
where $\Pi^{ }_T$ is to be inserted from \eq\nr{PiT}. There are
two structures; for the simpler one we proceed in analogy with 
\eqs\nr{dl_Id}, \nr{Id_2}: 
\ba
  \Tint{Q} \frac{q^2}{Q^4(Q-P)^2}
  & \to & 
  - \fr12 \frac{{\rm d}}{{\rm d}\lambda^2}
 \Tint{Q}  
  \frac{q^2}{[Q^2+\lambda^2][(Q-P)^2+\lambda^2]} 
 \nn & \to & 
  - \fr12 \frac{{\rm d}}{{\rm d}\lambda^2}
  \int_{\vec{q}} \frac{\pi q^2}{4 E_q^2}
 \delta(\omega - 2 E_q) 
 \bigl( 1+2 n^{ }_{E_q} \bigr)
 \nn & = & 
  - \fr12 \frac{{\rm d}}{{\rm d}\lambda^2}
 \biggl\{  
   \frac{(\omega^2-4\lambda^2)^{\fr32}}{64\pi\omega}
  \bigl( 1 + 2 n_{\frac{\omega}{2}}\bigr)
  \,\theta(\omega - 2\lambda) + \rmO(\epsilon)
 \biggr\}
 \nn & \to & 
 \frac{3\pi}{4(4\pi)^2}
  \bigl( 1 + 2 n_{\frac{\omega}{2}}\bigr)
 \;. \la{htl_nlo_1}
\ea
The second term is a bit more complicated, and it is 
useful to introduce an infrared regulator: 
\ba
 & & \hspace*{-2cm}
 - \Tint{Q}\int_z \frac{1}{Q^2(Q-P)^2}   \frac{q z}{iq_n + q z}
 \; \to \; 
 - \int_z \Tint{Q} \frac{1}{(q_n - p_n)^2 + q^2} 
   \frac{1}{q_n^2 + E_q^2} \frac{q z}{iq_n + q z}
 \nn & = & 
 - \int_{z,q} T\sum_{q_n} T\sum_{r_n} 
   \frac{\beta \delta_{r_n + p_n - q_n}}{(r_n^2+q^2)(q_n^2+E_q^2)}
   \fr12\biggl(
     \frac{q z}{iq_n + q z} + \frac{q z}{- iq_n + q z}
   \biggr) 
 \nn & = & - \int_{z,q}
 \int_0^\beta \! {\rm d}\tau \,
 \frac{q^2z^2 e^{i p_n\tau}}{E_q^2 - q^2z^2}
 \, T\sum_{r_n} \frac{e^{i r_n\tau}}{r_n^2 + q^2}
 \, T\sum_{q_n} 
 \biggl(
  \frac{e^{-i q_n\tau}}{q_n^2 + q^2z^2} - 
  \frac{e^{-i q_n\tau}}{q_n^2 + E_q^2} 
 \biggr)
 \;, \la{mat_sum}
\ea
where $E_q^2 \equiv q^2 + \lambda^2$; we symmetrized in 
$z\leftrightarrow -z$; represented the Kronecker-$\delta$
as an integral; and partial-fractioned the dependence on $q_n^2$. 
The sums can be carried out, e.g.\
\be
 T\sum_{r_n} \frac{e^{i r_n\tau}}{r_n^2 + q^2}
 = 
 \frac{n_q}{2q}\Bigl[ e^{q\tau} + e^{q(\beta-\tau)} \Bigr]
 \;, 
\ee
and subsequently also the integral over $\tau$, yielding
\ba
 I(p_n) & \equiv & - \int_{z,q}
 \frac{q^2 z^2}{\lambda^2 + q^2 (1-z^2)}
 \frac{n_q}{4q} \times 
 \nn & \times & 
 \biggl\{
   \frac{n_{qz}}{qz} \biggl[ 
       \Bigl( e^{q(1+z)\beta} - 1 \Bigr)
       \biggl( \frac{1}{-i p_n + q(1+z)} +  \frac{1}{i p_n + q(1+z)} \biggr) 
 \nn & & \hspace*{1cm} + \, 
       \Bigl( e^{q\beta} - e^{qz\beta} \Bigr)
       \biggl( \frac{1}{-i p_n + q(1-z)} +  \frac{1}{i p_n + q(1-z)} \biggr) 
    \biggr]
  \nn & & \hspace*{-3mm} - \,
   \frac{n^{ }_{E_q}}{E_q} \biggl[ 
       \Bigl( e^{(q+E_q)\beta} - 1 \Bigr)
       \biggl( \frac{1}{-i p_n + q + E_q} +  \frac{1}{i p_n + q + E_q} \biggr) 
 \nn & & \hspace*{1cm} + \, 
       \Bigl( e^{q\beta} - e^{E_q\beta} \Bigr)
       \biggl( \frac{1}{-i p_n + q - E_q} +  \frac{1}{i p_n + q - E_q} \biggr) 
    \biggr]
 \biggr\}  
 \;. \la{Ipn_n}
\ea
We can now set $p_n \to -i [\omega + i 0^+]$ and
take the imaginary part; for $\omega > \lambda > 0$, 
only three channels contribute. If we also substitute $z\to -z$
in one of them, and make use of the $\delta$-constraints in order to 
re-express the arguments of the Bose distributions, the expression
reduces to 
\ba
 \im I(-i[\,\omega + i 0^+])
 & = & 
 -\frac{\pi}{4}
 \int_{z,q}
 \frac{z^2}{\lambda^2 + q^2 (1-z^2)}
 \bigl( 1 + n_q + n_{\omega - q}\bigr) 
 \nn & & \quad \times \, 
 \Bigl[
   \;\frac{2}{z}\, \delta(\omega - q - qz) - 
   \frac{q}{E_q}\, \delta(\omega - q - E_q) 
 \Bigr]
 \;.
\ea
The integrals over $z$ can be carried out:
\ba
 \fr12 \int_{-1}^{+1} \! {\rm d}z \, 
 \frac{2z\, \delta(\omega - q - qz)}{\lambda^2 + q^2 (1-z^2)} \, 
  & = & 
 \frac{(\omega - q)\theta(q - \frac{\omega}{2})}
 {q^2(\lambda^2 + 2 \omega q - \omega^2)}
 \;, \\ 
 \fr12 \int_{-1}^{+1} \! {\rm d}z \, 
 \frac{z^2}{\lambda^2 + q^2 (1-z^2)}  
 & = & 
 - \frac{1}{q^2}
 \biggl(
  1 + \frac{E_q}{2q}\ln\frac{E_q - q}{E_q + q} 
 \biggr)
 \;.
\ea
The $q$-integral over $\delta(\omega - q - E_q)$ yields 
\be
 \int_0^\infty \! {\rm d}q \, \phi(q,E_q) \, \delta(\omega - q - E_q)
 = \left. \frac{E_q}{q + E_q} \; \phi(q,E_q) 
 \right|_{q = \frac{\omega^2 - \lambda^2}{2\omega}, \;
 E_q = \frac{\omega^2 + \lambda^2}{2\omega} }
 \;,
\ee
whereas the other $q$-integral is split into the ranges
$(\frac{\omega}{2},\omega)$ and $(\omega,\infty)$. In the 
former range, we add and subtract $1 + 2n_{\frac{\omega}{2}}$ from
the Bose distributions, whereas in the latter range, we write
$1+n_{\omega - q} = -n_{q - \omega}$ in order to have a positive
argument. In total, setting $\lambda\to 0$ whenever possible, 
\ba
 \im I(-i[\omega + i 0^+])
 & \approx & 
 -\frac{1}{8\pi}
 \biggl\{ 
  \int_{\frac{\omega}{2}}^{\omega} 
  \! {\rm d}q \, \frac{\omega - q}{2\omega q - \omega^2}
  \bigl( n_q + n_{\omega - q} - 2 n_{\frac{\omega}{2}} \bigr)
 \\ & & \qquad + \, 
  \int_{\omega}^{\infty} 
  \! {\rm d}q \, \frac{\omega - q}{2\omega q - \omega^2}
  \bigl( n_q - n_{q - \omega}  \bigr)
 \nn & & \qquad + \, 
  \int_{\frac{\omega}{2}}^{\omega} 
  \! {\rm d}q \, \frac{\omega - q}{\lambda^2 + 2\omega q - \omega^2}
  \bigl( 1 +  2 n_{\frac{\omega}{2}} \bigr)
 \nn & & \qquad + \, 
  \left. \frac{E_q}{q + E_q}
  \biggl( \frac{q}{E_q} + \fr12 \ln\frac{E_q - q}{E_q + q}  \biggr)
  \bigl( 1+ n_q + n^{ }_{E_q} \bigr)
  \right|_{q = \frac{\omega^2 - \lambda^2}{2\omega}}
 \biggr\} 
 \;.   \nonumber
\ea 
The last two terms contain a logarithmic divergence, but their
sum is finite: 
\ba
  \int_{\frac{\omega}{2}}^{\omega} 
  \! {\rm d}q \, \frac{\omega - q}{\lambda^2 + 2\omega q - \omega^2}
  & = &  
  \fr14\ln\frac{\omega^2}{\lambda^2} - \fr14
 + \rmO(\lambda\ln\lambda) 
  \;, \\ 
  \left. \frac{E_q}{q + E_q}
  \biggl( \frac{q}{E_q} + \fr12 \ln\frac{E_q - q}{E_q + q}  \biggr)
  \right|_{q = \frac{\omega^2 - \lambda^2}{2\omega}}
  & = &  
   \fr14 \ln\frac{\lambda^2}{\omega^2} + \fr12
  + \rmO(\lambda\ln\lambda) 
  \;. 
\ea
If we also write 
\ba
  n_q + n_{\omega - q} - 2 n_{\frac{\omega}{2}} 
  & = & 
  \bigl(1 + 2 n_{\frac{\omega}{2}} \bigr)
  \frac{n_q (1 + n_{\omega - q})}{n^2_{q-\frac{\omega}{2}} }
  \;, \\  
  n_q - n_{q- \omega}
  & = & 
  - \bigl(1 + 2 n_{\frac{\omega}{2}} \bigr)
  \frac{n_q (1 + n_{q-\omega})}{n^2_{\frac{\omega}{2}} }
  \;,
\ea
and combine then with the contribution in \eq\nr{htl_nlo_1}
as well as the leading-order term from \eq\nr{lo_htl}, 
the full result becomes 
\ba
 \frac{-\rho_\chi^\rmii{HTL}(\omega)}{16 d_A c_\chi^2}
 & = & \frac{\pi g^4}{(4\pi)^2} \bigl( 1 + 2 n_{\frac{\omega}{2} } \bigr)
 \biggl\{ \omega^4 - 4 \omega^2 \mE^2
 \biggl[
  \fr14
  - \, 2
  \int_{\frac{\omega}{2}}^{\omega} 
  \! {\rm d}q \, \frac{\omega - q}{2\omega q - \omega^2}
  \frac{n_q (1 + n_{\omega - q})}{n^2_{q-\frac{\omega}{2}} }
 \nn & & \hspace*{3cm} 
  -  \, 2 
  \int_{\omega}^{\infty} 
  \! {\rm d}q \, \frac{q - \omega}{2\omega q - \omega^2}
  \frac{n_q (1 + n_{q-\omega})}{n^2_{\frac{\omega}{2}} }   
 \biggr]
 \biggr\} 
 + \rmO(g^8) 
 \;. \hspace*{1cm} \la{nlo_htl}
\ea
In a systematic HTL-computation, we need
furthermore to replace the distribution functions 
through their ``classical'' limits, 
$(1+)\, n^{ }_{q} \to T/ q$ etc.; then the integrals
are trivially doable and we get \eq\nr{nlo_htl_compact}.
(For notational simplicity 
we however always display the prefactor
$1 + 2 n_{\frac{\omega}{2} }$ in its
``quantum'' form.)

We end by showing that additional terms in $\tilde G^\rmii{HTL}_\theta$, 
on the second row of \eq\nr{Gtheta_HTL}, give no $\omega$-dependent 
contribution in the ``naive'' limit. At leading order this is clear: 
both $\Delta^{ }_E(Q)$ and $\Delta^{ }_T(Q)$ get replaced with $Q^2$, 
so that we are faced with $2g^4 \Tinti{Q} 1$, which vanishes in 
dimensional regularization. At NLO, expanding in $\Pi^{ }_E$, $\Pi^{ }_T$
and substituting $Q\to P-Q$, we obtain
\ba
  \frac{\tilde G_\theta^\rmii{HTL}(P)}{4 d_A c_\theta^2 } +  
 \frac{\tilde G_\chi^\rmii{HTL}(P)}{16 d_A c_\chi^2 (D-3)} & = &
 -  4 g^4  \Tint{Q}
 \biggl\{ 
  \frac{\Pi^{ }_E(Q)}{Q^2}
  + (D-2) 
  \frac{\Pi^{ }_T(Q)}{Q^2}
 \biggr\} 
 \nn & = & 
 -  4 g^4 \mE^2\, \Tint{Q} \frac{1}{Q^2}
 \;,  \la{htl_extra}
\ea
where we inserted the self-energies from \eq\nr{PiT}.
This amounts to a ``contact contribution'': as discussed around 
\eqs\nr{contact_theta}, \nr{contact_chi}, at $\rmO(g^6)$ the 
two channels do differ in this respect. However, since
\eq\nr{htl_extra} is $P$-independent, there is no contribution
to the spectral function.  

%
\subsection{Resummed HTL computation}
\la{ss:resum_HTL}

When \eqs\nr{Gtheta_HTL}, \nr{Gchi_HTL} are kept in their 
unexpanded forms, the resulting spectral functions can be expressed 
as convolutions of two ``elementary'' spectral functions. 
More precisely, if we make use of the spectral representation,  
\be
 \frac{1}{\Delta^{ }_T(Q)} = 
 \int_{-\infty}^{\infty} \! \frac{{\rm d}\omega_1}{\pi} \, 
 \frac{\rho^{ }_T(\omega_1,q)}{\omega_1 - i q_n }
 \;, \quad 
 \frac{1}{\Delta^{ }_T(Q-P)} = 
 \int_{-\infty}^{\infty} \! \frac{{\rm d}\omega_2}{\pi} \, 
 \frac{\rho^{ }_T(\omega_2,q)}{\omega_2 - i (q_n - p_n)}
 \;, \la{spec_rep}
\ee
then (cf.\ e.g.\ ref.~\cite{bp_dilepton})
\ba
 & & \hspace*{-2cm}
 \im \biggl\{ 
   T \sum_{q_n} \frac{1}{\Delta^{ }_T(Q)\Delta^{ }_T(Q-P)}
 \biggr\}_{P \to (-i [\omega + i 0^+],\vec{0})}
 \nn 
 & = &  
 \bigl( 1 + 2 n_{\frac{\omega}{2}} \bigr)
 \int_{-\infty}^{\infty} \! \frac{{\rm d}q^0}{\pi} \, 
 \, \rho^{ }_T(q^0,q) \, \rho^{ }_T(\omega - q^0,q) 
 \, \frac{n_{q^0} n_{\omega - q^0} }{n^2_{\frac{\omega}{2}} }
 \;. \la{rho_rho}
\ea
Let us recall a derivation. 
Inserting \eq\nr{spec_rep} into the left-hand side
of \eq\nr{rho_rho}, the sum can be carried out like in \eq\nr{mat_sum}:
\ba
 T \sum_{q_n} \frac{1}{[\omega_1 - i q_n][\omega_2 - i (q_n - p_n)]}
 & = & 
 T \sum_{q_n} T \sum_{r_n} \frac{\beta \delta_{r_n + p_n - q_n}}
 {(\omega_1 - i q_n)(\omega_2 - i r_n)}
 \nn 
 & = & 
 \int_0^\beta \! {\rm d}\tau \, e^{i p_n\tau}
 \,
 \underbrace{T\sum_{q_n} \frac{e^{-i q_n\tau}}{\omega_1 - i q_n}}_
 {n_{\omega_1} e^{(\beta-\tau)\omega_1}}
 \,
 \underbrace{T\sum_{r_n} \frac{e^{i r_n\tau}}{\omega_2 - i r_n}}_
 {n_{\omega_2} e^{\tau \omega_2}}
 \nn 
 & = & n_{ \omega_1 } n_{\omega_2} \, 
 \frac{e^{\beta\omega_2} - e^{\beta\omega_1}}{ip_n + \omega_2 - \omega_1}
 \;. \la{rho_rho_der1}
\ea
The summations over $q_n$ and $r_n$ here are ``marginally'' 
convergent; their results are discontinuous at $\tau = 0$
and $\tau = \beta$, but there are no Dirac-$\delta$'s, so the 
subsequent integration is unproblematic. Setting $p_n\to -i[\omega + i 0^+]$
and taking the imaginary part yields 
$
 \im\bigl\{ {1}/[{\omega + \omega_2 - \omega_1 + i 0^+}] \bigr\} 
  = -\pi \delta(\omega + \omega_2 - \omega_1)
$,
so that the integral over $\omega_1$ can be  
carried out. Afterwards we write
\ba
 n_{\omega + \omega_2} n_{\omega_2} 
 \Bigl( e^{\beta(\omega + \omega_2)} - e^{\beta\omega_2}\Bigr)
 & = &  
 n_{\omega + \omega_2} \bigl( 1 + n_{\omega_2} \bigr)  
 \bigl( e^{\beta\omega} - 1\bigr)
 \nn 
 & = & - n_{\omega + \omega_2} n_{-\omega_2}
 \frac{ e^{\frac{\beta\omega}{2}} + 1 }{n_{\frac{\omega}{2}} }
 \;. \la{rho_rho_der2}
\ea
Substituting $\omega_2\to - q^0$ and making use of 
$\rho^{ }_T(-q^0,q) = - \rho^{ }_T(q^0,q)$ we recover \eq\nr{rho_rho}.

In order to make use of \eq\nr{rho_rho}, the objects 
appearing should be regular enough to allow for a faithful spectral 
representation; as discussed in \se\ref{ss:sum}, this requires that
the corresponding Euclidean correlators vanish as $p_n\to \infty$.
This is not the case with the terms on the second row of 
\eq\nr{Gtheta_HTL}; however, adding and subtracting $\Pi^{ }$'s
in the numerator, and shifting $Q\to P-Q$ where appropriate, 
it is easy to see that apart from $P$-independent terms the result
can be rewritten as 
\ba
 \frac{\tilde G_\theta^\rmii{HTL}(P)}{4 d_A c_\theta^2 } & = & 
 - 2 (D-2) g^4 p_n^2 
 \Tint{Q} \frac{q^2}{\Delta^{ }_T(Q) 
 \Delta^{ }_T(Q-P)}
 \nn & & \hspace*{-1.5cm} \; + \, 
 2 g^4  \Tint{Q}
 \biggl\{ 
  \frac{\Pi^{ }_E(Q)\Pi^{ }_E(Q-P)}{\Delta^{ }_E(Q) 
 \Delta^{ }_E(Q-P)}
 + (D-2)
   \frac{\Pi^{ }_T(Q) \Pi^{ }_T(Q-P)}{\Delta^{ }_T(Q) 
 \Delta^{ }_T(Q-P)}
 \biggr\} + \mbox{const} 
 \;. \la{Gtheta_HTL_2}  
\ea
Now all individual ``propagators'' decrease as a function
of the four-momentum. 

{}From \eqs\nr{Gchi_HTL}, 
\nr{spec_rep}, \nr{rho_rho}, \nr{Gtheta_HTL_2}  we observe that
the following spectral functions are needed: 
\ba
 & &  \rho^{ }_T (q^0,q) \equiv
 \im \biggl\{ \frac{1}{\Delta^{ }_T(q_n,q)} \biggr\}_{q_n\to -i[q^0+i0^+]}
 \;, \\
 & & 
 \hat\rho^{ }_T (q^0,q) \equiv
 \im \biggl\{ \frac{\Pi^{ }_T(q_n,q)}{\Delta^{ }_T(q_n,q)} 
   \biggr\}_{q_n\to -i[q^0+i0^+]}
 \;, \\ 
 & & 
 \hat\rho^{ }_E (q^0,q) \equiv
 \im \biggl\{ \frac{\Pi^{ }_E(q_n,q)}{\Delta^{ }_E(q_n,q)}
   \biggr\}_{q_n\to -i[q^0+i0^+]}
 \;. 
\ea
To determine these we recall that, 
for $D\to 4$, the $z$-integrals in \eq\nr{PiT} can be carried out: 
\ba
  \Pi^{ }_T(-i[q^0 + i 0^+],\vec{q}) & = & 
  \frac{\mE^2}{2}
  \biggl\{ \frac{(q^0)^2}{q^2} + \frac{q^0}{2q}
  \biggl[ 1 - \frac{(q^0)^2}{q^2} \biggr]
  \ln\frac{q^0 + q + i 0^+}{q^0 - q + i 0^+ } \biggr\}
  \;, \\ 
  \Pi^{ }_E(-i[q^0 + i 0^+],\vec{q}) & = & 
  \mE^2
  \biggl[ 1 - \frac{(q^0)^2}{q^2} \biggr]
  \biggl\{ 1 - \frac{q^0}{2q}
  \ln\frac{q^0 + q + i 0^+}{q^0 - q + i 0^+ } \biggr\}
  \;. 
\ea
Denoting 
\be
 \mathcal{Q} \equiv (q^0,\vec{q}) \;, \quad
 \mathcal{Q}^2 \equiv (q^0)^2 - q^2 \;, \quad
 \eta \equiv \frac{q^0}{q}
 \;, 
\ee
a straightforward computation leads to 
\ba
 \rho^{ }_T(\mathcal{Q}) & = & 
 \left\{
   \begin{array}{ll} 
      \displaystyle\frac{\Gamma^{ }_T(\eta)}
      {\Sigma^2_T(\mathcal{Q})+\Gamma^2_T(\eta)} \;,  &
      |\eta| < 1 \;, \\[3mm]
      \displaystyle
      \pi \mathop{\mbox{sign}} (\eta)
      \, \delta(\Sigma^{ }_T(\mathcal{Q})) \;, & 
      |\eta| > 1 \;, 
   \end{array} 
 \right.  \la{rho_T} \\
 \hat{\rho}^{ }_T(\mathcal{Q}) & = & 
 \left\{
   \begin{array}{ll} 
      \displaystyle\frac{\mathcal{Q}^2\, \Gamma^{ }_T(\eta)}
      {\Sigma^2_T(\mathcal{Q})+\Gamma^2_T(\eta)} \;,  &
      |\eta| < 1 \;, \\[3mm]
      \displaystyle
      \pi \mathop{\mbox{sign}} (\eta)\, \mathcal{Q}^2
      \, \delta(\Sigma^{ }_T(\mathcal{Q})) \;, & 
      |\eta| > 1 \;, 
   \end{array} 
 \right. \\
 \hat{\rho}^{ }_E(\mathcal{Q}) & = & 
 \left\{
   \begin{array}{ll} 
      \displaystyle\frac{q^2\, \Gamma^{ }_E(\eta)}
      {\Sigma^2_E(\mathcal{Q})+\Gamma^2_E(\eta)} \;,  &
      |\eta| < 1 \;, \\[3mm]
      \displaystyle
      \pi \mathop{\mbox{sign}} (\eta)\, q^2
      \, \delta(\Sigma^{ }_E(\mathcal{Q})) \;, & 
      |\eta| > 1 \;, 
   \end{array} 
 \right. \la{rho_E}
\ea
where we have introduced the well-known 
functions~\cite{htl1}--\cite{htl3}
\ba
 \Sigma^{ }_T(\mathcal{Q}) & \equiv & 
 -\mathcal{Q}^2 + \frac{\mE^2}{2}\biggl[
  \eta^2 + \frac{\eta(1-\eta^2)}{2}
  \ln\left| \frac{1+\eta}{1-\eta} \right| \biggr] 
 \;, \la{SigT} \\
 \Gamma^{ }_T(\eta) & \equiv & \frac{\pi \mE^2 \eta (1-\eta^2)}{4}
 \;, \la{GamT} \\
 \Sigma^{ }_E(\mathcal{Q}) & \equiv & 
  q^2 + \mE^2 \biggl[ 1 - \frac{\eta}{2}
  \ln\left| \frac{1+\eta}{1-\eta} \right| \biggr] 
 \;, \la{SigE} \\
 \Gamma^{ }_E(\eta) & \equiv & \frac{\pi \mE^2 \eta}{2}
 \;. \la{GamE} 
\ea
For integrals over the poles it is useful to note that
\ba
  \left. \frac{1}{|\partial_{q^0} \Sigma^{ }_T({\mathcal{Q})}|}
  \right|_{\Sigma^{ }_T(\mathcal{Q}) = 0}
  & = & 
  \frac{|q^0| (\eta^2-1)}{\mE^2 \eta^2 - q^2 (\eta^2 - 1)^2}
  \;, \\
  \left. \frac{1}{|\partial_{q} \Sigma^{ }_T({\mathcal{Q})}|}
  \right|_{\Sigma^{ }_T(\mathcal{Q}) = 0}
  & = & 
  \frac{q  (\eta^2-1)}{\mE^2 \eta^2 - 3 q^2 (\eta^2 - 1)^2}
  \;, \\
  \left. \frac{1}{|\partial_{q^0} \Sigma^{ }_E({\mathcal{Q})}|}
  \right|_{\Sigma^{ }_E(\mathcal{Q}) = 0}
  & = & 
  \frac{|q^0| (\eta^2-1)}{\mE^2 - q^2 (\eta^2 - 1)}
  \;, \\
  \left. \frac{1}{|\partial_{q} \Sigma^{ }_E({\mathcal{Q})}|}
  \right|_{\Sigma^{ }_E(\mathcal{Q}) = 0}
  & = & 
  \frac{q  (\eta^2-1)}{\mE^2 - 3 q^2 (\eta^2 - 1)}
  \;. 
\ea
The approximate pole positions can also be worked out in specific limits: 
$
 \Sigma^{ }_T(\mathcal{Q}) = 0
$
implies
$\eta \approx \frac{\mE}{\sqrt{3} q} + \frac{3\sqrt{3} q}{5\mE}$ 
for  $q \ll \mE$ and 
$\eta \approx 1 + \frac{\mE^2}{4q^2}$ for  $q \gg \mE$,  
whereas
$
 \Sigma^{ }_E(\mathcal{Q}) = 0
$
leads to $\eta \approx \frac{\mE}{\sqrt{3} q}
+ \frac{3\sqrt{3} q}{10\mE} $ for  $q \ll \mE$
and  $\eta \approx 1 + 2 e^{-2({q^2}/{\mE^2}+1)}$ for $q \gg \mE$.

%
\begin{figure}[t]

\centerline{%
~~\epsfysize=7.5cm\epsfbox{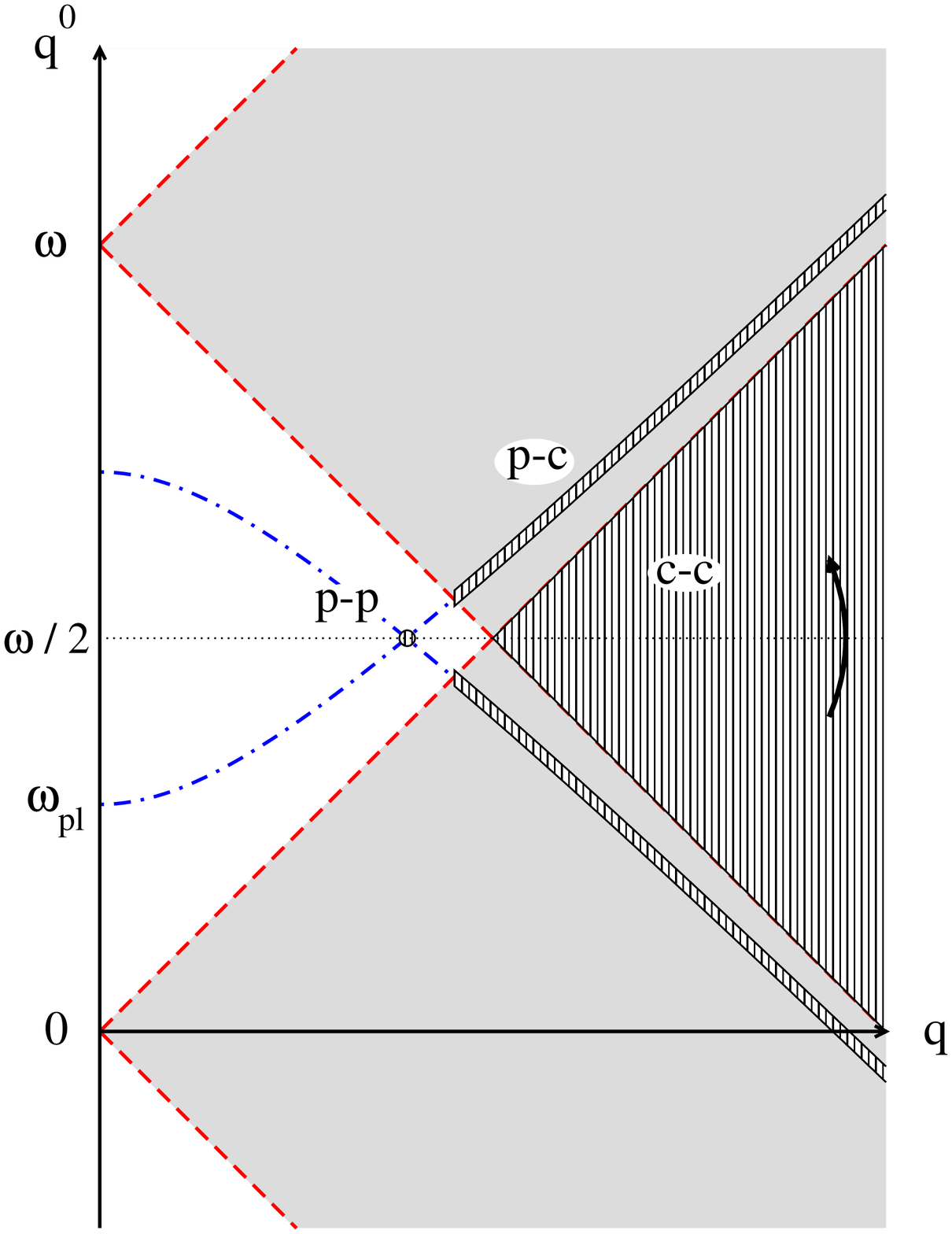}%
}

\caption[a]{\small 
An illustration of the phase space relevant for HTL-resummation. 
The shaded areas and the dash-dotted
blue lines indicate regions in which at least one of the spectral functions 
has non-zero support; the final contribution emerges from the part
where both are simultaneously non-zero, which happens within 
the hashed areas. For $\omega > 2 \omega_\rmi{pl}$, where 
$\omega_\rmi{pl} \equiv \mE / \sqrt{3}$, this leads to 
pole-pole (``p-p''), pole-cut (``p-c'') as well as cut-cut (``c-c'')
contributions (cf.\ ref.~\cite{bp_dilepton}). 
As indicated by the arrow, symmetry allows us
to restrict the integration to the range $q^0 \ge \omega/2$.
} 
\la{fig:dispersion}
\end{figure}
%

With the spectral functions at hand, we are left to insert them
to the imaginary parts of \eqs\nr{Gchi_HTL}, \nr{Gtheta_HTL_2}, 
rewritten through \eq\nr{rho_rho}: 
\ba
 \frac{\rho_\theta^\rmii{HTL}(\omega)}{4d_Ac_\theta^2} & = & 
 \frac{\pi g^4 \bigl( 1 + 2 n_{\frac{\omega}{2} } \bigr)}{(4\pi)^2}
 \biggl\{ 
   \frac{16}{\pi^2}
   \int_0^\infty \! {\rm d}q \, q^2 \int_{-\infty}^{\infty} \! {\rm d}q^0 \, 
 \nn & & \times
   \Bigl[ 2 \Bigl( \omega^2 q^2  + \mathcal{Q}^2((\omega - q^0)^2 - q^2) 
            \Bigr) 
             \rho^{ }_T(q^0,q) \rho^{ }_T(\omega - q^0,q)
 \nn & & \hspace*{5.2cm} + \, 
       \hat{\rho}^{ }_E(q^0,q) \hat{\rho}^{ }_E(\omega - q^0,q)
   \Bigr]
   \frac{n^{ }_{q^0}n^{ }_{\omega - q^0}}{n^2_{\frac{\omega}{2}}}
 \biggr\} 
 \;, \nn \la{HTL_theta_final} \\ 
 \frac{-\rho_\chi^\rmii{HTL}(\omega)}{16d_Ac_\chi^2} & = & 
 \frac{\pi g^4 \bigl( 1 + 2 n_{\frac{\omega}{2} } \bigr)}{(4\pi)^2}
 \biggl\{ 
   \frac{32 \omega^2}{\pi^2}
   \int_0^\infty \! {\rm d}q \, q^4 \int_{-\infty}^{\infty} \! {\rm d}q^0 \, 
   \rho^{ }_T(q^0,q)
   \rho^{ }_T(\omega - q^0,q)
   \frac{n^{ }_{q^0}n^{ }_{\omega - q^0}}{n^2_{\frac{\omega}{2}}}
 \biggr\} 
 \;, \nn \la{HTL_chi_final}
\ea
where we have identified the same prefactor
as in \eq\nr{nlo_htl}. Given the structures of 
\eqs\nr{rho_T}--\nr{rho_E}, these integrals include three types 
of contributions, illustrated in \fig\ref{fig:dispersion}
(the situation is similar to that in ref.~\cite{bp_dilepton}).
Taking $\rho_\chi^\rmii{HTL}$ as an example, 
the ``pole--pole'' term, which only contributes 
for $\omega > 2 \omega_\rmi{pl}$, 
where $\omega_\rmi{pl} \equiv \mE/\sqrt{3}$
denotes the ``plasmon frequency'', amounts to 
\ba
 & & \hspace*{-1cm}
   \frac{32 \omega^2}{\pi^2}
   \int_0^\infty \! {\rm d}q \, q^4 \int_{0}^{\infty} \! {\rm d}q^0 \, 
   \,\pi \delta (\Sigma^{ }_T(q^0,q))
   \, \pi \delta(\Sigma^{ }_T(\omega - q^0,q))
   \frac{n^{ }_{q^0}n^{ }_{\omega - q^0}}{n^2_{\frac{\omega}{2}}}
 \nn & = & 
 32 \omega^2 \int_0^\infty \! {\rm d}q \, q^4
   \frac{1}{|\partial^{ }_{q^0} \Sigma^{ }_T(q^0,q)|}
   \delta(\Sigma^{ }_T(\omega - q^0,q))
   \left.
   \frac{n^{ }_{q^0}n^{ }_{\omega - q^0}}{n^2_{\frac{\omega}{2}}}
   \right|_{\Sigma^{ }_T(q^0,q) = 0}
 \nn & = & 
 32 \omega^2  q^4 \left.
   \frac{1}{|\partial^{ }_{q^0} \Sigma^{ }_T(q^0,q)|}
   \frac{1}{| 
     -\partial^{ }_{q^0} \Sigma^{ }_T  \,  \frac{ {\rm d}q^0}{{\rm d}q} + 
      \partial^{ }_q \Sigma^{ }_T
   |} 
   \right|_{\Sigma^{ }_T(q^0,q) = 0 ,  \; q^0 = \frac{\omega}{2}}
 \nn & = & 
 16 \omega^2  q^4 \left.
   \frac{1}{|\partial^{ }_{q^0} \Sigma^{ }_T(\frac{\omega}{2},q)|}
   \frac{1}{| 
      \partial^{ }_q \Sigma^{ }_T(\frac{\omega}{2},q)
   |} 
   \right|_{\Sigma^{ }_T(\frac{\omega}{2},q) = 0}
 \;. 
\ea
Here we made use of the fact that along the curve on which 
$\Sigma^{ }_T(\mathcal{Q}) = 0$, 
$
 \frac{ {\rm d}q^0}{{\rm d}q} = 
 -\frac{\partial_q \Sigma^{ }_T}{\partial_{q^0}\Sigma^{ }_T}
 \;.
$

For the ``pole--cut'' contribution, making use of the reflection
symmetry with respect to the axis $q^0 = \frac{\omega}{2}$
(cf.\ \fig\ref{fig:dispersion}), we similarly get
\ba
 & & \hspace*{-1cm}
   \frac{64 \omega^2}{\pi^2}
   \int_0^\infty \! {\rm d}q \, q^4 \int_{0}^{\infty} \! {\rm d}q^0 \, 
   \,\pi \delta (\Sigma^{ }_T(q^0,q))
   \, \frac{\Gamma^{ }_T(\omega - q^0,q)}
      {\Sigma^{2}_T(\omega - q^0,q)+\Gamma^{2}_T(\omega - q^0,q)}
   \frac{n^{ }_{q^0}n^{ }_{\omega - q^0}}{n^2_{\frac{\omega}{2}}}
 \nn & = & 
 \frac{64 \omega^2}{\pi} \int_{q_\rmii{min}}^\infty \! {\rm d}q \, q^4
   \frac{1}{|\partial^{ }_{q^0} \Sigma^{ }_T(q^0,q)|}
   \, \frac{\Gamma^{ }_T(\omega - q^0,q)}
      {\Sigma^{2}_T(\omega - q^0,q)+\Gamma^{2}_T(\omega - q^0,q)}
   \left.
   \frac{n^{ }_{q^0}n^{ }_{\omega - q^0}}{n^2_{\frac{\omega}{2}}}
   \right|_{\Sigma^{ }_T(q^0,q) = 0}
  \hspace*{-1cm} \;,\hspace*{1cm}  \la{pc}
\ea
where the value of $q_\rmi{min}$ is determined from 
$\Sigma^{ }_T(\omega - q_\rmi{min},q_\rmi{min}) = 0$ for 
$\omega > \omega_\rmi{pl}$, and 
$\Sigma^{ }_T(\omega + q_\rmi{min},q_\rmi{min}) = 0$ for 
$\omega < \omega_\rmi{pl}$. 
Finally, the ``cut--cut'' term reads
\ba
 & & \hspace*{-1cm}
   \frac{64 \omega^2}{\pi^2}
   \int_{\frac{\omega}{2}}^\infty \! {\rm d}q \, q^4 
   \int_{\frac{\omega}{2}}^{q} \! {\rm d}q^0 \, 
   \, \frac{\Gamma^{ }_T(q^0,q)}
      {\Sigma^{2}_T(q^0,q)+\Gamma^{2}_T(q^0,q)}
   \, \frac{\Gamma^{ }_T(\omega - q^0,q)}
      {\Sigma^{2}_T(\omega - q^0,q)+\Gamma^{2}_T(\omega - q^0,q)}
   \frac{n^{ }_{q^0}n^{ }_{\omega - q^0}}{n^2_{\frac{\omega}{2}}}
  \;. \nn \la{cc}
\ea

In a systematic HTL-computation, we are
furthermore to replace the Bose distributions 
in \eqs\nr{pc}, \nr{cc} through their ``classical'' limits, 
$n^{ }_{q^0} \to T/ q^0$ etc., so that 
$
   {n^{ }_{q^0}n^{ }_{\omega - q^0}} / {n^2_{\frac{\omega}{2}}}
   \to \frac{\omega^2}{4 q^0(\omega - q^0)}
$.
Thereby the HTL-results 
only feel the scale $\frac{\omega}{\mE}$; 
this is in analogy with the thermal corrections, 
which only depend on $\frac{\omega}{\pi T}$. 

\newpage


\end{document}